\documentclass[a4paper,11pt, oneside,titlepage=false]{scrbook}
\usepackage[utf8]{inputenc}

\usepackage{tikz-cd}
\usepackage{xcolor}
\usepackage{amssymb}
\usepackage{amsmath}
\usepackage{amsthm}
\usepackage{thmtools}

\usepackage{stmaryrd} 
\usepackage{verbatim}
\usepackage{ccicons}
\usepackage{enumitem}

\usepackage{bussproofs}
\newcommand{\proofskip}{0.5em}
\newcommand{\rulelabel}[1]{\hypertarget{#1}{\textsf{\scriptsize #1}}}

\newenvironment{bproof}
{\leavevmode\hbox\bgroup}
{\DisplayProof\egroup}
\newcommand{\binaryRule}[4]{\begin{bproof}
\AxiomC{\ensuremath{#1}}
\AxiomC{\ensuremath{#2}}
\RightLabel{\rulelabel{#4}}
\BinaryInfC{\ensuremath{#3}}
\end{bproof}\vspace{\proofskip}}

\usepackage[colorinlistoftodos,prependcaption,textsize=tiny]{todonotes}

\newcommand{\plan}[1]{}
\newcommand{\BA}[1]{}
\newcommand{\KW}[1]{}

\usepackage[style=numeric,backend=biber]{biblatex}
\usepackage{listings}
\lstMakeShortInline|
\protected\def\ARROW{\ensuremath{\to}}
\DeclareUnicodeCharacter{2192}{\ARROW}
\protected\def\EX{\ensuremath{\exists}}
\DeclareUnicodeCharacter{2203}{\EX}
\protected\def\FORALL{\ensuremath{\forall}}
\DeclareUnicodeCharacter{2200}{\FORALL}
\protected\def\AND{\ensuremath{\wedge}}
\DeclareUnicodeCharacter{2227}{\AND}
\protected\def\OR{\ensuremath{\vee}}
\DeclareUnicodeCharacter{2228}{\OR}
\protected\def\NOT{\ensuremath{\neg}}
\DeclareUnicodeCharacter{00AC}{\NOT}
\protected\def\TURNSTYLE{\ensuremath{\vdash}}
\DeclareUnicodeCharacter{22A2}{\TURNSTYLE}
\protected\def\TIMES{\ensuremath{\times}}
\DeclareUnicodeCharacter{00D7}{\TIMES}
\protected\def\LAMBDA{\ensuremath{\lambda}}
\DeclareUnicodeCharacter{03BB}{\LAMBDA}

\usepackage{hyperref}
\usepackage[capitalize]{cleveref}

\theoremstyle{plain}
\newtheorem{thm}{Theorem}
\crefname{thm}{theorem}{theorems}
\Crefname{thm}{Theorem}{Theorems}
\newtheorem{prop}[thm]{Proposition}

\newtheorem{lemma}[thm]{Lemma}

\newtheorem*{reading*}{Further Reading}

\theoremstyle{definition}
\newtheorem{rem}[thm]{Remark}
\newtheorem{dfn}[thm]{Definition}
\crefname{dfn}{Definition}{Definitions}
\Crefname{dfn}{Definition}{Definitions}

\newtheorem{exa}[thm]{Example}
\crefname{exa}{Example}{Examples}
\Crefname{exa}{Example}{Examples}
\newtheorem{exer}[thm]{Exercise}
\crefname{exer}{Exercise}{Exercises}
\Crefname{exer}{Exercises}{Exercises}

\newtheorem{intu}[thm]{Intuition}
\newtheorem{explanation}[thm]{Explanation}

\newtheorem{nota}[thm]{Notation}

\newtheorem{solution}[thm]{Solution}

\newcommand{\cfont}[1]{\ensuremath{\mathsf{#1}}}
\newcommand{\Cat}[1]{\mathcal{#1}}
\newcommand{\CC}{\Cat{C}}
\newcommand{\DD}{\Cat{D}}
\newcommand{\EE}{\Cat{E}}
\newcommand{\Catb}[1]{\mathbf{#1}}
\newcommand{\List}{\Catb{List}}
\newcommand{\BinTree}{\Catb{BinTree}}
\newcommand{\Maybe}{\Catb{Maybe}}
\newcommand{\SET}{\Catb{Set}}
\newcommand{\PTSET}{\Catb{PtSet}}
\newcommand{\FINSET}{\Catb{FinSet}}
\newcommand{\CAT}{\Catb{Cat}}
\newcommand{\POS}{\Catb{Pos}}

\newcommand{\PREtoCAT}{\Catb{Pre2Cat}}
\newcommand{\MONtoCAT}{\Catb{Mon2Cat}}
\newcommand{\MON}{\Catb{Mon}}
\newcommand{\HASK}{\Catb{Hask}}

\newcommand{\COQ}{\Catb{Coq}}
\newcommand{\MAT}{\Catb{Mat}}
\newcommand{\SKELFINSET}{\Catb{Fin}}
\newcommand{\REL}{\Catb{Rel}}
\newcommand{\ALG}[1]{\Cat{A}lg(#1)}
\newcommand{\COALG}[1]{\Cat{C}o\Cat{A}lg(#1)}
\newcommand{\Ob}[1]{{#1}_0}
\newcommand{\Hom}[3][]{\cfont{hom}_{#1}(#2,#3)}
\newcommand{\CHom}[3]{{#1}(#2,#3)}
\newcommand{\Id}[1][]{\cfont{Id}_{#1}}
\newcommand{\Comp}{\cdot}
\newcommand{\NatTrans}[3]{#1 : #2 \Rightarrow #3}
\newcommand{\op}[1]{\ensuremath{{#1}^\text{op}}}
\newcommand{\inl}{\ensuremath{\iota_l}}
\newcommand{\inr}{\ensuremath{\iota_r}}
\newcommand{\projl}{\ensuremath{\pi_l}}
\newcommand{\projr}{\ensuremath{\pi_r}}
\newcommand{\intoproduct}[2]{\ensuremath{\langle #1, #2 \rangle}}
\newcommand{\outofcoproduct}[2]{\ensuremath{[ #1, #2 ]}}

\newcommand{\productmap}[2]{\ensuremath{ #1 \times #2 }}
\newcommand{\coproductmap}[2]{\ensuremath{ #1 + #2 }}
\newcommand{\rec}{\ensuremath{\mathsf{rec}}}

\newcommand{\Initalg}[1]{\ensuremath{\mu^{#1}}}
\newcommand{\Terminalcoalg}[1]{\ensuremath{\nu^{#1}}}
\newcommand{\Inv}[1]{#1 ^{-1}}
\newcommand{\catam}[1]{\llparenthesis #1 \rrparenthesis}
\newcommand{\anam}[1]{\llbracket #1 \rrbracket} 
\newcommand{\In}{\ensuremath{\mathsf{in}}}
\newcommand{\Out}{\ensuremath{\mathsf{out}}}

\newcommand{\nil}{\ensuremath{\mathsf{nil}}}
\newcommand{\cons}{\ensuremath{\mathsf{cons}}}

\newcommand{\co}[2]{\ensuremath{#2 \circ #1}}

\newcommand{\NN}{\ensuremath{\mathbb{N}}}
\newcommand{\Zero}{\ensuremath{\mathsf{zero}}}
\newcommand{\Succ}{\ensuremath{\mathsf{succ}}}

\newcommand{\Bool}{\ensuremath{\mathsf{Bool}}}
\newcommand{\QQ}{\ensuremath{\mathbb{Q}}}

\newcommand{\Leaf}[1]{\ensuremath{\mathsf{leaf}(#1)}}
\newcommand{\Branch}[2]{\ensuremath{\mathsf{branch}(#1, #2)}}

\newcommand{\Colist}{\ensuremath{\mathsf{Colist}}}
\newcommand{\Stream}{\ensuremath{\mathsf{Stream}}}
\newcommand{\tail}{\ensuremath{\mathsf{tail}}}
\newcommand{\head}{\ensuremath{\mathsf{head}}}
\newcommand{\nats}{\ensuremath{\mathsf{nats}}}
\newcommand{\zip}{\ensuremath{\mathsf{zip}}}
\newcommand{\map}{\ensuremath{\mathsf{map}}}
\newcommand{\filter}{\ensuremath{\mathsf{filter}}}

\newcommand{\const}{\ensuremath{\mathsf{const}}}

\title{Category Theory for Programming}
\author{Benedikt Ahrens \and Kobe Wullaert}
\date{}

\begin{document}

\maketitle

\vspace*{\fill}
We thank Csanád Farkas, Arnoud van der Leer, Niyousha Najmaei, and Pepijn Vunderink for their contributions to these notes.
\vspace*{\fill}

\noindent
This work is licensed under a Creative Commons \href{https://creativecommons.org/licenses/by-sa/4.0/}{``Attribution-ShareAlike 4.0 International (CC BY-SA 4.0)''} license. \href{https://creativecommons.org/licenses/by-sa/4.0/}{\ccbysa}

\noindent
The source code for this document can be obtained from
\begin{center}
\url{https://github.com/benediktahrens/CT4P}.
\end{center}

\newpage

\paragraph*{Abstract}
In these lecture notes, we give a brief introduction to some elements of category theory.
The choice of topics is guided by applications to functional programming.
Firstly, we study initial algebras, which provide a mathematical characterization of datatypes and recursive functions on them.
Secondly, we study monads, which give a mathematical framework for effects in functional languages.
The notes include many problems and solutions.

\newpage
\tableofcontents
\newpage

\chapter{Introduction}

\section{About Category Theory}
\emph{Category theory} is a mathematical area of endeavour and language developed to reconcile and unify mathematical phenomena from different disciplines.
It was developed from the 1940s on, in particular by Samuel Eilenberg and Saunders Mac Lane.
Category theory studies objects by studying the way they interact with other objects of the same kind.
Its power arises from \emph{abstraction}:
by boiling down constructions to their essence, analogous situations can be formally identified using category theory.

The objects we can study using category theory are diverse;
while the origins of category theory lie in pure mathematics, it has now found applications throughout mathematics, computer science, and other areas of endeavor.
In these notes we introduce some basic concepts from category theory.
The choice of topics is guided by applications in the area of \emph{(functional) programming}; firstly, the structuring of effectful programs using \emph{monads} (see \cref{sec:monads}), and, secondly, the mathematical justification and derivation of \emph{recursion principles} such as \verb!fold! for the datatype of lists (see \cref{sec:initial-algs}).
We also study infinite datatypes (see \cref{sec:coinductive}).

\section{Learning Material on Category Theory}
\label{sec:material}

There are many topics of category theory not discussed in these notes.
Below, we list some resources that are freely available;
a very extensive list of resources is given at \url{https://www.logicmatters.net/categories/}.

\begin{itemize}
\item Pierce's book \cite{pierce} (available for free) gives a brief introduction to category theory with some applications to computing.

\item Leinster's book \cite{leinster} (available for free online, under a free license) gives a concise introduction to category theory.
It is a good resource for the basic concepts, but does not feature many examples from computer science.

\item The rather substantial textbook by Barr and Wells \cite{barr-wells} (available for free online) covers a lot more than we are going to discuss in these notes.

\item The Catsters \cite{catsters} provide a lecture series on category theory on YouTube.
  
\item A list of resources on category theory is maintained at \url{https://www.logicmatters.net/categories/}.
\end{itemize}

Throughout these notes, pointers to further sources, such as textbooks and research articles, are given.

\chapter{Brief Summary of Logical Foundations}
\label{sec:notation}

In these notes, we use set theory, \emph{informally}, as the language in which we define the concepts of category theory.

\section{About set theory}

\begin{description}
\item [Sets/Collections] If $X$ is a set and $x$ is an element of $X$, we write $x \in X$.
  To define a set, we usually specify its elements, in an informal way.
  For instance, we consider the set $\NN$ of all natural numbers, with, e.g., $5 \in \NN$ and $37 \in \NN$.
  The set $\QQ$ is the set of all rational numbers, with, e.g., $2/3 \in \QQ$.

  We also consider a ``set of all sets''. Its elements are all sets.
  Here, we should be careful, as the set of all sets cannot be an element of itself; see the Barber paradox.\footnote{\url{https://en.wikipedia.org/wiki/Barber_paradox}}
  To fix this issue, we consider \emph{collections} in addition to \emph{sets};
  the ``set'' of all sets is not a set, but a collection instead.

  Of course, the collection of all collections gives rise to the same problem;
  so, when considering foundations of mathematics very formally, one usually has an infinite hierarchy of sets or collections (or types, e.g., in Martin-Löf type theory).
  
\item [Functions] If $X$ and $Y$ are sets, we can consider the set of functions from $X$ to $Y$. This set is denoted $X \to Y$. We write $f \in X \to Y$, but also $f : X \to Y$, to say that $f$ is a function from $X$ to $Y$.
  If $x \in X$, then $f(x) \in Y$ denotes the image of $x$ under $f$.
  When defining a function, we use the notation $x \mapsto f(x)$ to describe the image of $x$ under $f$. For instance, we define the function
  \begin{align*}
    \_/3 : \NN &\to \QQ
    \\
    n & \mapsto n/3
  \end{align*}
  This function takes as input a natural number, say, $n \in \NN$, and gives as output the rational number $n/3 \in \QQ$.
  
\item [Cartesian Product] Given two sets $X$ and $Y$, we can form a new set $X\times Y$, the \emph{(cartesian) product} of $X$ and $Y$. Its elements are pairs, written $(x,y)$, of an element $x \in X$ and $y \in Y$.
  For instance, we can write addition of natural numbers as a function
  \begin{align*}
    (+) : \NN\times \NN &\to \NN
    \\
    (m,n) & \mapsto m + n
  \end{align*}
\item[(Disjoint) Union] Given two sets $X$ and $Y$, we can form a new set $X + Y$, the disjoint union of $X$ and $Y$. Its elements are those of $X$ and those of $Y$, with information about whether an element ``comes from'' $X$ or $Y$. (To be more concrete, one could say that elements of $X+Y$ are of the form $(0,x)$ for $x\in X$ or $(1,y)$ for $y\in Y$.)
\end{description}

\section{About logic}

We also use logical connectives in the definition of categorical concepts.

\begin{description}
\item[Definitions] When we write $x := y$, we define a new mathematical object $x$ to be given by $y$.
  For instance, we could define a function doubling its input as follows:
  \begin{align*}
    \mathsf{double} &: \NN \to \NN
    \\
    \mathsf{double} &:=  x \mapsto x + x
  \end{align*}
  
\item[Equality] Given a set $X$ and $a,b\in X$, we can ask whether $a$ is equal to $b$; we write $a = b$ to assert that $a$ is equal to $b$.
  To \emph{prove} that two elements $a$ and $b$ of a set are equal, we have to understand what $a$ and $b$ are. For numbers $a$ and $b$, we intuitively know when they are equal.
  When proving that two functions $f, g : X \to Y$ are equal, we use the \emph{axiom of function extensionality}: to show that $f = g$, it suffices to show that for any possible input $x \in X$, the functions $f$ and $g$ agree on that input, that is, $f(x) = g(x)$.
  When showing that two monoid morphisms are equal, it suffices to show that their underlying functions are equal.
  When showing that two matrices are equal, it suffices to show that they are equal in each component.
\item [Implication]
  Given propositions $A$ and $B$, we write $A \implies B$ to say that $A$ implies $B$.
  
  More generally, if $X$ is a set and $P$ and $Q$ are predicates on $X$, we write $P\implies Q$ to express that if $P(x)$ holds for an element $x\in X$, then also $Q(x)$  holds for the element $x$.
  As an example, consider the predicates on natural numbers $n \in \NN$, ``$n$ is divisible by $4$'' and ``$n$ is even''.
  Then ``divisible by $4$'' $\implies$ ``is even''.

  Moreover, we write $P \iff Q$ if $P\implies Q$ and $Q\implies P$.
\item If $X$ is a set and $P$ is a predicate on $X$, we write:
\begin{description}
\item[forall] $\forall x\in X, P(x)$ to express that for every element in $X$, the property $P$ holds.
\item[exists] $\exists x\in X, P(x)$ to express that there exists at least one element in $X$ for which the property holds.
  For instance, consider the predicate
  \[  \mathsf{even}(n) := \exists k \in \NN, n = 2\times k.\]
\item[exists uniquely] $\exists! x\in X, P(x)$ to express that there exists a unique element in $X$ for which the property holds.
\end{description}
\end{description}

As an example, we can write the axiom of function extensionality as
\[ \big(\forall x \in X, f(x) = g(x)\big) \implies f = g \]

\section{About mathematical writing}

Mathematical text is typically structured into numbered ``environments'', for instance, ``definitions'', ``lemmas'', ``theorems'', and ``proofs''.
We briefly discuss each of these environments.

\begin{description}
\item[Definition] A definition introduces a new mathematical concept. You can think of a definition as an abbreviation, e.g., for a large structure consisting of many components.
  For instance, in \cref{dfn:category}, we define a category to be a structure consisting of objects, morphisms, etc.
  From then on, we use ``category'' as a shorthand for such a ``bundle'' of data satisfying the properties of a category.

  In the beginning, \textbf{it is important to learn definitions by heart.} Later, when you have understood the concepts well, you will be able to reproduce their definitions without explicitly learning them by heart.
\item[Example] An example is, usually, an instance of a definition. In \cref{example:set} (and \cref{lemma:set-category}), we look at specific data (sets and functions between them) and establish that they form a category.
\item[Lemma/Theorem/Proposition] All of these environments serve to formulate, precisely, a mathematical statement.
  The precise choice of environment signifies the status of the statement.
  A lemma is often of technical nature, a theorem is important, and a proposition is less important.
\item[Exercise] You are encouraged to think hard about every exercise in these notes. Some exercises are accompanied by solutions. It is worth first writing down your own solution, and then comparing it with the given solution.
\end{description}

\chapter{Categories}
\label{sec:categories}

\begin{reading*}
The definition of categories is also given in \cite[\S 2.1]{barr-wells}. Plenty of examples of categories are given in \cite[\S\S 2.3--2.5]{barr-wells}.

  The definition of categories is also given in \cite[\S 1.1]{leinster}, together with some examples.
  There, also isomorphisms are discussed, which we define in \cref{sec:isos}.

  The tutorial \cite{pierce} features the definition of categories in \cite[\S 2.1]{pierce}.
  It also introduces the notion of ``diagram'', which we do not use in the present notes.
\end{reading*}

\section{Definition}

\begin{dfn}\label{dfn:category}
  A \textbf{category} $\CC$ consists of the following data:
\begin{enumerate}
\item A collection of objects, denoted by $\Ob{\CC}$;
\item For any given objects $X,Y \in \Ob{\CC}$, a collection of morphisms from $X$ to $Y$, denoted by $\Hom[\CC]{X}{Y}$ (or $\Hom{X}{Y}$ when the category $\CC$ is clear, or $\CHom \CC X Y$ or $X \to Y$);
\item For each object $X\in \Ob{\CC}$, a morphism $\Id[X] \in \Hom[C]{X}{X}$, called the \emph{identity morphism} on $X$;
\item A binary operation
\[
(\co{}{})_{X,Y,Z} : \Hom{Y}{Z} \to \Hom X Y \to \Hom X Z,
\]
called the \emph{composition}, and written infix without the indices $X,Y,Z$ as in $\co{f}{g}$.
\end{enumerate}
Moreover, this data should satisfy the following properties:
\begin{enumerate}
\item (\textbf{Left unit law}) For any morphism $f \in \Hom X Y$, we have 
\[
 \co{\Id[X]} {f} = f.
\]
\item (\textbf{Right unit law}) For any morphism $f \in \Hom X Y$, we have 
\[
  \co f {\Id[Y]} = f.
\]
\item (\textbf{Associative law}) For any morphisms $f\in \Hom X Y$, $g\in \Hom Y Z$ and $h\in \Hom Z W$, we have
\[
     \co {(\co f g)}{h} =  \co f {(\co g  h)}.
\]
\end{enumerate}
\end{dfn}


\begin{nota} Let $\CC$ be a category.
\begin{itemize}
\item We write $X\in\CC$ instead of $X\in \Ob{\CC}$. 
\item Let $X,Y\in \CC$ be objects. A morphism $f\in\CHom{\CC}{X}{Y}$ can be visualized as \[ X \xrightarrow{f} Y. \]
\item Let $X,Y, Z \in \Ob{\CC}$ objects in $\CC$ and consider the following morphisms:
\[
f\in\CHom{C}{X}{Y}, \quad g\in\CHom{C}{Y}{Z}, \quad h\in\CHom{C}{X}{Z}.
\]
These morphisms can be visualized as a triangle:
\[
\begin{tikzcd}
X \arrow[r, "f"] \arrow[dr,swap, "h"] & Y \arrow[d, "g"] \\
& Z
\end{tikzcd}
\]
We say that such a triangle \textbf{commutes} if $h = \co{f}{g}$.
\item Let $X,Y_1,Y_2, Z \in \Ob{\CC}$ objects in $\CC$ and consider the following morphisms:
\[
f_1\in\CHom{C}{X}{Y_1}, \quad f_2\in\CHom{C}{X}{Y_2}, \quad g_1\in\CHom{C}{Y_1}{Z}, \quad g_2\in\CHom{C}{Y_2}{Z}.
\]
These morphisms can be visualized as a square:
\[
\begin{tikzcd}
X \arrow[r, "f_2"] \arrow[d,swap, "f_1"] & Y_2 \arrow[d, "g_2"] \\
Y_1 \arrow[r, swap, "g_1"] & Z 
\end{tikzcd}
\]
We say that such a square \textbf{commutes} if $\co{f_1}{g_1} = \co{f_2}{g_2}$.
\end{itemize}
\end{nota}

\section{Examples of categories}

\begin{exa}\label{example:set} The \textbf{category of sets}, denoted by $\SET$, is the category specified by the following data:
\begin{itemize}
\item An object is a set.
\item If $X$ and $Y$ are sets, then is $\CHom \SET X Y$ the set of all functions from $X$ to $Y$.
\item The identity morphism $\Id[X]$ (on $X\in\Ob{\SET}$) is the identity function on $X$, i.e.
\begin{align*}
  \Id[X] : X &\to X
  \\
  x &\mapsto x.
\end{align*}
\item The composition of functions is given by the usual composition of functions, i.e. for $f\in \CHom \SET X Y$ and $g\in \CHom \SET Y Z$, the composition of $f$ and $g$ is:
\begin{align*}
  g \circ f : X&\to Z
  \\
  x &\mapsto g(f(x)).
\end{align*}
\end{itemize}
\end{exa}
\begin{lemma}\label{lemma:set-category}
  The data of $\SET$ satisfies the properties of a category; hence $\SET$ is indeed a category.
\end{lemma}
\begin{proof}
  We first show that the left unit law holds. Let $X,Y\in \mathbf{Set}$ be sets and $f\in \CHom \SET X Y$ a function. We have to show that $\co {\Id[X]}{f} = f$; hence it suffices to show that they are pointwise equal.
  To show this, we fix an arbitrary $x \in X$, and compute
\[
   (f\circ \Id[X])(x) = f\left(\Id[X](x)\right) = f(x),
\]
where the first (resp. second) equality holds by definition of the composition (resp. identity morphism).\\
That the right unit law holds is analogous. To show that the associator law holds, let $X,Y,Z,W\in\mathbf{Set}$ and $f\in \CHom \SET X Y, g\in \CHom \SET Y Z$ and $h\in \CHom \SET Z W$. We have to show $h\circ (g\circ f) = (h\circ g)\circ f$; hence it suffices again to show that they are pointwise equal.
To show this, we fix an arbitrary $x \in X$, and compute
\begin{eqnarray*}
  \left(h\circ (g\circ f)\right)(x) &=& h\left((g\circ f)(x)\right)
  \\ 
                                    &=& h(g(f(x)))
  \\ 
                                    &=& (h\circ g)(f(x))
  \\ 
                                    &=& \left((h\circ g)\circ f\right)(x),
\end{eqnarray*}
where the first (resp. second, third, fourth) equality holds by definition of the composition of $h$ and $g\circ f$ (resp. composition of $g$ and $f$, composition of $h$ and $g$, composition of $h\circ g$ and $f$).
\end{proof}

We are now going to describe the category whose collection of objects is given by collection of Coq types:
\begin{exa}\label{exa:coq-cat}
  Consider the following data: 
\begin{itemize}
\item An object is a Coq type (of some fixed universe).
\item If $X$ and $Y$ are Coq types, then is $\CHom \COQ X Y$ the function type $X\to Y$.
\item The identity morphism $\Id[X]$ (on $X\in \Ob{\COQ}$) is the identity function on $X$, i.e.
\begin{lstlisting}
Definition idfun {X} : X -> X := fun x => x.
\end{lstlisting}
\item The composition of functions is given by the composition of functions:
\begin{lstlisting}
Definition compfun {X Y Z} (f : X -> Y) (g : Y -> Z) : X -> Z
:= fun x => g (f x).
\end{lstlisting}
\end{itemize}
  Try it out, e.g., on \url{https://jscoq.github.io/scratchpad.html}:
\begin{lstlisting}
Eval compute in (compfun (fun x => x + 1) (fun x => x * 3) 5).
\end{lstlisting}  
\end{exa}

\begin{exer}
  Prove (on paper) that the data defined in \cref{exa:coq-cat} defines a category.
  That is, show that it satisfies the axioms of a category.
  You might need to use the \textbf{axiom of functional extensionality}:
\begin{lstlisting}
Axiom functional_extensionality: forall {A B} (f g : A -> B),
  (forall x, f x = g x) -> f = g.
\end{lstlisting}
\end{exer}

\begin{exa}
  We repeat the definitions of \cref{exa:coq-cat} in Haskell instead of Coq.
  Does this data satisfies the axioms of a category?

  Due to Haskell allowing for the |undefined| value in each type, the situation is slightly more complicated; consider the following two functions:
\begin{lstlisting}
undef1 :: a -> a
undef1 = undefined

undef2 :: a -> a
undef2 = \x -> undefined
\end{lstlisting}
These are not equal by definition, but we have $\Id \Comp$ |undef1| $=$ |undef2|.
So by the right unit law, we must have that |undef1| = |undef2| (as morphisms in our sought category).
\end{exa}


\begin{exer}
  Read the Haskell wiki page on the category $\HASK$ \cite{haskell-wiki-hask}.
\end{exer}

However, when considering functions to be equal when they are \textbf{pointwise} equal, we can define a category of Haskell types:
\begin{dfn}\label{example:hask} The \textbf{category of Haskell types}, denoted by $\HASK$, is the category specified by the following data:
\begin{itemize}
\item An object is a Haskell type.
\item If $X$ and $Y$ are Haskell types, then is $\CHom \HASK X Y$ the collection of functions modulo the equivalence relation $\sim$ defined by identifying pointwise equal functions:
\[
f \sim g :\iff \forall x : X, f(x) = g(x).
\]
i.e. a morphism in $\HASK$ is an equivalence class of (Haskell) functions.
\item The identity morphism $\Id[X]$ (on $X\in\HASK$) is the equivalence class of the identity function on $X$.
\item The composition of (Haskell) functions is given by the equivalence class of the composition of functions, i.e., for $f\in \CHom \HASK X Y$ and $g\in \CHom \HASK Y Z$, the composition of $f$ and $g$ is the equivalence class of:
\[g\circ f : X\to Z: \lambda x. g(f(x)).\]
\end{itemize}
\end{dfn}

\begin{exa}\label{example:posetcategories}
Recall that a \textit{preordered set} $(X,\leq)$ consists of a set $X$ together with a binary relation $(\leq)$ on $X$ which satisfies the following properties:
\begin{itemize}
\item \textbf{Reflexivity}: $\forall x\in X: x\leq x$.
\item \textbf{Transitivity}: $\forall x,y,z\in X: \left(x\leq y \wedge y\leq z\right) \implies x\leq z$.
\end{itemize}

  Let $(X,\leq)$ be a preordered set. We define the category $\PREtoCAT(X,\leq)$ as follows:
\begin{itemize}
\item The objects are the elements of $X$.
\item Let $x,y \in X$ be elements. The hom-set $\Hom x y$ consists of a unique element if $x\leq y$ and is empty otherwise.
\item  We define an identity morphism for each $x\in X$.
  By reflexivity (i.e., $x\leq x$), we have that $\Hom x x$ consists of a unique element, which we take to be the identity.
\item We define, for each $x,y,z\in X$, a composition operator
\[
\Hom y z \to \Hom x y \to \Hom x z.
\]
By definition of the hom-sets, we only have to define it in case $x\leq y$ and $y\leq z$.
But then, by transitivity (i.e. if $x\leq y$ and $y\leq z$, then $x\leq z$), we have that $\Hom x z$ consists of a unique element; that unique element is the composite.
\end{itemize}
We are now going to show that the axioms of a category holds.
To show the right unit law, we have to show that for each $x,y\in X$ and $f\in \Hom x y$, we have $\co{\Id[x]}{f} = f$.
This indeed holds since every hom-set has a unique element, but both $\co{\Id[x]}{f}$ and $f$ live in the same hom-set; hence they must be equal.
The proof that left unit law and associator law hold are analogous.
\end{exa}

\begin{exer}[\cref{sol:post_antisymmetry}]\label{exer:post_antisymmetry}
  A \textbf{partially ordered set} (poset) is a preordered set $(X,\leq)$ satisfying the following additional axiom:
  \begin{itemize}
  \item \textbf{Antisymmetry}: $\forall x,y\in X: (x\leq y \wedge y\leq x) \implies x=y$.
  \end{itemize}
  What does this axiom say about $\PREtoCAT(X,\leq)$?
\end{exer}

\begin{rem}
  To understand a definition in category theory, it is very helpful to think about what the definition means in a preordered set, viewed as a category. 
\end{rem}

\begin{exa}\label{example:poset} The category of posets, denoted by $\POS$, is the category specified by the following data:
\begin{itemize}
\item An object is a poset $(X,\leq)$.
\item A morphism from a poset $(X,\leq_X)$ to $(Y,\leq_Y)$ consists of a function $f:X\to Y$ such that the following property holds:
\[
\forall x_1, x_2 \in X: x_1\leq_X x_2 \implies f(x_1)\leq_Y f(x_2).
\]
\item The identity morphism on $(X,\leq_X)$ is the identity function on $X$.
\item The composition given by the composition of functions.
\end{itemize}

Before we can show that this data satisfies the axioms of a category, notice that the identity function is a morphism of posets and that the composition of poset morphisms is again a poset morphism, indeed: If $x_1\leq_X x_2$, then we also have $\Id[X](x_1) \leq_X \Id[X](x_2)$ because $\Id[X](x) = x$. If $f\in\CHom{\POS}{(X,\leq_X)}{(Y,\leq_Y)}$ and $g\in\CHom{\POS}{(Y,\leq_Y)} {(Z,\leq_Z)}$ are morphisms of posets, and $x_1,x_2\in X$, we have 
\[
  x_1\leq_X x_2 \implies f(x_1)\leq_Y f(x_2) \implies g(f(x_1))\leq_Z g(f(x_2)),
\]
where the first (resp. second) implication holds by $f$ (resp. $g$) being a morphism of posets. So our data is indeed well-defined.

The axioms of a category are satisfied by this data; the proof is exactly the same proof as showing that $\SET$ is a category because the identity and composition are defined in the same way.
\end{exa}



\begin{lemma}\label{lemma:uniqueid} Let $\CC$ be a category. For any object $X\in\CC$, $\Id[X]$ is the unique morphism which satisfies the following property: For any $Y\in\CC$ and $f\in\CHom \CC X Y$, we have 
\[
\co{\Id[X]} f = f.
\]
\begin{proof}
Assume $\tilde{\Id[X]}$ also satisfies this property, in particular we have $\co {\tilde{\Id[X]}} {\Id[X]} = \Id[X]$. However, by the right unit law, we also must have $\co{\tilde{\Id[X]}}{\Id[X]} = \tilde{\Id[X]}$. Hence, $\Id[X] = \tilde{\Id[X]}$.
\end{proof}
\end{lemma}

\begin{exa}\label{exa:monoidofrationalnumbers} In this example we are going to define a category which captures the multiplication of the rational numbers. Let $\CC$ be the category defined by the following data:
\begin{itemize}
\item There is a unique object $\star$.
\item The (only) hom-set is given by
\[
\Hom{\star}{\star} = \mathbb{Q},
\]
i.e. each morphism corresponds with a rational number.
\item The composition is defined by the multiplication of rational numbers:
\[
\mathbb{Q} \to\mathbb{Q}\to\mathbb{Q} : (p,q)\mapsto p\cdot q.
\]
\item The identity morphism (of $\star$) is given by $1$.
\end{itemize}
That $\CC$ is indeed a category follows because for each $p\in\mathbb{Q}$, we have $p\cdot 1 = p = 1\cdot p$ (which shows the unit laws) and by associativity of multiplication, i.e. $(p\cdot q)\cdot h = p\cdot (h \cdot q)$ (which shows the associativity of the composition).
\end{exa}
The construction in \cref{exa:monoidofrationalnumbers} uses no specific properties of the rational numbers, only that it has a multiplication which is associative and such that there is a special element which does not change an element when it is multiplied with this special element. Hence, \cref{exa:monoidofrationalnumbers} can be generalized as follows:
\begin{dfn}\label{monoidcategory}
Recall that a monoid is a set $M$ equipped with binary operation $m : M \to M \to M$ which is associative, i.e. 
\[
\forall x,y,z\in M: m(x,m(y,z)) = m(m(x,y),z),
\]
and such that there is an identity element, i.e. 
\[
\exists e\in M: \forall x\in M: m(e,x)=x=m(x,e).
\]
Let $(M,m,e)$ be a monoid. The category $\MONtoCAT(M,m,e)$ is defined by the following data:
\begin{itemize}
\item There is a unique object $\star$.
\item The (only) hom-set is given by 
\[
\Hom{\star}{\star} = M.
\]
\item The identity morphism on $\star$ is the identity element $e$.
\item The composition of morphisms $x$ and $y$ is given by $\co{x}{y} := m(x,y)$.
\end{itemize}
\end{dfn}

That for each monoid $(M,m,e)$, $\MONtoCAT(M,m,e)$ is indeed a category, follows directly by the properties of being a monoid. Indeed, the axioms of a category become precisely:
\begin{enumerate}
\item $\forall x\in M: m(x,e)=x$,
\item $\forall x\in M: m(e,x)=x$,
\item $\forall x,y,z\in M: m(m(x,y),z) = m(x,m(y,z))$.
\end{enumerate}

\begin{rem} Notice that this category illustrates that there is no relation between the collection of objects and the hom-sets since there is now only one object and the collection of the hom-set can be as small or as large as possible.
In fact, we can associate a different number of categories to a single monoid. We can for example consider an arbitrary set of objects $I$ and the defining the hom-sets as follows:
\[
\Hom{i}{j} := 
\begin{cases}
M ,\quad \text{ if } i=j,\\
\emptyset, \quad \text{ if } i\not=j.
\end{cases}
\]
\end{rem}

\begin{exer}[\cref{sol:categories_coming_from_monoids}]\label{exer:categories_coming_from_monoids}
  Let $\CC$ be a category. When does $\CC$ ``come from a monoid'', that is, when is there a monoid $(M,m,e)$ such that $\CC$ of the form $\MONtoCAT(M,m,e)$?
\end{exer}

\begin{exer}[\cref{sol:category_of_monoids}]\label{exer:category_of_monoids}
  Define a category $\MON$ whose objects are monoids, i.e. define a suitable notion of morphism between monoids and moreover show that this indeed defines a category.
\end{exer}

\begin{exer}[\cref{sol:opposite}]\label{exer:opposite}
  Let $\CC$ be a category. Define a category $\op\CC$ such that
  \begin{itemize}
  \item the objects of $\op\CC$ are the same as those of $\CC$; and
  \item the morphisms $\CHom {\op\CC} X Y$ are morphisms $\CHom \CC Y X$.
  \end{itemize}
  The category $\op\CC$ is called the \textbf{opposite (category)} of $\CC$.
\end{exer}

\begin{exer} Let $G$ be a directed graph. Then $G$ induces a category $\mathbf{Graph}(G)$ as follows:
\begin{itemize}
\item The collection of objects $\Ob{\mathbf{Graph}(G)}$ is the set of vertices of $G$. 
\item The morphisms between object are the (directed) paths, that is, finite sequences of composable edges, between them.
\item For each object $x$ (i.e. vertex), the identity morphism on $x$ is the \textit{identity path}.
\item The composition of morphisms is the composition of paths.
\end{itemize}
We call $\mathbf{Graph}(G)$ the \textbf{category generated by $G$}. Show that $\mathbf{Graph}(G)$ is indeed a category.
\end{exer}

\begin{exer} Argue why the morphisms are chosen to be paths and it is not sufficient to just take the edges. 
\end{exer}

\begin{exa}\label{exa:graph_terminalcat} Consider the following graph $G$:
\[
\begin{tikzcd}
x
\end{tikzcd}
\]
i.e. the graph with only object vertex and no edges. The category generated by $G$ is the category generated is the so-called \textit{terminal category}, that is, the category with a single object and a single morphism (the identity morphism of the unique object). 
The terminal category is denoted by $\bullet$.
\end{exa}

\begin{exa}\label{exa:graph_intervalcat} Consider the following graph $G$:
\[
\begin{tikzcd}
x \arrow[r] & y
\end{tikzcd}
\]
The category generated by $G$ is the category generated is the so-called \textit{interval category}, that is, the category with two objects and, besides the identity morphisms, a unique morphism (living in $\Hom{x}{y}$).
\end{exa}

In the following example we use the following notation: 
\begin{itemize}
\item If $f$ is a morphism in a category, we denote $f^{2} := \co{f}{f}, f^{3} := \co{f}{f^2}$, etc.
\item We also label the edges in order to refer to them.
\end{itemize}
\begin{exa}\label{exa:graph_xy_yx} Consider the following graph $G$:
\[
\begin{tikzcd}
x \arrow[r, "f", bend left] & y \arrow[l, "g", bend left]
\end{tikzcd}
\]
The category generated by $G$ consists of the following data:
\begin{itemize}
\item The collection of objects is $\{x,y\}$.
\item The hom-sets are given as follows:
\begin{itemize}
\item $\Hom{x}{x}$ contains
\[
\Id[x], \co{f}{g}, (\co{f}{g})^2, (\co{f}{g})^3, \cdots,
\]
But these are not the only ones, we also have that each of these can be precomposed or postcomposed with $\Id[x]$, however, by the unit laws, we know that these don't give us any \textit{new} morphisms. The same remark holds for the associativity law. This comment also holds for the upcoming hom-sets.
\item $\Hom{y}{y}$ contains
\[
\Id[y], \co{g}{f}, (\co{g}{f})^2, (\co{g}{f})^3, \cdots,
\]
\item $\Hom{x}{y}$ contains  
\[
f, \co{f}{(\co{g}{f})}, \co{f}{(\co{g}{f})^2}, \co{f}{(\co{g}{f})^3}, \cdots
\]
\item $\Hom{y}{x}$ contains
\[
g, \co{g}{(\co{f}{g})}, \co{g}{(\co{f}{g})^2}, \co{g}{(\co{f}{g})^3}, \cdots
\] 
\end{itemize}
\end{itemize}
\end{exa}

\begin{exa}\label{exa:graph_yx_yz_zw} Consider the following graph $G$:
\[
\begin{tikzcd}
& w & \\
x & & z \arrow[lu] \\
& y \arrow[lu] \arrow[ru] &
\end{tikzcd}
\]
The category generated by $G$ has four objects (namely $x,y,z,w$) and the hom-sets are: 
\begin{itemize}
\item $\Hom{y}{x}, \Hom{y}{z}$ and $\Hom{z}{w}$ are singleton sets, 
\item $\Hom{x}{y}, \Hom{z}{y}, \Hom{x}{w}, \Hom{w}{x}$ and $\Hom{w}{z}$ are all empty. 
\item $\Hom{y}{w}$ consists of the path $y\to z\to w$.
\item For each vertex $v$, we have that $\Hom{v}{v}$ consists only of the identity path on $v$.
\end{itemize}
\end{exa}

\begin{exer}[\cref{sol:connection_graphs_preordersets}] \label{exer:connection_graphs_preordersets}
Describe the connection between the categories generated by graphs and the categories associated to preordered sets. What does the property of anti-symmetry correspond to under this connection with graphs?
\end{exer}

\begin{exer} Define a category $\Catb{Aut}$ whose objects are (deterministic finite) automata. 
\end{exer}

\begin{exer}[\cref{sol:categories_with_natural_numbers}] \label{exer:categories_with_natural_numbers}
  In this exercise, we study several different categories which all have the set of natural numbers as their collection of objects.
  Define in detail the categories sketched below:
  \begin{enumerate}
  \item The category $\POS(\NN, \leq)$ generated by the preorder on natural numbers given by the ``less than or equal'' relation (the category that looks like this: $0 \to 1 \to 2 \to \ldots$).
  \item The category $\SKELFINSET$, where a morphism $f : m \to n$ is a function from the ``standard finite set'' $[m]$ to the standard finite set $[n]$. Here, $[m] := \{0,\ldots,m-1\}$.
  \item The category $\MAT$, where a morphism $f : m \to n$ is a real matrix of dimension $n \times m$ (with $n$ rows and $m$ columns).
    Such matrices can represent linear maps between real vector spaces.
    Composition in this category is given by matrix multiplication.

    More generally for each field $\mathbb{F}$, one can define $\MAT_\mathbb{F}$, where a morphism $f : m \to n$ is a an $n \times m$ matrix over the field $\mathbb{F}$.
  \end{enumerate}

  Can you think of other categories that have natural numbers as their collection of objects? What about a category, where a morphism $f : m \to n$ is a monotonously increasing (decreasing) function from $[m]$ to $[n]$?
\end{exer}

\begin{exer}[\cref{sol:category_of_relations}] \label{exer:category_of_relations}
  Define a category $\REL$ of sets and binary relations. Recall that given sets $X$ and $Y$ a binary relation $R$ is a subset of the cartesian product $X \times Y$ of the sets.
\end{exer}

\chapter{Special Morphisms in a Category}

\section{Isomorphisms}
\label{sec:isos}

\begin{reading*}
  In this section, we study properties of arrows in a category.
  More information on this topic is given in \cite[\S 2.7]{barr-wells}.

  Also, \cite[\S 2.2]{pierce} briefly discusses isomorphisms.
\end{reading*}

\begin{dfn}[Isomorphism]
  Given a category $\CC$, objects $a,b \in \Ob{\CC}$ and a morphism $f : a \to b$ in $\CC$, we say that $f$ is an \textbf{isomorphism} when there is a morphism $g : b \to a$ (in the other direction!) such that $f \Comp g = \Id$ and $g \Comp f = \Id$.
  We write $f : a \cong b$ for a morphism $f$ that is an isomorphism.

  In this case, we call $g$ the \textbf{inverse} of $f$ and $f$ the inverse of $g$. (The latter is justified by \cref{exer:inverse-iso}.)
\end{dfn}

\begin{exer}[\cref{sol:inverse-iso}]\label{exer:inverse-iso}
  Show that if $f : a \to b$ is an isomorphism with inverse $g : b \to a$, then $g$ is an isomorphism with inverse $f$.
\end{exer}

\begin{exer}[\cref{sol:inverse_uniqueness}]\label{exer:inverse_uniqueness}
  Show that a morphism $f : a \to b$ in $\CC$ is an isomorphism \textbf{in at most one way}, that is, show that its inverse is unique if it exists.
\end{exer}

\begin{exer}[\cref{sol:compofiso}]\label{exer:compofiso}
  Show that the composition of two isomorphisms is an isomorphism.
\end{exer}

\begin{exer}[\cref{sol:iso-bool}]\label{exer:iso-bool}
  Consider the datatype
\begin{lstlisting}
data BW = Black | White
\end{lstlisting}
Construct two (different!) isomorphisms between |BW| and the type |Bool| of booleans.
\end{exer}

\begin{exer}[\cref{sol:iso_in_sets}]\label{exer:iso_in_sets}
  Describe the isomorphisms in $\SET$.
\end{exer}

\begin{exer}[\cref{sol:iso_in_pos}]\label{exer:iso_in_pos}
  Describe the isomorphisms in $\POS$.
\end{exer}

\begin{exer}[\cref{sol:iso_in_posetcategory}]\label{exer:iso_in_posetcategory}
  Let $(X,\leq)$ be a poset.
  Describe the isomorphisms in $\POS(X,\leq)$.
\end{exer}

\begin{exer}
  Describe the isomorphisms in $\MON$.
\end{exer}

\begin{exer} Let $\mathcal{G}$ be the category generated by the following graph:
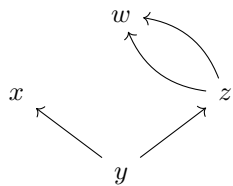
\[
\begin{tikzcd}
& w & \\
x & & z \arrow[lu, bend left] \arrow[lu,bend right] \\
& y \arrow[lu] \arrow[ru] &
\end{tikzcd}
\]
Show that the only isomorphisms in $\mathcal{G}$ are the identity morphisms (i.e. the identity paths).
\end{exer}

\begin{exer}[\cref{sol:iso_in_posetcategory}] \label{exer:iso_in_cats_of_nats}
  Describe the isomorphisms in $\POS(\NN,\leq)$, $\SKELFINSET$ and $\MAT$.
\end{exer}

\section{Sections and Retractions}
\label{sec:sections}

\begin{dfn}[Section, Retraction]
  A pair $(s,r)$ of morphisms $s : a \to b$ and $r : b \to a$ in $\CC$ is called a \textbf{section-retraction pair} if $\co{s}{r} = \Id[a]$.

  In such a case, we call $s$ a section and $r$ a retraction.
\end{dfn}

\begin{rem}
  Note that a morphism can be a retraction in more than one way, that is, there can be more than one section $s$ such that $\co{s}{r} = \Id$.
\end{rem}

Intuitively, a section-retraction pair $(s,r)$ of morphisms $s : a \to b$ and $r : b \to a$ in a category $\CC$ provides a way for $a$ to ``live inside'' $b$.
Note that for a given $a$ and $b$ there can be many ways for $a$ to live inside $b$.

\begin{exer}[\cref{sol:section-retraction-bool-int}]\label{exer:section-retraction-bool-int}
  Construct two different section-retraction pairs between the type |Bool| of booleans and the type |Int| of integers (e.g., in Haskell).
\end{exer}

\begin{exer}
 Show that the type |Maybe a| is a retract of the type |[a]|. 
 
 Hint: The idea is that |Nothing| corresponds to the empty list |[]| and that |Just x| corresponds to the one-element list |[x]|. Make this idea precise by writing back and forth functions between these types so that they exhibit |Maybe a| as a retract of |[a]|. 
\end{exer}

\section{Monomorphisms and Epimorphisms}
\label{sec:mono-epi}

\begin{reading*}
See also \cite[p. 134]{leinster} and \cite[\S\S 2.8--2.9]{barr-wells}.
Also, \cite[\S 2.2]{pierce} briefly discusses monomorphisms and epimorphisms.
\end{reading*}

From undergraduate mathematics courses you know what injective and surjective functions between sets are.
The definitions of ``injective'' and ``surjective'' do not carry over to any category (though they do for categories that are, in some sense, ``similar'' to the category of sets).
In this section, we study two properties of morphisms in a category that, in the category of sets, are equivalent to ``injective'' and ``surjective'', respectively.

\begin{dfn}[Monomorphism]
  Let $f : a \to b$ be a morphism in $\CC$. We say that $f$ is a \textbf{monomorphism} if, for any two morphisms $g_1, g_2 : z \to a$, like in the following diagram,
  \begin{center}
    \begin{tikzcd}
    z \arrow[r, "g_2"', shift right] \arrow[r, "g_1", shift left] & a \arrow[r, "f"] & b
    \end{tikzcd}
  \end{center}
  we have
  \[ \co{g_1}{f} = \co{g_2}{f} \text{ implies } g_1 = g_2 .\]
\end{dfn}

\begin{exer}[\cref{sol:mono-inj}]\label{ex:mono-inj}
  In the category of sets, show that a morphism $f : X \to Y$ is a monomorphism if and only if it is injective.
\end{exer}

\begin{dfn}[Epi]
  Let $f : a \to b$ be a morphism in $\CC$. We say that $f$ is an \textbf{epimorphism} if, for any two morphisms $g_1, g_2 : b \to z$, like in the following diagram,
  \begin{center}
    \begin{tikzcd}
    a \arrow[r, "f"] & b \arrow[r, "g_2"', shift right] \arrow[r, "g_1", shift left] & z
    \end{tikzcd}
  \end{center}
  we have
  \[ \co{f}{g_1} = \co{f}{g_2} \text{ implies } g_1 = g_2 .\]
\end{dfn}

\begin{exer}\label{ex:epi-surj}
  In the category of sets, show that a morphism $f : X \to Y$ is an epimorphism if and only if it is surjective.
\end{exer}

\begin{exer}[\cref{sol:sections_in_set_injective}]\label{exer:sections_in_set_injective}
  In the category of sets, show that if $(s,r)$ is a section-retraction pair, then the section $s$ is injective.
  Hint: you can use \cref{ex:mono-inj}.
\end{exer}
\begin{exer}
  In the category of sets, show that if $(s,r)$ is a section-retraction pair, then the retraction $r$ is surjective.
    Hint: you can use \cref{ex:epi-surj}.
\end{exer}

\begin{exer}[\cref{sol:iso_to_monoepi}]\label{exer:iso_to_monoepi}
  Show that any isomorphism $f : a\cong b$ (in some arbitrary category $\CC$) is both a monomorphism and an epimorphism.
\end{exer}

\begin{exer}[\cref{sol:counterexample_monoepi_not_iso}]\label{exer:counterexample_monoepi_not_iso}
  Show that the converse of \cref{exer:iso_to_monoepi} does not hold in general, i.e. give an example of a category where there exists a morphism which is both an epi- and a monomorphism, but which is not an isomorphism.

Hint: Consider a preordered set.
\end{exer}

\begin{exer} Let $\mathcal{G}_1$ (resp. $\mathcal{G}_2$ and $\mathcal{G}_3$) be the category generated by the following graph:
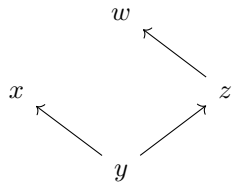
\[
\begin{tikzcd}
& w & \\
x & & z \arrow[lu] \\
& y \arrow[lu] \arrow[ru] &
\end{tikzcd}
\]
resp.
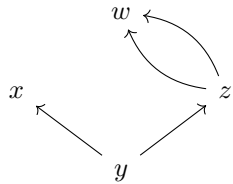
\[
\begin{tikzcd}
& w & \\
x & & z \arrow[lu, bend left] \arrow[lu,bend right] \\
& y \arrow[lu] \arrow[ru] &
\end{tikzcd}
\]
resp.
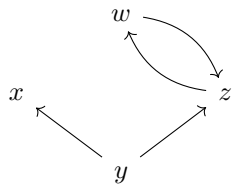
\[
\begin{tikzcd}
& w \arrow[rd,bend left] & \\
x & & z \arrow[lu, bend left] \\
& y \arrow[lu] \arrow[ru] &
\end{tikzcd}
\]

Describe the mono- and epimorphisms in these categories.
\end{exer}

\begin{exer} Describe the monomorphisms, epimorphisms and isomorphisms in the category generated by the following graph:
\[
\begin{tikzcd}
x \arrow[r] & y
\end{tikzcd}
\]
\end{exer}

\chapter{Universal Properties}\label{sec:universal}

In category theory, we study objects of a category by studying the ``interactions'' they have with other objects in the category.
What does this mean?
The only interactions we know between objects are the morphisms of the category.
Given an object of a category, we can ask what the morphisms out of this object, and into this object, are.
More generally, we can ask what the morphisms out of, or into, an object are that make some given diagrams commute.
We usually are interested in objects such that there exists a unique morphism out of it, or into it, such that a given diagram commutes.
We say that this object satisfies a ``universal property''.

This description is very vague and abstract; it will hopefully be clearer once we have looked at some specific universal properties.
The first universal property is that of \emph{initiality}.

\begin{reading*}
  On initial and terminal objects, see also \cite[\S 2.7.16]{barr-wells} and \cite[p. 48ff]{leinster}.

  Products and coproducts, other special limits and colimits, and the general definition of limits and colimits, are discussed in \cite[\S\S 5.1, 5.2]{leinster}.

  Pierce's tutorial discusses the (co)limits defined here in \cite[\S\S 2.3--2.4]{pierce}, and further (co)limits in  \cite[\S\S 2.5--2.7]{pierce}.

\end{reading*}

\section{Initial Objects}
\label{sec:initial-objects}

We say that an object of a category $\CC$ is \emph{initial} if it has a unique morphism to any object in the category:

\begin{dfn}
  Let $\CC$ be a category. An object $A \in \Ob{\CC}$ is \textbf{initial} if there is exactly one morphism from $A$ to any object $B \in \Ob{\CC}$.
\end{dfn}

\begin{explanation}[about \textbf{Unique Existence}]
  When you need to show that there exists a unique thing with some property, there are two things to prove:
  \begin{description}
  \item[Existence] You need to construct a thing and show that it has the desired property.
  \item[Uniqueness] You need to show that any (abstract) thing that has the desired property is equal to the thing you have constructed. Equivalently, you can show that any two abstract things with the desired property are equal.
  \end{description}
\end{explanation}

We look at some specific categories, and try to identify initial objects in them.
These examples might seem a tad boring to you;
indeed, in ``simple'' categories, initial objects are usually quite simple as well.
However, in more complicated categories---that is, in categories where the objects and morphisms are complicated---an initial object can be very interesting; see, for instance, \cref{sec:initial-algs}.

\begin{exer}
  Does the category $\bullet$ (the terminal category defined in \cref{exa:graph_terminalcat}) have an initial object?
\end{exer}

\begin{exer}
  Does the category 
  \[ 
  \begin{tikzcd}
  	A & B
  \end{tikzcd}  
   \] 
  have an initial object?
\end{exer}

\begin{exer}
  Does the category $A \to B$ have an initial object?
\end{exer}

\begin{exer}
  Does the category $A \rightleftarrows B$ have an initial object? (Here, the morphism $A \to B$ is inverse to the morphism $B \to A$.)
\end{exer}

\begin{exer}
  Does the category $A \rightrightarrows B$ have an initial object?
\end{exer}

\begin{exer}
  Does the category 
  \[
  \begin{tikzcd}
  A \arrow[r] \arrow[loop, swap, looseness=4, "f"] & B
  \end{tikzcd}
  \]
  have an initial object? (Here, the morphism $f$ is different from $\Id[A]$).
\end{exer}

\begin{exer}[\cref{sol:initial_set}]\label{exer:initial_set}
  Identify an initial object in the category $\SET$ of sets.
  Prove that it is indeed initial.
\end{exer}

\begin{exer}
  Identify an initial object in the category $\COQ$ of Coq types.
  Prove that it is indeed initial.
\end{exer}

\begin{exer}[\cref{sol:initial_posetcat}]\label{exer:initial_posetcat}
  Let $(X,\leq)$ be a poset. Describe what an initial object looks like in  $\POS(X,\leq)$.
\end{exer}

\begin{exer}[\cref{sol:initial-unique}]\label{exer:initial-unique}
  Let $A$ and $A'$ be initial objects in $\CC$. Construct an isomorphism $i : A \cong A'$.
\end{exer}

\begin{exer}[\cref{sol:initiality_preserved_by_iso}]\label{exer:initiality_preserved_by_iso}
  Let $A$ be an initial object in $\CC$, and let $A'$ be isomorphic to $A$ (via an isomorphism $i : A \cong A'$).
  Show that $A'$ is an initial object of $\CC$.
\end{exer}

\begin{rem}
  \Cref{exer:initial-unique} shows that initial objects in a category $\CC$ are \textbf{essentially unique}, that is, they are \textbf{unique up to (unique) isomorphism}.

  This justifies using the determinate article: we will say that $A$ is \textbf{the} initial object of $\CC$.

  This is more generally the case for any object with a universal property, see, e.g., \cref{exer:terminal-unique,exer:product-unique}.
\end{rem}

\begin{exer}[\cref{sol:cat-without-initial}]\label{exer:cat-without-initial}
  Construct a category that does not have an initial object.
\end{exer}

\begin{exer}[\cref{sol:initial_pointset}]\label{exer:initial_pointset} Let $\PTSET$ be the category of pointed sets, that is the category whose objects are pairs $(X,x)$ with $X$ a set and $x\in X$ and a morphism from $(X,x)$ to $(Y,y)$ is defined as a function $f:X\to Y$ such that $f(x)=y$. Identify an initial object in $\PTSET$.
\end{exer}

\begin{exer}[\cref{sol:initial_cats_of_nats}]\label{exer:initial_cats_of_nats}
Do the categories $\POS(\NN, \leq)$, $\SKELFINSET$ and $\MAT$ have initial objects? If yes, what does an initial object look like in these categories?
\end{exer}

\begin{exer}[\cref{sol:initial_rel}] \label{exer:initial_rel}
Identify an initial object in the category $\REL$ and show that it is initial.
\end{exer}

\begin{rem}
  The concept of initial object seems trivial and boring in the categories considered above.
  However, in complicated categories, initial objects can be complicated and exciting;
  we will see this in \cref{sec:initial-algs}.
\end{rem}

\section{Terminal Objects}
\label{sec:terminal-objects}

\begin{dfn}
  Let $\CC$ be a category. An object $B \in \Ob{\CC}$ is \textbf{terminal} (or \textbf{final}) if there is exactly one morphism to $B$ from any object $A \in \Ob{\CC}$.
  A terminal object in a category is denoted by \textbf{1}.
\end{dfn}

\begin{exer}
  Does the category $\bullet$ have a terminal object?
\end{exer}

\begin{exer}
  Does the category $A \to B$ have a terminal object?
\end{exer}

\begin{exer}
  Does the category $A \rightleftarrows B$ have a terminal object?
\end{exer}

\begin{exer}
  Does the category $A \rightrightarrows B$ have a terminal object?
\end{exer}

\begin{exer}[\cref{sol:terminal_set}]\label{exer:terminal_set}
  Identify a terminal object in the category $\SET$ of sets.
  Prove that it is indeed terminal.
\end{exer}

\begin{exer}
  Identify a terminal object in the category $\COQ$ of Coq types.
  Prove that it is indeed terminal.
\end{exer}

\begin{exer}[\cref{sol:terminal_posetcat}]\label{exer:terminal_posetcat}
  Let $(X,\leq)$ be a poset. Describe what a terminal object looks like in  $\POS(X,\leq)$.
\end{exer}

\begin{exer}[\cref{sol:terminal-unique}]\label{exer:terminal-unique}
  Let $B$ and $B'$ be terminal objects in $\CC$. Construct an isomorphism $i : B \cong B'$.
\end{exer}

\begin{exer}[\cref{sol:terminality_preserved_by_iso}]\label{exer:terminality_preserved_by_iso}
  Let $B$ be a terminal object in $\CC$, and let $B'$ be isomorphic to $B$ (via an isomorphism $i : B \cong B'$).
  Show that $B'$ is a terminal object of $\CC$.
\end{exer}

\begin{exer}[\cref{sol:terminal_iff_initial_op}]\label{exer:terminal_iff_initial_op}
  Show that $\CC$ has a terminal object if and only if $\op\CC$ has an initial object.
\end{exer}

\begin{exer}[\cref{sol:cat-without-terminal}]\label{exer:cat-without-terminal}
  Construct a category that does not have an terminal object.
\end{exer}

\begin{exer}[\cref{sol:terminal_cats_of_nats}]\label{exer:terminal_cats_of_nats}
  Do the categories $\POS(\NN, \leq)$, $\SKELFINSET$ and $\MAT$ have terminal objects? If yes, what does a terminal object look like in these categories?
\end{exer}

\begin{exer}[\cref{sol:terminal_rel}] \label{exer:terminal_rel}
  Identify a terminal object in the category $\REL$ and show that it is terminal.
\end{exer}

\section{(Binary) Products}
\label{sec:products}

\begin{dfn}\label{def:binproduct}
  Let $\CC$ be a category and let $A,B \in \Ob\CC$ be objects of $\CC$.

  A triple $(P,\projl : P \to A ,\projr : P \to B)$ is called a \textbf{product of $A$ and $B$} if for any triple $(Q,q_1 : Q \to A, q_2 : Q \to B)$ there is exactly one morphism $f : Q \to P$ such that the following diagram commutes:
  \[
    \begin{tikzcd}
      &
      Q \ar[ld, "q_1"'] \ar[rd, "q_2"] \ar[d, dashed, "f"]
      &
      \\
      A
      &
      P \ar[l, "\projl"] \ar[r, "\projr"']
      &
      B
    \end{tikzcd}
  \]
  If $A$ and $B$ have a specified product $(P,\projl : P \to A ,\projr : P \to B)$, then the object $P$ is often called $A \times B$.
  The morphism $f : Q \to A \times B$ determined by $(Q, q_1, q_2)$ is denoted by $\intoproduct { q_1} {q_2}$.
\end{dfn}

\begin{exer}
  Does the category $\bullet$ have products?
\end{exer}

\begin{exer}
  Does the category $A \to B$ have products?
\end{exer}

\begin{exer}
  Does the category $A \rightleftarrows B$ have products?
\end{exer}

\begin{exer}
  Does the category $A \rightrightarrows B$ have products?
\end{exer}

\begin{exer}[\cref{sol:product-represent}]\label{exer:product-represent}
  Let $\CC$ be a category, let $A,B\in\Ob{\CC}$ and let $(A\times B,\projl,\projr)$ be a product of $A$ and $B$ in $\CC$. Fix an object $X\in\Ob{\CC}$. Construct an isomorphism between the set $\CC(X, A\times B)$ and the set $\CC(X,A)\times\CC(X,B)$.  
\end{exer}

\begin{exer}[\cref{sol:product_set}]\label{exer:product_set}
  Identify a product of sets $X$ and $Y$ in the category $\SET$ of sets.
  Prove that it is indeed a product.
\end{exer}

\begin{exer}
  Identify a product of types $A$ and $B$ in the category $\COQ$ of Coq types.
  Prove that it is indeed a product.
\end{exer}

\begin{exer}[\cref{sol:product_posetcat}]\label{exer:product_posetcat}
  Let $(X,\leq)$ be a poset. Describe what a product looks like in  $\POS(X,\leq)$.
\end{exer}

\begin{exer}[\cref{sol:product_cats_of_nats}]\label{exer:product_cats_of_nats}
  Do the categories $\POS(\NN, \leq)$, $\SKELFINSET$ and $\MAT$ have products? If yes, describe what a product looks like in these categories. 
\end{exer}

\begin{exer}[\cref{sol:product_rel}] \label{exer:product_rel}
  Identify a product of sets $X$ and $Y$ in the category $\REL$. Prove that it is a product. 
\end{exer}

\begin{exer}[\cref{sol:product-unique}]\label{exer:product-unique}
  Given two products of $A$ and $B$ in a category $\CC$, construct an isomorphism between them, that is, between their underlying objects.
\end{exer}

\begin{exer}[\cref{sol:product_preserved_by_iso}]\label{exer:product_preserved_by_iso}
  Given a product $(P,\projl : P \to A ,\projr : P \to B)$ of $A$ and $B$ in $\CC$, and an object $P'$ that is isomorphic to $P$ via an isomorphism $i : P \cong P'$, construct a product with object $P'$ of $A$ and $B$.
\end{exer}

\begin{exer}[\cref{sol:product_with_terminal}]\label{exer:product_with_terminal} Let $\CC$ be a category and $T\in\Ob{\CC}$ a terminal object.
  For any object $A\in \Ob{\CC}$, construct a product of $A$ and $T$.

  Hint: to form an idea what the object $A \times T$ should be, solve the exercise first in a specific category, e.g., in the category of sets or in a category coming from a preordered set.
\end{exer}

\begin{exer}[\cref{sol:product_iff_terminal_in_subcategory}]\label{exer:product_iff_terminal_in_subcategory} Let $\CC$ be a category and $A,B\in\Ob{\CC}$ be objects. Show that the product of $A$ and $B$ exists if and only if the following category has a terminal object:
\begin{itemize}
\item The objects are triples $(P,p_l: P\to A, p_r:P\to B)$.
\item A morphism from $(P,p_l,p_r)$ to $(Q,q_l,q_r)$ is a morphism $f : P \to Q$ such that the following diagram commutes:
\[
\begin{tikzcd}
& P \arrow[ld,swap, "p_l"] \arrow[rd, "p_r"] \arrow[d, "f"] & \\
A & Q \arrow[l, "q_l"]  \arrow[r,swap, "q_r"] & B
\end{tikzcd}
\]
\item The composition and identity are inherited from the structure of $\CC$.
\end{itemize}
\end{exer}

\begin{exer}[\cref{sol:product_of_morphisms}]\label{exer:product_of_morphisms}
  Let $\CC$ be a category with a choice of product $(A\times B, \projl, \projr)$ for any two objects $A,B\in \Ob{\CC}$.
  Given morphisms $f : A \to C$ and $g : B \to D$ in $\CC$, construct a morphism
  \[ f \times g : A \times B \to C \times D.\]
\end{exer}

\begin{exer}[\cref{sol:swap_binary_product}]\label{exer:swap_binary_product}
  Let $\CC$ be a category with a choice of product $(A\times B, \projl, \projr)$ for any two objects $A,B\in \Ob{\CC}$.
  For any $A, B \in \Ob\CC$, construct an isomorphism
  \[ A \times B \cong B \times A. \]
\end{exer}

\begin{exer}[Equational reasoning for products]
  Let $\CC$ be a category with binary products.
  Consider the following objects and morphisms in $\CC$.

  \[
    \begin{tikzcd}[column sep=large]
      &
      &
      A \ar[r, "h"]
      &
      C \ar[r, "h'"]
      &
      E
      \\
      Y \ar[r, "j"]
      &
      Z \ar[ru, "f"] \ar[rd, "g"'] 
      &
      A\times B \ar[u, "\pi_A"'] \ar[d, "\pi_B"]
      &
      C \times D  \ar[u] \ar[d]
      &
      E \times F  \ar[u] \ar[d]
      \\
      &
      &
      B \ar[r,"k"]
      &
      D \ar[r,"k'"]
      &
      F
    \end{tikzcd}
  \]
  
  Prove the following equations:
  \begin{align}
    \co{j}{\intoproduct{f}{g}} &= \intoproduct{\co j f}{\co j g}
    \\
    \intoproduct{\co f h}{\co g k} &= \co {\intoproduct f g} {(\productmap h k)} 
    \\
    \productmap{(\co h {h'})}{(\co k {k'})} &= \co {(\productmap h k)}{(\productmap {h'} {k'})} 
    \\
    \intoproduct{\pi_A}{\pi_B} &= \Id[A \times B]
  \end{align}
\end{exer}

\section{(Binary) Coproducts}
\label{sec:coproducts}

\begin{dfn}
   Let $\CC$ be a category and let $A,B \in \Ob\CC$ be objects of $\CC$.

  A triple $(C,\inl : A \to C,\inr : B \to C)$ is called a \textbf{coproduct of $A$ and $B$} if for any triple $(D,i_l : A \to D, i_r : B \to D)$ there is exactly one morphism $f : C \to D$ such that the following diagram commutes:
  \[
    \begin{tikzcd}
      A \ar[r, "\inl"] \ar[rd, "i_l"']
      &
      C  \ar[d, dashed, "f"]
      &
      B \ar[l, "\inr"'] \ar[ld, "i_r"]
      \\
      &
      D 
    \end{tikzcd}
  \]
  If $A$ and $B$ have a specified coproduct $(C,\inl : A \to C,\inr : B \to C)$, then the object $C$ is often called $A + B$.
  The morphism $f : A + B \to D$ determined by $(D, i_l, i_r)$ is denoted by $\outofcoproduct{i_l}{i_r}$.
  
\end{dfn}

\begin{exer}
  Does the category $\bullet$ have coproducts?
\end{exer}

\begin{exer}
  Does the category $A \to B$ have coproducts?
\end{exer}

\begin{exer}
  Does the category $A \rightleftarrows B$ have coproducts?
\end{exer}

\begin{exer}
  Does the category $A \rightrightarrows B$ have coproducts?
\end{exer}

\begin{lemma}
\label{lemma}
Given category $\CC$ with choice of coproduct, objects $A,B\in \Ob{\CC}$, their coproduct $(A+B,\inl,\inr)$ and a morphism $f:A+B\to A+B$. If $\co {\inl} {f}=\inl$ and $\co {\inr} {f}=\inr$, then $f=\Id[A+B]$.
\end{lemma}
\begin{proof}
By definition of coproduct $A+B$ we know there is a unique $g: A+B \to A+B$, s.t. $\co {\inl} {g}=\inl$ and $\co {\inr} {g}=\inr$. Since it is unique, $f$ and $g$ must be the same morphism. Furthermore, we know that the identity morphism $\Id[A+B]: A+B \to A+B$ exists, and satisfies the equations $\co {\inl} {\Id[A+B]} = \inl$ and $\co {\inr} {\Id[A+B]} = \inr$. Again because of uniqueness of $g$, $\Id[A+B]$ must be the same as $g$, therefore $f=g=\Id[A+B]$.
\end{proof}

\begin{exer}[\cref{sol:coproduct-represent}] \label{exer:coproduct-represent}
Let $\CC$ be a category, let $A,B\in\Ob{\CC}$ and let $(A+B,\inl,\inr)$ be a coproduct of $A$ and $B$ in $\CC$. Fix an object $X\in\Ob{\CC}$. Construct an isomorphism between the set $\CC(A+B,X)$ and the set $\CC(A,X)\times\CC(B,X)$.
\end{exer}

\begin{exer}\label{exer:coproduct_set}
  Identify a coproduct of sets $X$ and $Y$ in the category $\SET$ of sets.
  Prove that it is indeed a coproduct.
\end{exer}

\begin{exer}
  Identify a coproduct of types $A$ and $B$ in the category $\COQ$ of Coq types.
  Prove that it is indeed a coproduct.
\end{exer}

\begin{exer}\label{exer:coproduct_posetcat}
  Let $(X,\leq)$ be a poset. Describe what a coproduct looks like in  $\POS(X,\leq)$.
\end{exer}

\begin{exer}[\cref{sol:coproduct_cats_of_nats}]\label{exer:coproduct_cats_of_nats}
  Do the categories $\POS(\NN, \leq)$, $\SKELFINSET$ and $\MAT$ have coproducts? If yes, describe what a coproduct looks like in these categories. 
\end{exer}

\begin{exer}\label{exer:coproduct_rel}
  Identify a coproduct of sets $X$ and $Y$ in the category $\REL$. Prove that it is a coproduct. 
\end{exer}

\begin{exer}\label{exer:coproduct-unique}
  Given two coproducts of $A$ and $B$ in a category $\CC$, construct an isomorphism between them, that is, between their underlying objects.
\end{exer}

\begin{exer}\label{exer:coproduct_preserved_by_iso}
  Given a coproduct $(C,\inl : A \to C ,\inr : B \to C)$ of $A$ and $B$ in $\CC$, and an object $C'$ that is isomorphic to $C$ via an isomorphism $i : C \cong C'$, construct a coproduct with object $C'$ of $A$ and $B$.
\end{exer}

\begin{exer}\label{exer:coproduct_with_initial} Let $\CC$ be a category and $I\in\Ob{\CC}$ an initial object.
  For any object $A\in \Ob{\CC}$, construct a coproduct of $A$ and $I$.

  Hint: To form an idea what the object $A + I$ should be, solve the exercise first in a specific category, e.g., in the category of sets or in a category coming from a preordered set.
\end{exer}

\begin{exer}\label{exer:coproduct_iff_initial_in_subcategory} Let $\CC$ be a category and $A,B\in\Ob{\CC}$ be objects. Show that the coproduct of $A$ and $B$ exists if and only if the following category has an initial object:
\begin{itemize}
\item The objects are triples $(C,c_l: A\to C, c_r:B\to C)$.
\item A morphism from $(C,c_l,c_r)$ to $(D,d_l,d_r)$ is a morphism $f : C \to D$ such that the following diagram commutes:
\[
    \begin{tikzcd}
      A \ar[r, "c_l"] \ar[rd, "d_l"']
      &
      C  \ar[d, "f"]
      &
      B \ar[l, "c_r"'] \ar[ld, "d_r"]
      \\
      &
      D
    \end{tikzcd}
  \]
\item The composition and identity are inherited from the structure of $\CC$.
\end{itemize}
\end{exer}

\begin{exer}\label{exer:coproduct_of_morphisms}
  Let $\CC$ be a category with a choice of coproduct $(A + B, \inl, \inr)$ for any two objects $A,B\in \Ob{\CC}$.
  Given morphisms $f : A \to C$ and $g : B \to D$ in $\CC$, construct a morphism
  \[ f + g : A + B \to C + D.\]
\end{exer}

\begin{exer}[\cref{sol:swap_binary_coproduct}]\label{exer:swap_binary_coproduct}
  Let $\CC$ be a category with a choice of coproduct $(A+ B, \inl, \inr)$ for any two objects $A,B\in \Ob{\CC}$.
  For any $A, B \in \Ob\CC$, construct an isomorphism
  \[ A + B \cong B + A. \]
\end{exer}

\begin{exer}[Equational reasoning for coproducts]
  Let $\CC$ be a category with binary coproducts.
  Consider the following objects and morphisms in $\CC$.

  \[
    \begin{tikzcd}[column sep=large]
      E \ar[r, "h'"] \ar[d]
      &
      A \ar[r, "h"] \ar[d, "\iota_A"]
      &
      C  \ar[d] \ar[rd, "f"]
      \\
      E + F
      &
      A+ B
      &
      C + D 
      &
      Y \ar[r, "j"]
      &
      Z
      \\
      F \ar[r, "k'"] \ar[u]
      &
      B \ar[u, "\iota_B"] \ar[r,"k"]
      &
      D \ar[u] \ar[ru, "g"']
    \end{tikzcd}
  \]
  
  Prove the following equations:
  \begin{align}
    \co{\outofcoproduct{f}{g}}{j} &= \outofcoproduct{\co f j}{\co g j}
    \\
    \outofcoproduct{\co h f}{\co k g} &= \co {(\coproductmap h k)} {\outofcoproduct f g}
    \\
    \coproductmap{(\co {h'} h)}{(\co {k'} k)} &= \co {(\coproductmap {h'} {k'}) }{(\coproductmap h k)}
    \\
    \outofcoproduct{\iota_A}{\iota_B} &= \Id[A + B]
  \end{align}
\end{exer}

\begin{exer}
  Consider the category with one object and the rational numbers $\mathbb{Q}$ as morphisms.
  Can you construct an initial object in this category? A terminal object? Products? Coproducts?
\end{exer}

\begin{rem}
  Initial and terminal objects and products and coproducts are special cases of \textbf{limits and colimits}.
  We are not studying, in these notes, the general notion of (co)limit.
  However, the examples above should suffice for you to understand, in your own time, other (co)limits, such as
  \begin{itemize}
  \item pullbacks and pushouts;
  \item products and coproducts of families of objects (not just of pairs of objects); and
  \item equalizers and coequalizers.
  \end{itemize}
\end{rem}

\chapter{Functors}\label{sec:functors}
An important aspect in computer programming is the transformation of data. For example, if you have a data type $X$, then one can consider also the data type $\List(X)$ of lists with values in $X$. If one thinks of the objects in a category to be data types, then we can ask even more. If $f:X\to Y$ is a function (between the data types), then this also induces  a function from the $X$-valued lists to the $Y$-valued lists as follows:
\begin{align}\label{eqn:function_on_list}
  \List(f) : \List(X)&\to \List(Y)
  \\
  [x_1,\ldots x_n] &\mapsto [f(x_1),\ldots,f(x_n)]. \notag
\end{align}

\begin{rem}
  The ``\ldots'' above are informal --- a formal definition would define $\List(f)$ by structural recursion on lists, of course.
\end{rem}

A \textit{functor} formalizes this phenomenon:
\begin{dfn} Let $\CC$ and $\DD$ be categories. A \textbf{functor} $F$ from $\CC$ to $\DD$ consists of the following data:
\begin{itemize}
\item A function 
\[
\Ob{\CC} \to \Ob{\DD},
\]
written as $X\mapsto F(X)$.
\item For each $X,Y\in \Ob{\CC}$, a function
\[
\CHom{\CC}{X}{Y} \to \CHom{\DD}{F(X)}{F(Y)},
\]
written as $f\mapsto F(f)$.
\end{itemize}
Moreover, this data should satisfy the following properties:
\begin{itemize}
\item (\textbf{Preserves composition}) For $f\in \Hom[\CC]{X}{Y}$ and $g\in \Hom[\CC]{Y}{Z}$, we have $F(\co f g) = \co {Ff}{Fg}$.
\item (\textbf{Preserves identity}) For $X\in\CC$, we have $F(\Id[X]) = \Id[F(X)]$.
\end{itemize}
\end{dfn}

\begin{rem}[Explanation]
	A function from a set $X$ to a set $Y$ maps every element of $X$, to an element in $Y$.
	Analogously, a \emph{functor} from a category $\CC$ to a category $\DD$ maps every object of $\CC$ to an object in $\DD$.
	(This is the first piece of data in the definition of a functor; made formal as a function between the underlying types/collections/sets of objects.)
	However, a category does not only consist of a set of objects.
	(Indeed, between every two objects $c_1, c_2 \in \Ob{\CC}$, we have a set of morphisms; denoted $\CHom{\CC}{x}{y}$).
	Hence, a functor, as a ``morphism'' between categories, should also map every morphism in $\CC$ to a morphism in $\DD$ (while suitably preserving domain and codomain of the morphism).
	(This is the second piece of data in the definition of a functor.)
	
	In the examples and exercises below, the first piece (i.\,e.,\, the function between the collection of objects) is always given.
	(Why? Because between fixed categories, there may be many functors, or none.)
	While the second piece of data is not uniquely determined by the function between the object-sets, there is often a canonical (obvious) choice.
	Below, we ask the reader to find the canonical choice.
	A functor is completely determined by the function (or action) on objects and morphisms.
	However, in order to guarantee that these functions are well-behaved (w.\,r.\,t.,\, to the structure given by the categories), we furthermore require that the composition and identity are preserved.
They need to be checked to complete the definition of a functor.
	(Often, these properties hold by construction; that is, by unfolding the data, provided in the second piece.)
\end{rem}

\begin{exa} \label{example:functor_list} The \textbf{list-functor} (on sets), denoted by $\List$, is the functor from $\SET$ to $\SET$ defined by the following data:
\begin{itemize}
\item The function on objects is given by:
\[
\Ob{\SET}\to \Ob{\SET}: X\mapsto \List(X).
\]
\item For each $X,Y\in\SET$, the function on morphisms is given by
\[
\CHom{\SET}{X}{Y} \to \CHom{\SET}{\List(X)}{\List(Y)}: f\mapsto \List(f),
\]
where $\mathbf{List}(f)$ is given in \cref{eqn:function_on_list}.
\end{itemize}

\end{exa}

\begin{exer}
  Show that $\List$ is a  functor, that is, show that it preserves identity and composition of functions.
  Hint: use structural induction on lists.
\end{exer}

\begin{exer}
  Consider the function $\Ob{\Maybe} : \Ob\SET \to \Ob\SET$ sending a set $X$ to $X + \{*\}$.
  For any two sets $X$ and $Y$ and $f : X \to Y$, define a function
  \[ \Maybe(f) : \Ob\Maybe X \to \Ob\Maybe Y\]
  and show that this assignment satisfies the functor laws.
\end{exer}

\begin{exer}[\cref{sol:functor_prod_on_left}]\label{exer:functor_prod_on_left}
  Let $A \in \Ob\SET$.
  Construct a functor $(\times A) : \SET \to \SET$ that, on objects, is given by
  \[ (\times A) X := X \times A. \]
\end{exer}

\begin{exer}
  Let $\CC$ be a category with chosen products, and let $A \in \Ob\CC$.
  Construct a functor $(\times A) : \CC \to \CC$ that, on objects, is given by
  \[ (\times A) X := X \times A. \]
\end{exer}

\begin{exer}
  Let $A \in \Ob\SET$.
  Construct a functor $(+ A) : \SET \to \SET$ that, on objects, is given by
  \[ (+ A) X := X + A. \]
\end{exer}

\begin{exer}
  Let $\CC$ be a category with chosen coproducts, and let $A \in \Ob\CC$.
  Construct a functor $(+ A) : \CC \to \CC$ that, on objects, is given by
  \[ (+ A) X := X + A. \]
\end{exer}

\begin{exer}
  Let $R \in \Ob\SET$ be a set.
  Construct a functor $(R \to) : \SET \to \SET$ that, on objects, is given by
  \[ (R \to) X := R \to X. \]
\end{exer}

\begin{exer}
  Let $\CC$ be a category and let $R \in \Ob\CC$.
  Construct a functor $\CHom \CC R - : \CC \to \SET$ that, on objects, is given by
  \[ (\CHom \CC R -) X := \CHom \CC R X. \]
\end{exer}

\begin{exer}\label{ex:poset_functors}
  Let $(X,\leq_X)$ and $(Y,\leq_Y)$ be preordered sets.
  Describe the functors from $\PREtoCAT(X,\leq_X)$ to $\PREtoCAT(Y,\leq_Y)$.
  Before writing out the definitions, what would you expect the answer to be?
\end{exer}

\begin{exer}
  Let $\CC$ be a category with chosen products $(A\times B, \pi_A, \pi_B)$ for any two objects $A$ and $B$.
  Construct a functor
  \[ (\times) : \CC\times \CC \to \CC\]
  from the product category $\CC\times \CC$ to $\CC$.
  The objects of $\CC\times \CC$ are pairs of objects in $\CC$, and morphisms $\CHom{(\CC\times\CC)}{(X,X')}{(Y,Y')}$ are pairs $(f : X \to Y, f' : X' \to Y')$ of morphisms in $\CC$.
\end{exer}

\begin{exer}
  Let $\CC$ be a category with chosen coproducts $(A + B, \iota_A, \iota_B)$ for any two objects $A$ and $B$.
  Construct a functor
  \[ (+) : \CC\times \CC \to \CC\]
  from the product category $\CC\times \CC$ to $\CC$.
\end{exer}

\begin{exer}
\label{exer:const-functor}
Let $\CC$ and $\DD$ be categories and $d : \DD$.
Construct a functor $\const_d : \CC \to \DD$ that, on objects, is given by $\const_d(c) := d$.
\end{exer}

\begin{exer}\label{ex:monoid_functors}
  Let $(M,m,e)$ be a monoid and let $\MONtoCAT(M,m,e)$ be its corresponding category as defined in \cref{monoidcategory}.
  Describe the functors from $\MONtoCAT(M,m,e)$ to $\SET$.
\end{exer}

\begin{reading*}
  We do not discuss here whether/when/how (co)limits can be transported along functors.
  You can find some information on this in \cite[\S 5.3]{leinster}.
\end{reading*}

\section{Categories as Objects of a Category?}
Notice that a functor is a function between categories which preserves the structure of a category. So by the \textit{philosophy} of category theory, this would define a category whose objects are categories and whose morphisms are functors. In order to make this precise, we would also need a \textit{identity functor} and we should have a \textit{composition of functors}.

\begin{exa}\label{example:functor_id} Let $\CC$ be a category. The \textbf{identity functor on $\CC$}, denoted by $\Id[\CC]$, is the functor specified by the following data:
\begin{itemize}
\item The function on objects is given by
\[
\Ob{\CC}\to \Ob{\CC}: X\mapsto X.
\]
\item For each $X,Y\in\CC$, the function on morphisms is given by
\[
\CHom \CC X Y\to \CHom \CC X Y: f\mapsto f.
\]
\end{itemize}
\end{exa}

\begin{exer} Show that $\Id[\CC]$ (defined in \cref{example:functor_id}) satisfies the properties of a functor, i.e. $\Id[\CC]$ is indeed a functor.
\end{exer}

\begin{exa}\label{example:functor_comp} Let $\CC,\DD$ and $\EE$ be  categories and $F:\CC\to\DD$ and $G:\DD\to\EE$ functors. The \textbf{composition functor of $F$ with $G$}, denoted by $F\Comp G$, is the functor specified by the following data:
\begin{itemize}
\item The function on objects is given by
\[
\Ob{\CC}\to \Ob{\EE}: X\mapsto G(F(X)).
\]
\item For each $X,Y\in\CC$, the function on morphisms is given by
\[
\CHom \CC X Y\to \CHom{\EE}{G(F(X))}{G(F(Y))}: f\mapsto G(F(f)).
\]
\end{itemize}
\end{exa}

\begin{exer} Show that $F\Comp G$ (defined in \cref{example:functor_comp}) satisfies the properties of a functor, i.e. $F\Comp G$ is indeed a functor.
\end{exer}

\begin{dfn} The \textbf{category of categories}, denoted by $\CAT$, is the category specified by the following data:
\begin{itemize}
\item An object is a category.
\item If $\CC, \DD\in\CAT$ are categories, then is $\CHom \CAT \CC \DD$ the collection of all functors from $\CC$ to $\DD$.
\item The identity morphism on a category $\CC$ is the identity functor on $\CC$ defined in \cref{example:functor_id}.
\item The composition of morphisms, i.e. functors, is the composition of functors defined in \cref{example:functor_comp}.
\end{itemize} 
\end{dfn}

\begin{exer} Show that $\CAT$ satisfies the property of a category, i.e. $\CAT$ is indeed a category.
\end{exer}

\begin{rem}
  When showing that $\CAT$ is a category, one is forced to consider \emph{equality of objects} when showing that two functors are equal.
  This goes against the spirit of category theory, where we only ever consider \emph{equality of (parallel) morphisms}.
  We want to consider two objects ``the same'' when they are \emph{isomorphic}, not when they are \emph{equal}. Of course, any two equal objects are isomorphic to each other, but not the other way round; for instance, in the category of sets, the cartesian product $A \times B$ is isomorphic to $B \times A$, but they are not equal.

  To stay within the spirit of category theory, one can instead consider $\CAT$ as a \textbf{bicategory}.%
  \footnote{See, e.g.,   \url{https://ncatlab.org/nlab/show/bicategory\#detailedDefn} for a definition of bicategories.}
  In a bicategory, one has one more layer of things: objects, morphisms, and 2-cells between parallel morphisms. One also calls objects ``0-cells'' and morphisms ``1-cells'', for consistency.
  Importantly, in a bicategory, the laws concerning 1-cells (as stated in \cref{dfn:category}) do not hold up to equality, but only up to isomorphism of 2-cells.

  An important example is the bicategory given by the following data, which we only list partially:
  \begin{enumerate}
  \item 0-cells are categories;
  \item 1-cells are functors;
  \item 2-cells are natural transformations (see \cref{sec:nat-trans});
  \item composition and identity of 1-cells is composition and identity of functors.
  \end{enumerate}

  We do not delve into bicategories in these notes; an introductory text is, for instance, Leinster's \cite{leinster:basic-bicats}.
\end{rem}

\chapter{Inductive Datatypes and Initial Algebras}
\label{sec:initial-algs}

In this chapter, we discuss how inductive datatypes as used in functional programming give rise to initial objects in suitable categories.

We first look at some example datatypes in \cref{sec:examples}.
Afterwards, we study a general framework that allows us to discuss many datatypes at once, in \cref{sec:datatypes-as-initial}.

\section{Examples}
\label{sec:examples}

In \cref{sec:universal}, it was discussed how the empty type, the unit type, the product type, and the sum type can be interpreted in a category.
Those type constructors were completely determined by how they interact with the other objects in the category.
Furthermore, as \cref{exer:coproduct_iff_initial_in_subcategory} illustrates, the aforementioned type constructors can all be described as initial objects.
It turns out that these observations are not specific to those type constructors, but this principle applies to many data types one uses in a functional language.
That this is indeed the case will be illustrated in this section with a couple of examples.
First, we look at the type of natural numbers (\cref{exer:nat-initial}), then we will see that the same can be done for more complex types such as those used in the development of programming languages (\cref{exer:aexp}).

\begin{exer}\label{exer:nat-initial}
  Consider the datatype
\begin{lstlisting}[mathescape=true]
data $\NN$ ::=
| $\Zero$ : $\NN$
| $\Succ$ : $\NN \to \NN$
\end{lstlisting}
This type is defined recursively.
First, $\NN$ contains an element called $\Zero$.
Second, for every element (natural number) $n$ there is a new element $\Succ(n)$, which corresponds to $n + 1$.
That is, $\NN$ is closed under the successor.
Alternatively, we can say that $\NN$ is the minimal type/set which is closed under these constructors.

Since $\NN$ is defined recursively, this implies that functions out of $\NN$ can also be defined recursively.
That is, this inductive datatype comes with a recursion principle to define functions from the natural numbers to any other type $X$,
\[
  \binaryRule
  {z \in X}
  {s : X \to X}
  {\rec(z,s) : \NN \to X}
  {rec}
\]
such that
\begin{equation}\label{eq:computation-rules-nat}
  \rec(z,s)(\Zero) = z \quad\text{ and }\quad \rec(z,s)(\Succ(n)) = s (\rec(z,s)(n)).
\end{equation}

The recursion principle can be stated more generally.
Indeed, $z \in X$ is a proposition $P$ dependent on $z$, that is $P(z) := z \in X$.
Furthermore, the function $s : X \to X$ is used to witness that if $n \in \NN$ is mapped to $x_n \in X$, then $\Succ(n)$ is mapped to $x_{\Succ(n)} \in X$. 
Hence, $s : X \to X$ can be rewritten as $\forall n \in \NN, P(n) \Rightarrow P(\Succ(n))$.
Hence, the recursion principle can be stated more generally via its induction principle:
\[
  \binaryRule
  {P(\Zero)}
  {\forall n : \NN, P(n) \Rightarrow P(\Succ (n))}
  {\forall n : \NN, P(n)}
  {ind}
\]

The recursion principle can be summarized as saying that for every set $X$ together with a chosen element $z \in X$ and a function $s : X \to X$, there is a unique function $f$ from $\NN$ to $X$ such that $f(\Zero) = z$ and $f(\Succ(n)) = s(f(n))$.
The unique existence means precisely that $\NN$, together with its constructors, is initial among those triples $(X, z, s)$.
This is made precise by considering the following category:
\begin{itemize}
\item Objects are triples $(X, z \in X, s : X \to X)$ with $X$ a set;
\item Morphisms from $(X, z \in X, s : X \to X)$ to $(X', z' \in X', s' : X' \to X')$ are functions
  $f : X \to X'$ such that the following diagrams commute:
  \[
    \begin{tikzcd}
      1 \ar[r, "z"] \ar[rd, "z'"']
      &
      X \ar[d, "f"]
      \\
      &
      X'
    \end{tikzcd}
    \quad
    \begin{tikzcd}
      X \ar[r, "s"] \ar[d, "f"']
      &
      X \ar[d, "f"]
      \\
      X' \ar[r, "s'"]
      &
      X'
    \end{tikzcd}
  \]
\item Composition and identity are given by composition of functions in $\SET$.
  (Check that this is well-defined, that is, that the composition of two functions making the above diagrams commute makes the right diagrams commute again.)
\end{itemize}

We now show that the recursion and induction principles can be used to show that the natural numbers, with $\Zero$ and $\Succ$, are an initial object in this category.

\begin{enumerate}
\item Draw the equations of \cref{eq:computation-rules-nat} as diagrams in the category of sets.
\item Show that the triple $(\NN, \Zero, \Succ)$ is an initial object in this category.
\end{enumerate}

\end{exer}

The next exercise is very similar to \cref{exer:nat-initial}.

\begin{exer}\label{exer:aexp}
  Consider the datatype
\begin{lstlisting}[mathescape=true]
data Exp ::=
| Int     : $\mathbb{Z}$ $\to$ Exp
| Plus    : Exp $\times$ Exp $\to$ Exp
| Squared : Exp $\to$ Exp
\end{lstlisting}
  and consider the following category:
  \begin{itemize}
  \item Objects are quadruples $(X, I : \mathbb{Z} \to X, P : X \to X \to X, S : X \to X)$ with $X$ a set/type;
  \item Morphisms from $(X, I, P, S)$ to $(X', I', P', S')$ are functions
    $f : X \to X'$ such that the following diagrams commute:
    \[
      \begin{tikzcd}
        \mathbb{Z} \ar[r, "I"] \ar[rd, "I'"']
        &
        X \ar[d, "f"]
        \\
        &
        X'
      \end{tikzcd}
      \quad
      \begin{tikzcd}
        X\times X \ar[r, "P"] \ar[d, "f \times f"']
        &
        X \ar[d, "f"]
        \\
        X' \times X' \ar[r, "P'"]
        &
        X'
      \end{tikzcd}
      \quad
      \begin{tikzcd}
        X \ar[r, "S"] \ar[d, "f"']
        &
        X \ar[d, "f"]
        \\
        X' \ar[r, "S'"]
        &
        X'
      \end{tikzcd}
    \]
  \item Composition and identity are given by composition of functions in $\SET$. (Check that this is well-defined.)
  \end{itemize}

We show that the datatype of expressions, together with its constructors, forms an initial object in the category defined above.

\begin{enumerate}
\item Write down the recursion and induction principles for |Exp|.
\item Write down the computation rules (the analogs of \cref{eq:computation-rules-nat}) for |Exp|, and draw them as diagrams.
\item Show that the quadruple consisting of the type |Exp| together with the functions |Int|, |Plus|, and |Squared|, is an initial object in this category. 
\end{enumerate}
  
\end{exer}

\begin{exer}
  Define an ``evaluation'' function $\mathsf{ev} : \mathsf{Exp} \to \mathbb{Z}$ that associates to any expression its ``semantics'' or ``denotation'', that is, the integer the expression denotes.
  Specifically, define this function as a morphism in the category of \cref{exer:aexp}.
\end{exer}

\begin{exer}[\cref{sol:natlist_is_initial}]\label{exer:natlist_is_initial}
  Consider the datatype
\begin{lstlisting}[mathescape=true]
  data NatList ::=
  | nil     : NatList
  | cons    : $\NN$ $\to$ NatList $\to$ Natlist
\end{lstlisting}

  \begin{enumerate}
  \item Write down the recursion and induction principles for |Natlist|.
  \item  Write down the computation rules (the analogs of \cref{eq:computation-rules-nat}) for |NatList|, and draw them as diagrams.
  \item Construct a suitable category $\Cat{L}$ and show that ($\mathsf{NatList}, \nil, \cons$) is initial in that category.
  \end{enumerate}
  
\end{exer}

\section{Datatypes as Initial Algebras}
\label{sec:datatypes-as-initial}

\begin{reading*}
  This chapter is strongly inspired by Varmo Vene's Ph.D.\ thesis \cite[Chapter 2]{vene_phd}.
  A good explanation of recursion on lists is given in Graham Hutton's tutorial paper \cite{DBLP:journals/jfp/Hutton99}.
  The tutorial on (co)algebras and (co)induction by Jacobs and Rutten \cite{jacobs-rutten-tutorial} provides an excellent overview to the categorical view on inductive and coinductive datatypes.
\end{reading*}

In \cref{sec:examples}, it was discussed how inductively defined data types naturally give rise to a category where the data type is the initial object in that category, and that the initiality correspond to the recursion principle.
Even though different types lead to different categories, those categories are all tuples consisting of a set together with morphisms (modeling the constructor) into the set.
In this section, we generalize the construction of those categories.
In particular, this generalization allows us to compute the recursion principle for inductively defined data types.

To model the constructors, observe that multiple functions $A_0 \to X, \cdots, A_n \to X$ with the same codomain can be equivalently described as one function $(A_0 + \cdots + A_n) \to X$ (see \cref{exer:coproduct-represent}).
For example, an object in the category which models the natural numbers is equivalently a pair $(X \in \SET, [z,s] : 1 + X \to X)$.


\begin{dfn} Let $F:\CC\to \CC$ be an endofunctor. An \textbf{$F$-algebra} consists of the following data:
\begin{enumerate}
\item An object $X\in \Ob \CC$.
\item A morphism $\phi \in \CHom{\CC}{F(X)}{X}$.
\end{enumerate}
\end{dfn}

\begin{intu}
An algebra is roughly a set equipped with some operations, such as multiplication.
The \emph{arities}, that is, the inputs, of the operations are determined by the functor $F$.
An important class of functors are \emph{polynomial functors} built using the coproduct $(+)$ and the product $(\times)$.
Intuitively, the different summands of a polynomial functor each correspond to one datatype constructor, whereas the use of the product indicates that a constructor takes several inputs.
\end{intu}

\begin{exa}\label{exa:nno_initial_alg_maybe}
 Let $\Maybe : \SET \to \SET$ be the endofunctor given on objects by $\Maybe(X) := 1 + X$. 
 A $\Maybe$-algebra is a pair $(X,\phi)$ of a set $X$ and a function $\phi : 1 + X \to X$.
 By the universal property of the coproduct, $\phi$ is given, equivalently,
 by two functions $z$ and $s$ as follows.
 \begin{align*}
    z : 1 &\to X
    \\
    s : X &\to X
 \end{align*}
Here, think of ``$z$'' standing for ``zero'', and ``$s$'' standing for ``successor''.
\end{exa}

\begin{exa}\label{example:algebra_of_monoids} Let $F$ be the endofunctor induced by:
\begin{align*}
  F: \SET &\to \SET
  \\
  X &\mapsto 1 + (X\times X).
\end{align*}
An $F$-algebra consists of a set $X\in\SET$ together with a function $\phi:\CHom{\SET}{1 + (X\times X)}{X}$. Since $\phi$ is function from the disjoint union, we have that $\phi$ corresponds uniquely to two functions:
\begin{align*}
e &: 1\to X
\\
m &: X\times X\to X.
\end{align*}
This is precisely the data of a monoid.

Conversely, if $(M,m,e)$ is a monoid, then this induces a function
\[
1 + (M\times M) \xrightarrow{\phi = [\phi_e,\phi_m]} M,
\]
defined by pattern matching as follows:
\begin{align*}
\phi_e : 1 &\to M
\\
\star &\mapsto e
\\
\phi_m : M\times M &\to M 
\\
(x,y) &\mapsto m(x,y)
\end{align*}

The pair $(M, \phi)$ is an $F$-algebra.

\end{exa}

\begin{rem}
 Note that a monoid $(M,m,e)$ also satisfies some laws.
 The laws are not expressed in \cref{example:algebra_of_monoids}.
 To incorporate the laws, one studies instead algebras of a monad.
\end{rem}

\begin{dfn}\label{dfn:alg-hom}
Let $F:\CC\to \CC$ be an endofunctor and $(X,\phi)$ and $(Y,\psi)$ be $F$-algebras. 
A \textbf{($F$-algebra) homomorphism} from $(X,\phi)$ to $(Y,\psi)$ is a morphism $f\in \CHom{\CC}{X}{Y}$ such that the following diagram commutes:
\[
\begin{tikzcd}
FX
\arrow[r, "\phi"] 
\arrow[d,swap, "Ff"]
& X
\arrow[d, "f"] 
\\
FY
\arrow[r, swap, "\psi"] 
& Y
\end{tikzcd}
\]
\end{dfn}

\begin{exer} 
  Let $F$ be the endofunctor defined as in \cref{example:algebra_of_monoids}, i.e., the endofunctor whose algebras correspond to monoids. 
  Describe the $F$-algebra homomorphisms.
\end{exer}

\begin{dfn}\label{definition:category_of_Falgebras} Let $F:\CC\to \CC$ be an endofunctor. The \textbf{category of $F$-algebras}, denoted by $\ALG{F}$, is defined by the following data:
\begin{itemize}
\item The objects are the $F$-algebras.
\item The morphisms are the $F$-algebra homomorphisms.
\item The identity on $(X,\phi)$ is given by the identity $\Id[X]$ in $\CC$.
\item The composition is given by the composition of morphisms in $\CC$.
\end{itemize}
\end{dfn}

\begin{exer}
  \begin{enumerate}
  \item[]
  \item Draw two diagrams to illustrate \cref{definition:category_of_Falgebras}.
  \item Show that $\ALG{F}$ satisfies the properties of a category.
  \end{enumerate}
\end{exer}

We are interested in \textbf{initial objects of $\ALG{F}$}, if they exist.
We call these ``initial $F$-algebras''.
For a general endofunctor $F$, an initial $F$-algebra does not need to exist;
but for many interesting choices of $F$, such an initial object does exist.
Before coming to the general definition (see \cref{dfn:initial-alg}),
we consider an example.

\begin{exer}
  Consider the functor $\Maybe : \SET \to \SET$.
  \begin{enumerate}
  \item Show that the initial $\Maybe$-algebra is given by the pair $(\NN, [\Zero,\Succ])$, 
    where $\NN$ is the set of natural numbers, and $\Zero : 1 \to \NN$ and $\Succ : \NN\to\NN$ 
    are the function picking out zero and the successor function, respectively.
  \item Given any other $\Maybe$-algebra $(X,[z,s])$, unpack what it means for the square of \cref{dfn:alg-hom} to commute.
  \item Compare the data from the previous exercise to a definition of a function $f : \NN \to X$ by pattern matching (e.g., in Haskell).
  \end{enumerate}
\end{exer}

\begin{dfn}\label{dfn:initial-alg}
  Let $F:\CC\to\CC$ be an endofunctor. An \textbf{initial $F$-algebra} (if it exists) is an initial object in $\ALG{F}$.
  Unfolding the definition, this means that it is an $F$-algebra $(\Initalg{F}, \In)$ such that for any $F$-algebra $(X,\phi)$, there exists a unique morphism $\catam{\phi} \in \CHom{\CC}{\Initalg{F}}{X}$ such that the following diagram commutes:
\begin{equation}\label{eq:initial-f-alg}
\begin{tikzcd}
{F\Initalg{F}} 
\arrow[r, "\In"] 
\arrow[d,swap, "F\catam{\phi}"]
& {\Initalg{F}} 
\arrow[d, "\catam{\phi}"] 
\\
FX
\arrow[r, swap, "\phi"] 
& X
\end{tikzcd}
\end{equation}
The morphism $\catam{\phi}$ is called the \emph{catamorphism} generated by $\phi$.
\end{dfn}

\begin{exer}
Let $F : \CC \to \CC$ be an endofunctor.
The initial $F$-algebra, if it exists, is unique up to isomorphism.
\end{exer}

\begin{exer}[\cref{sol:in_catamorphism_id}]\label{exer:in_catamorphism_id}
  Let $F:\CC\to\CC$ be an endofunctor and let $(\Initalg F, \In)$ be an initial algebra.
  Show that
  \[\catam{\In} = \Id[\Initalg{F}].\]
\end{exer}

\begin{exer}\label{exer:bool_as_initial_algebra}
  Let $\mathbf{Bool}$ be the inductive data type generated by the following two constructors:
\begin{lstlisting}
    True  : Bool
    False : Bool
\end{lstlisting}
  Define an endofunctor $F:\SET\to \SET$ such that the $F$-algebras can be described as triples $(X, b_1, b_2)$ with $X$ a set and $b_1,b_2\in X$.
  
  Moreover, show that $(\mathsf{Bool}, \mathsf{True}, \mathsf{False})$ is an initial object in $\ALG{F}$.
\end{exer}

Observe that $\mathbf{Bool}$ is the coproduct $1 + 1$.
Hence, more generally:
\begin{exer}\label{exer:coproduct_as_initial_algebra}
  The disjoint union (i.e., the coproduct) of two sets $X$ and $Y$ can also be described as an inductive data type; indeed, it is generated by the following two constructors:
\begin{lstlisting}
    f : X -> X+Y
    g : Y -> X+Y
\end{lstlisting}
  Define an endofunctor $F:\SET\to \SET$ such that the $F$-algebras can be described as triples $(C, i_l, i_r)$ with $C$ a set and $i_l : X\to C, i_r : Y\to C$ be functions.
 
  Moreover, show that $(X + Y, \inl, \inr)$ is the initial object in $\ALG{F}$, where $X+Y$ is the disjoint union of $X$ and $Y$ (i.e. the coproduct in $\SET$) and $\inl:X\to X+Y, \inr:Y\to X+Y$ the canonical inclusions.
\end{exer}

\begin{exer}\label{exer:conatural_numbers_are_not_initial}
Let $\mathbb{N}^{c}$ be the conatural numbers, i.e. $\mathbb{N} + \{\infty\}$. Consider the endofunctor $\Maybe : \SET \to \SET$ (defined in \cref{exa:nno_initial_alg_maybe}), i.e. the functor given on objects by
\begin{align*}
  \Ob{\Maybe} : \Ob{\SET} &\to \Ob{\SET}
  \\
  X &\mapsto 1 + X.
\end{align*}
The functions
\begin{align*}
zero^{c} : \mathbf{1}&\to \mathbb{N}^{c} \\ 
           \star&\mapsto 0,\\
succ^{c} : \mathbb{N}^{c}&\to \mathbb{N}^{c} \\
          x & \mapsto 
\begin{cases}
n+1,\quad  \text{ if } x := n\in\mathbb{N},\\
\infty,\quad  \text{ if } x := \infty.\\
\end{cases}
\end{align*}
form a $\Maybe$-algebra. However, show that $(\mathbb{N}^{c},zero^{c},succ^{c})$ is not an initial $\Maybe$-algebra.
\end{exer}

\begin{exer}[\cref{sol:initialalg_for_idfun_with_initialob}]\label{exer:initialalg_for_idfun_with_initialob} Let $\CC$ be a category with an initial object $\bot$. Show that $(\bot, \Id[\bot])$ is the initial algebra for the identity (endo)functor on $\CC$.
\end{exer}

\begin{exer}\label{exer:initialalg_for_list}
  Let $A$ be a set and define $F$ to be the functor induced by 
\begin{align*}
  F_A:\SET &\to\SET
\\
  X &\mapsto 1 + (A\times X).
\end{align*}

\begin{enumerate}
\item \label{enum:list-alg} Show that an $F_A$-algebra consists of a triple $(X,n,c)$, where $X$ is a set, $n\in X$ is an element of $X$, and $c : A \times X \to X$ is a function.
\item Show that the initial $F_A$-algebra is given by the set $\List(A)$ of $A$-valued lists, together with constants $\nil \in \List(A)$ and $\cons : A \times \List(A) \to \List(A)$.
\item Given any other $F_A$-algebra $(X,n,c)$, unpack what it means for the square of \cref{dfn:alg-hom} to commute.
Compare it to a definition of a function $f : \List(A) \to X$ by pattern matching.
\end{enumerate}

\end{exer}

\begin{rem}
  In the case of lists, the operator $\catam{\_}$ is also known as |fold|, which in Haskell is defined as follows:
  \begin{lstlisting}
    fold            :: (a → b → b) → b → ([a] → b)
    fold f v   []   =  v
    fold f v (x:xs) =  f x (fold f v xs)
  \end{lstlisting}
  Compare the input of |fold| to the data of an $F_A$-algebra given in \cref{enum:list-alg} of \cref{exer:initialalg_for_list}.
\end{rem}

\begin{exer}\label{exer:list-functions-as-fold}
  Define the following functions as a catamorphism, that is, using |fold|.
  In each case, draw the diagram corresponding to Diagram~\ref{eq:initial-f-alg} of \cref{dfn:initial-alg}.
  \begin{enumerate}
  \item |sum :: [Int] →  Int|
  \item |product :: [Int] →  Int|
  \item |and :: [Bool ] →  Bool|
  \item |or :: [Bool ] →  Bool|
  \item |(++) :: [a] →  [a] →  [a]|
  \item |length :: [a] →  Int|
  \item |reverse :: [a] →  [a]|
  \item |map :: (a →  b) →  ([a] →  [b])|
  \item |filter :: (a →  Bool ) →  ([a] →  [a])|
  \end{enumerate}
  Solutions are given in \cite[\S2]{DBLP:journals/jfp/Hutton99}.

  Hint: a systematic approach to reformulating functions on lists defined by explicit recursion in terms of |fold| is described in \cite[\S3.3]{DBLP:journals/jfp/Hutton99}.
\end{exer}

\begin{exer}\label{exer:initialalg_for_btree} Let $A$ be a set and define $F$ to be the functor induced by 
\[
F_A:\SET\to\SET : X\mapsto A + (X\times X).
\]
Show that the initial $F_A$-algebra is given by the set $\BinTree(A)$ of $A$-valued binary trees.
\end{exer}

\begin{rem} Notice that in \cref{exer:initialalg_for_list} and \cref{exer:initialalg_for_btree}, we can consider the functor $F_A$ as a bifunctor where we vary $A$, i.e.
\[
F: \SET\to \SET\to \SET: (A,X)\mapsto F_A(X).
\]
In particular, under the assumption that for every $A\in\SET$ the initial $F_A$-algebra exists, we can wonder if the assignment 
\[
\Ob{\SET} \to \Ob{\SET} : A\mapsto \mu F_A ,
\]
can be extended to a functor. The following exercise answers this question positively for arbitrary categories.
\end{rem}

\begin{exer}[\cref{sol:initialalg_for_bifunctor_functor}]\label{exer:initialalg_for_bifunctor_functor} Let $F:\CC\to\CC\to\CC$ be a bifunctor such that for any object $A\in\CC$, the initial algebra for the functor induced by 
\[
F_A : \CC\to\CC : X\mapsto F(A,X),
\]
exists. Show how
\[
\Ob{\CC} \to \Ob{\CC} : A\mapsto \mu F_A ,
\]
induces a functor.
\end{exer}

\section{Fusion Property}\label{sec:fusion}

The fusion property of \cref{exer:fusion-property} can be used to ``fuse'' a composition of functions into one function, possibly leading to more efficient code.

We are going to exemplify this using the datatypes of lists and of natural numbers.
Recall that $(\NN, [\Zero,\Succ])$ is the initial $\Maybe$-algebra.
We also write $(+1)$ for $\Succ$.

\begin{exer}[\textbf{Fusion property}, \cref{sol:fusion-property}]\label{exer:fusion-property}
  Let $F:\CC\to\CC$ be an endofunctor and let $(\Initalg F, \In)$ be an initial algebra. Show that
  for $F$-algebras $(X,\phi)$ and $(Y,\psi)$ and $f\in\CHom{\CC}{X}{Y}$, we have 
\[
\co{\phi}{f} = \co{F(f)}{\psi} \implies \co{\catam{\phi}}{f} = \catam{\psi}.
\]
This is summarized in the following diagram:
\begin{equation}\label{eq:initial-f-alg-composition}
\begin{tikzcd}
F\Initalg{F} 
\arrow[r, "\In"] 
\arrow[d,swap, "F\catam{\phi}"]
\ar[dd, bend right=60, "F\catam{\psi}"']
&
\Initalg{F}
\arrow[d, "\catam{\phi}"]
\ar[dd, bend left=60, "\catam{\psi}"]
\\
FX
\arrow[r, swap, "\phi"]
\ar[d, "Ff"'] 
&
X \ar[d, "f"]
\\
FY \ar[r, "\psi"']
&
Y
\end{tikzcd}
\end{equation}
\end{exer}

\begin{thm}[\textbf{Lambek's theorem}]
  Let $F:\CC\to\CC$ be an endofunctor and let $(\mu^F, \In)$ be an initial algebra. Then, $\In$ is an isomorphism whose inverse is given by $\Inv{\In} = \catam{F(\In)}$.
\end{thm}

\begin{exer} 
  Prove Lambek's theorem. 
\end{exer}

\begin{exer}[\cref{sol:list-concat-nil}, see also \cite{DBLP:journals/scp/Malcolm90}]
  \label{exer:list-concat-nil}
  Consider the function |(++) :: [a] →  [a] →  [a]| defined in \cref{exer:list-functions-as-fold},
  and |l :: [a]|.
  \begin{enumerate}
  \item Show that |nil ++ l = l|.
    
  \item Show that |l ++ nil = l|.
  \end{enumerate}
\end{exer}

\begin{exer}[{\cite[\S3.1]{DBLP:journals/jfp/Hutton99}}]
  Show that |(+1) . sum = fold (+) 1|, by showing that |(+1) . sum| makes Diagram~\ref{eq:initial-f-alg} of \cref{dfn:initial-alg} commute.
\end{exer}

\begin{reading*}
  The content of this section is very much inspired by \cite[\S3.2]{DBLP:journals/jfp/Hutton99}.
  You are strongly encouraged to read that section before reading the present section.

  The fusion property is called ``promotion theorem'' in Malcolm's work~\cite{DBLP:journals/scp/Malcolm90}.
\end{reading*}

\begin{exer}[See also \cite{DBLP:journals/scp/Malcolm90}]
  Consider the function |(++) :: [a] →  [a] →  [a]| defined in \cref{exer:list-functions-as-fold}.
  Show that |(l ++ m) ++ n = l ++ (m ++ n)| for any |l, m, n :: [a]|.
\end{exer}

\begin{exer}[{\cite[\S3.2]{DBLP:journals/jfp/Hutton99}}, \cref{sol:sum_plus_one_as_fold}]\label{exer:sum_plus_one_as_fold}
  Show that |(+1) . sum = fold (+) 1|, by using the fusion property of \cref{exer:fusion-property}.
\end{exer}

\begin{exer}
  Prove that |map g . map f = map (g . f)| using the fusion property.
\end{exer}

\begin{reading*}
  In \cref{sec:initial-algs,sec:fusion}, we have only looked at functions defined via \emph{iteration}, that is, functions defined as \emph{catamorphisms} of some $F$-algebra.
  While \emph{primitive recursive} functions can always be expressed as a catamorphism via tupling (see, e.g., \cite[\S4]{DBLP:journals/jfp/Hutton99} and \cite[\S3.1]{vene_phd}), it is more natural specify them via a slightly more sophisticated universal property, explained in detail in Vene's dissertation~\cite[Chapter~3]{vene_phd}.

  The interested reader might also study the tutorial by Fokkinga~\cite{Fokkinga_homo-cata} or the paper by Meijer et al.~\cite{DBLP:conf/fpca/MeijerFP91}.
  The paper \cite{DBLP:journals/scp/Malcolm90} by Malcolm contains more examples.
  
  A guide to further literature on recursion operators is given in \cite[\S6]{DBLP:journals/jfp/Hutton99}.
\end{reading*}

\chapter{Terminal Coalgebras and Coinductive Datatypes}\label{sec:coinductive}

\begin{reading*}
  We only give a brief introduction to (terminal) coalgebras in this section.
  A more systematic exploration of the topic is given in~\cite{DBLP:conf/fpca/MeijerFP91}.

  The paper \cite{DBLP:journals/scp/Malcolm90} by Malcolm contains further examples.
\end{reading*}

In \cref{sec:initial-algs}, we introduce (initial) algebras to model inductive types.
In this section, we introduce the dual notion thereof.
That is, we introduce (terminal) coalgebras which allows us to define coinductive data types.

\begin{dfn} Let $F:\CC\to \CC$ be an endofunctor. An \textbf{$F$-coalgebra} consists of the following data:
\begin{itemize}
\item An object $X\in \CC$.
\item A morphism $\phi \in \CHom{\CC}{X}{F(X)}$.
\end{itemize}
\end{dfn}
Notice that an $F$-algebra consists of a morphism $F(X)\to X$, while an $F$-coalgebra consists of a morphism $X\to F(X)$ in the other direction.

\begin{dfn} Let $F:\CC\to \CC$ be an endofunctor and $(X,\phi)$ and $(Y,\psi)$ be $F$-coalgebras. A \textbf{($F$-coalgebra) homomorphism} from $(X,\phi)$ to $(Y,\psi)$ is a morphism $f\in \CHom{\CC}{X}{Y}$ such that the following diagram commutes:
\[
\begin{tikzcd}
X
\arrow[r, "\phi"] 
\arrow[d,swap, "f"]
& FX
\arrow[d, "F(f)"] 
\\
Y
\arrow[r, swap, "\psi"] 
& FY
\end{tikzcd}
\]
\end{dfn}

\begin{exer} Define the category $\COALG{F}$ of $F$-coalgebras analogously to the category $\ALG{F}$ of $F$-algebras (as in \cref{definition:category_of_Falgebras}).
\end{exer}

\begin{dfn} Let $F$ be an endofunctor on $\CC$. The \textbf{terminal $F$-coalgebra} is the terminal object in $\COALG{F}$ which we denote by $(\Terminalcoalg{F}, \Out)$.
  
For $(X,\phi)$ an arbitrary $F$-coalgebra, we denote the unique morphism $(X,\phi) \to (\Terminalcoalg{F}, \Out)$ by $\anam{\phi}$, and we call a morphism of this form an \textit{anamorphism} (instead of a catamorphism as in \cref{dfn:initial-alg}).
\end{dfn}

\begin{exer} Spell out what it means for a coalgebra to be the terminal coalgebra.
\end{exer}

\begin{exer} Let $F:\CC\to\CC$ be an endofunctor and let $(\Terminalcoalg{F}, \Out)$ be a terminal coalgebra. Show that the following properties holds:
\begin{enumerate}
\item $\Id = \anam{\Out}$.
\item For $F$-algebras $(X,\phi)$ and $(Y,\psi)$ and $f\in\CHom{\CC}{X}{Y}$, we have 
\[
  \co f \psi   = \co \phi {F(f)} \implies \co{f}{\anam{\psi}} = \anam{\phi}.
\]
This is summarized in the following diagram:
\[
\begin{tikzcd}
X
\arrow[r, "\phi"] 
\arrow[d,swap, "f"]
\ar[dd, "\anam{\phi}"', bend right = 60]
&
FX
\arrow[d, "Ff"]
\ar[dd, "F\anam{\phi}", bend left = 60]
\\
Y
\arrow[r, swap, "\psi"]
\ar[d, "\anam{\psi}"']
&
FY
\ar[d, "F\anam{\psi}"]
\\
\Terminalcoalg{F}
\ar[r, "\Out"]
&
F\Terminalcoalg{F}
\end{tikzcd}
\]
\end{enumerate} 
\end{exer}

\begin{thm}[\textbf{Dual of Lambek's theorem}] Let $F:\CC\to\CC$ be an endofunctor and let $(\Terminalcoalg{F}, \Out)$ be a terminal coalgebra. Then $\Out$ is an isomorphism whose inverse is given by $\Inv{\Out} = \anam{F(\Out)}$.
\end{thm}

\begin{exer}
  Prove the dual of Lambek's theorem. 
\end{exer}

\begin{exer}\label{exer:terminalalg_for_idfun_with_terminalob} Let $\CC$ be a category with a terminal object $\top$. Show that $(\top, \Id[\top])$ is a terminal coalgebra for the identity (endo)functor on $\CC$.
\end{exer}

\begin{exer}[\cref{sol:conatural_numbers_terminal_coalgebra}] \label{exer:conatural_numbers_terminal_coalgebra} Let $F$ be the functor induced by 
\[
F:\SET\to\SET : X\mapsto 1 + X.
\]
Show that the terminal $F$-coalgebra is given by the following data:
\begin{itemize}
\item The underlying object is given by the set $\mathbb{N} + \{\infty\}$ of natural numbers with infinity.
\item The underlying function is given by the predecessor defined as follows: 
\begin{align*}
  \mathbb{N} + \{\infty\} &\to 1 + \mathbb{N} + \{\infty\}
  \\
  0 &\mapsto \star
  \\
  s(n) &\mapsto n
  \\
  \infty &\mapsto \infty
\end{align*}
where $\star$ is the unique element of $1$.
\end{itemize}
\end{exer}

\begin{exa}[Streams]
  The codata type of streams over a given set $A$ is given by the terminal coalgebra $(\Terminalcoalg{F_A}, \Out)$ of the functor $F_A (X) := A \times X$.
  We write $\Stream(A)$ for $\Terminalcoalg{F_A}$.
  The functions $\head : \Stream(A) \to A$ and $\tail : \Stream(A) \to \Stream(A)$
  are given by
  \begin{align*}
    \head &= \co{\Out}{\projl} : \Stream(A) \to A
    \\
    \tail &= \co{\Out}{\projr} : \Stream(A) \to \Stream(A)
  \end{align*}
  respectively. Put differently (recall the definition of the product in \cref{def:binproduct}), we have
  \[
    \Out = \langle \head, \tail \rangle : \Stream(A) \to A \times \Stream(A)
  \] 
  Given any two functions
  \begin{align*}
    h : C \to  A
    \\
    t : C \to C
  \end{align*}
  the anamorphism $\anam{\langle h, t \rangle }$ is the unique solution
  $f : C \to  \Stream(A)$
  of the equation system
  \begin{align*}
    \co{f} \head &= h
    \\
    \co f \tail &=  \co t f
  \end{align*}
  that is, the unique function $f : C \to \Stream(A)$ making the following square commute:
  \[
    \begin{tikzcd}[column sep = large]
      C
      \ar[r, "{\langle h, t \rangle}"]
      \ar[d, "\anam{\langle h, t \rangle}"']
      &
      A \times C
      \ar[d, "\Id \times \anam{\langle h, t \rangle}"]
      \\
      \Stream(A)
      \ar[r, "{\langle \head, \tail \rangle}"]
      &
      A \times \Stream(A).
    \end{tikzcd}
  \]
  
\end{exa}

\begin{exer}[\cref{sol:stream-of-nats}]\label{exer:stream-of-nats}
  Define, as an anamorphism, the function $\nats : \NN \to \Stream(\NN)$ which returns the stream of all natural
  numbers starting with the natural number given as the argument.

  What do you need to change to obtain instead the stream listing \emph{every other} number starting with the given one?
\end{exer}

\begin{exer}[\cref{sol:zip}]\label{exer:zip}
  Define, as an anamorphism, the function $\zip : \Stream(A) \times \Stream(B) \to \Stream(A \times B)$ that zips the argument streams together.
\end{exer}

\begin{exer}
  Define, as an anamorphism, the function $\map : (A \to B) \to \Stream(A) \to \Stream(B)$.
\end{exer}

\begin{exer}
  For a fixed set $A$, consider the functor given on objects by $F(X) := 1 + A \times X$.
  Its terminal $F$-coalgebra is the datatype $\Colist(A)$ of potentially infinite lists over elements in $A$, with suitable destructors $\head$ and $\tail$ and an ``exception'' in case the colist is empty.

  Define a function $\Colist(A) \to  \mathbb{N} + \{\infty\}$ counting the elements in a colist.  
\end{exer}

\begin{exer}
  Try to define $\filter : (A \to \Bool) \to \Stream(A) \to \Stream(A)$.
  What goes wrong?

  One might think that $\filter : (A \to \Bool) \to \Stream(A) \to \Colist(A)$ works instead.
  Does it?

  What about $\filter : (A \to \Bool) \to \Stream(A) \to \Stream(A+1)$?
  
\end{exer}


\chapter{Natural Transformations}
\label{sec:nat-trans}

Natural transformations are often considered the heart of category theory, and for good reason 
— the very subject was originally developed to express such transformations.
They provide the language to express many of the field's fundamental constructions.
Moreover, in the context of programming, natural transformations are crucial: they serve as the mathematical model for parametric polymorphism when a category is used to represent a programming language.

\section{Definition and First Examples}

\begin{dfn} Let $F,G: \CC\to\DD$ be functors. A \textbf{natural transformation} $\alpha$ from $F$ to $G$ consists of the following data:
\begin{itemize}
\item For each $X\in \Ob{\CC}$, a morphism $\alpha_X \in \CHom{\DD}{F(X)}{G(X)}$.
\end{itemize}
Moreover, this data should satisfy the following \textit{naturality condition}:\\
For each $f\in \CHom{\CC}{X}{Y}$, the following diagram should commute:
\begin{center}
\begin{tikzcd}
F(X) \arrow[r, "\alpha_X"] \arrow[d,swap, "F(f)"] & G(X) \arrow[d, "G(f)"]\\
F(Y) \arrow[r,swap, "\alpha_Y"] & G(Y)
\end{tikzcd}
\end{center}
\end{dfn}

\begin{nota} If $F,G:\CC\to\DD$ are functors, a natural transformation $\alpha$ from $F$ to $G$ is denoted as $\NatTrans{\alpha}{F}{G}$ or 
\begin{center}
\begin{tikzcd}[column sep=huge]
\CC
  \arrow[bend left=50]{r}[name=U,label=above:$\scriptstyle F$]{}
  \arrow[bend right=50]{r}[name=D,label=below:$\scriptstyle G$]{} &
\DD
  \arrow[shorten <=5pt, Rightarrow,to path={(U) -- node[label=right:$\alpha$] {} (D)}]{}
\end{tikzcd}
\end{center}
\end{nota}

\begin{exer}
Construct a natural transformation from the $\Maybe$ functor to the $\List$ functor, and conversely.
\end{exer}

\begin{exa} (\textbf{Currying}) Let $X$ be a set. Let $F := \SET(X, -)\times X : \SET\to\SET$ be the functor induced by the following data (on objects):
\[
Y\mapsto \SET(X,Y)\times X.
\]
The evaluation defines a natural transformation $\NatTrans{ev}{F}{\Id[\SET]}$ as follows:
\[
ev_Y : \SET(X,Y) \times X \to Y : (f,x) \mapsto f(x).
\]
Show that this indeed satisfies the naturality condition.
\end{exa}

\begin{exer}
  Given two preordered sets $(X,\leq_X)$ and $(Y, \leq_Y)$ and two functors $f, g : \PREtoCAT(X,\leq_X) \to \PREtoCAT(Y,\leq_Y)$ between their associated categories, what is a natural transformation from $f$ to $g$?
\end{exer}

Recall that for every object $d : \DD$ in a category gives rise to the constant functor (see \cref{exer:const-functor}).

\begin{exer}
  Can you define a natural transformation that counts the occurrences of elements from $X$ in a list?
  That is, for any set $X$, we would like to have a function of type $\List(X) \to X \to \NN$.
\end{exer}

\begin{exer}
Can you define a natural transformation that counts the number of elements in a list?
That is, for any set $X$, construct a function $len_X : \List(X) \to \NN$ and show that this is natural in $X$.
\end{exer}

\begin{exer}
  Let $(M,m,e)$ be a monoid and let $\MONtoCAT(M,m,e)$ be its corresponding category. Recall from \cref{ex:monoid_functors} that a functor from $\MONtoCAT(M,m,e)$ to $\SET$ is a set $X$ together with an action of $M$ on $X$, i.e. a function $\mu: M\times X\to X$ such that 
  \[
    \forall x\in X: \mu(e,x) = x, \quad \forall n_1,n_2\in M, x\in X: \mu(n_1, \mu(n_2,x)) = \mu(m(n_1,n_2), x).
  \]
  We will call a set $ X $ with an action of $ M $ on $ X $ an $ M $-set.
  Describe the natural transformations between $M$-sets.
\end{exer}

\begin{exer}
  Let $(X,\leq_X)$ and $(Y,\leq_Y)$ be posets.
  Recall from \cref{ex:poset_functors} that a functor between posets corresponds with an order-preserving function, i.e. $x_1 \leq_X x_2 \implies f(x_1) \leq_Y f(x_2)$.
  Describe the natural transformations between order-preserving functions.
\end{exer}

\begin{exer}
Let $\CC$ be a category and $\mathsf{bool}$ be the category with $2$ distinct objects $a,b$ and no non-trivial morphisms.
For every two objects $x, y \in \CC$, there is a functor $J : \mathsf{bool} \to \CC$ with $J(a) := x$ and $J(b) := y$.

Assume $x + y$ is a coproduct of $x$ and $y$.
Construct a natural transformation $\alpha$ from $J$ to the constant functor $\mathsf{const}_{(x + y)} : \mathsf{Bool} \to \CC$.
Furthermore, can you describe the universal property of the coproduct in terms of $\alpha$?
\end{exer}

\section{Functor Categories}

\begin{dfn}\label{dfn:nattrans_id} Let $F:\CC\to\DD$ be a functor. The \textbf{identity natural transformation} $\Id[F]$ on $F$ is given by the following data:
\[
\forall X\in\Ob{\CC}: (\Id[F])_{X} := \Id[F(x)].
\]
\end{dfn}

\begin{exer} Show that for any functor $F:\CC\to\DD$, the identity natural transformation $\Id[F]$ satisfies the properties of a natural transformation.
\end{exer}

\begin{dfn}\label{dfn:nattrans_comp} Let $F,G,H: \CC\to\DD$ be functors and $\NatTrans{\alpha}{F}{G}$, $\NatTrans{\beta}{G}{H}$ be natural transformations. The \textbf{(vertical) composition} of $\alpha$ and $\beta$ is the natural transformation $\Comp{\alpha}{\beta}$ is given by the following data:
\[
\forall X\in\Ob{\CC}: (\co{\alpha}{\beta})_{X} := \co{\alpha_X}{\beta_X}.
\]
\end{dfn}

\begin{exer} Show that for any functors $F,G,H: \CC\to\DD$ and $\NatTrans{\alpha}{F}{G}$, $\NatTrans{\beta}{G}{H}$ natural transformations, the (vertical) composition of $\alpha$ and $\beta$ satisfies the properties of a natural transformation.
\end{exer}

\begin{dfn} Let $\CC,\DD$ be categories. The \textbf{category of functors} or the \textbf{functor category} from $\CC\to\DD$, denoted by $Fun(\CC,\DD)$ or $[\CC,\DD]$, is given by the following data:
\begin{itemize}
\item An object is a functor $F:\CC\to\DD$.
\item Given functors $F, G:\CC\to\DD$, a morphism from $F$ to $G$ is a natural transformation $\NatTrans{\alpha}{F}{G}$.
\item The identity morphism on $F$ is given by the identity natural transformation $\Id[F]$ defined in \cref{dfn:nattrans_id}.
\item The composition of $\alpha$ and $\beta$ is given by the composition $\co{\alpha}{\beta}$ defined in \cref{dfn:nattrans_comp}.
\end{itemize}
\end{dfn}

\begin{exer} Show that for any two categories $\CC$ and $\DD$, the functor category from $\CC$ to $\DD$ satisfies the properties of a category.
\end{exer}

\begin{exer}
Let $\CC$ be the category generated by two objects $(V, E)$ and with $2$ morphisms $s, t : E \to V$.
Can you describe the category $[\CC, \SET]$?
\end{exer}

\begin{dfn}
A natural transformation $\alpha : F \Rightarrow G$ is a \textbf{natural isomorphism} if for each $X\in\Ob{\CC}$, we have that $\alpha_X$ is an isomorphism in $\DD$.
\end{dfn}

\begin{exer}
Let $F, G : \CC \to \DD$ be functors.
Show that a natural isomorphism from $F$ to $G$ is equivalent to an isomorphism between $F$ and $G$ in the functor category $[\CC, \DD]$.
\end{exer}

\begin{dfn}\label{dfn:nattrans_horcomp} Let $F,G : \CC\to\DD$ and $\tilde{F},\tilde{G}:\DD\to\EE$ be functors and $\NatTrans{\alpha}{F}{G}, \NatTrans{\beta}{\tilde{F}}{\tilde{G}}$ be natural transformations. The \textbf{horizontal composition} (also called the \textbf{Godement product}) of $\alpha$ and $\beta$, denoted by $\beta \bullet \alpha$, is defined as:
\begin{equation}\label{eqn:nattrans_horcomp}
\forall X\in \Ob{\CC}: (\beta\bullet\alpha)_X := \co{\tilde{F}(\alpha_X)}{\beta_{G(X)}}.
\end{equation}
\end{dfn}

\begin{exer} Show that $\alpha\bullet\beta$ (defined as in \cref{dfn:nattrans_horcomp}), is indeed a natural transformation.
\end{exer}

\begin{exer} Show the following property: 
\[
\forall X\in \Ob{\CC}: (\beta\bullet\alpha)_X = \co{\beta_{F(X)}}{\tilde{G}(\alpha_X)}.
\]
Hint: Write the equality as a (not-known commutative) square.
\end{exer}

\section{Equivalence of Categories}
Recall that objects $X,Y\in\Ob{\CC}$ are isomorphic if there exist morphisms $f\in\CHom{C}{X}{Y}$ and $g\in\CHom{C}{Y}{X}$ such that $\co{f}{g} = \Id[X]$ and $\co{g}{f} = \Id[Y]$. So in particular we have the notion of an isomorphism in the category $\CAT$ of categories. Spelled out, this means categories $\CC$ and $\DD$ are isomorphic if there exist functors $F:\CC\to\DD$ and $G:\CC\to\DD$ such that $\co{F}{G}= \Id[\CC]$ and $\co{G}{F} = \Id[\DD]$.\\
However, the following exercise shows that isomorphism of categories is not the correct notion of \textit{equivalence/sameness} between categories:\\
Let $\FINSET$ be the category whose objects are given by finite sets and whose morphisms are given by functions\footnote{Notice that the objects of $\FINSET$ form a subset of the objects of $\SET$, but given any two finite sets $X,Y \in \Ob{\FINSET}$, we have $\CHom{\FINSET}{X}{Y}$ = $\CHom{\SET}{X}{Y}$. We say in this case that $\FINSET$ is a (full) subcategory of $\SET$.}. That this is a category follows since $\SET$ is a category.\\
Let $\Catb{FinOrd}$ be the category whose objects are given by sets of the form 
\[
[n] := \left\{0,1,\cdots,n-1\right\},
\]
and whose morphisms are given by functions between these sets.\\
Every finite set $X$ is always in bijection with a set of the form $[n]$ (where $n$ is the size $\vert X\vert$ of $X$). For each set $X$, we fix a bijection $\phi^X: X\to [\vert X\vert]$. Consequently, we have a functor:
\begin{dfn} Let $U: \FINSET\to \Catb{FinOrd}$ be the functor specified by the following data:
\begin{itemize}
\item For $X\in \Ob{\FINSET}$, we define $U(X) := [\vert X\vert]$.
\item For $f\in \CHom{\FINSET}{X}{Y}$, we define $U(f): [\vert X\vert]\to [\vert Y\vert]$ as the unique function such that the following diagram commutes:
\begin{center}
\begin{tikzcd}
X \arrow[r, "\phi^X"] \arrow[d,swap, "f"] & {[\vert X\vert]} \arrow[d, "U(f)"] \\
Y \arrow[r,swap, "\phi^Y"] & {[\vert Y\vert]}
\end{tikzcd}
\end{center}
\end{itemize}
\end{dfn}

\begin{exer} Show that $U: \FINSET\to \Catb{FinOrd}$ is indeed a functor. In particular, you have to show that $U$ is well-defined on the morphisms.
\end{exer}

In order to show that $U$ is not an isomorphism, one can use the following lemma/exercise:
\begin{exer} Show that a functor $F:\CC\to\DD$ is an isomorphism if and only if $F$ satisfies the following properties:
\begin{itemize}
\item $F$ is injective on objects, i.e. 
\[
\forall X,Y\in\Ob{\CC}: F(X) = F(Y) \implies X=Y.
\]
\item $F$ is surjective on objects, i.e. 
\[
\forall Y\in\Ob{\DD}: \exists X\in\Ob{\CC} : F(X) = Y.
\]
\item $F$ is faithful, i.e. the following functions are injective
\[
\forall X,Y\in\Ob{\CC}: \CHom{\CC}{X}{Y} \xrightarrow{F_{X,Y}} \CHom{\DD}{F(X)}{F(Y)} : f\mapsto F(f)
\]
\item $F$ is full, i.e. for all $X,Y\in \Ob{\CC}$, $F_{X,Y}$ is surjective.
\end{itemize}
\end{exer}

\begin{exer} Show that $U$ is not an isomorphism, i.e. state which part of an isomorphism fails and give a concrete example that it fails.
\end{exer}

\begin{rem} The problem with $U$ (in the sense that it is not an isomorphism) is that multiple (finite) sets are mapped to the same set. For this reason, a good notion of equivalence between categories should not be injective on objects. Also, which is not clear from this example, we should also weaken the condition of $F$ being surjective on objects. Instead, we need that $F$ is \textbf{essentially surjective on objects}:
\[
\forall Y\in\Ob{\DD}: \exists X\in\Ob{\CC} : F(X) \cong Y.
\]
\end{rem}
Motivated by the remark, we define:
\begin{dfn} Categories $\CC$ and $\DD$ are \textbf{equivalent} if there exists a pair of functors $(F:\CC\to\DD, G:\DD\to\CC)$ such that there exists natural isomorphisms 
\[
\co{F}{G} \to \Id[\CC], \quad \co{G}{F} \to \Id[\DD]
\]
\end{dfn}

So, although $U$ is not an isomorphism, it does induce an equivalence of categories:
\begin{exer} Show that $U$ induces an equivalence of categories.
\end{exer}

\begin{exer} Let $\CC$ be the category whose objects are categories with a unique object and whose morphisms are functors between these one-object categories, i.e. $\CC$ is the (full) subcategory of $\CAT$ generated by the categories with a unique object. Show that $\CC$ is equivalent to the category $\MON$ of monoids.
\end{exer}
\begin{exer}
In the previous exercise, what happens if we do not consider $\CC$ to consist of those categories with a unique object, but with a unique object up to isomorphism? In other words, consider $\CC$ as the category whose objects are categories $\DD$ which satisfy the following property: 
\[
\forall X,Y \in \Ob{\DD}: X\cong Y.
\]
\end{exer}

The following exercise gives a characterization of a functor being an equivalence. However, in order to show this, one has to use the axiom of choice which means (informally) that if the following property holds:
\[ 
\exists x: P(x),
\]
then we can fix some $x$ such that $P(x)$ holds.
\begin{exer} Show that a functor $F:\CC\to\DD$ induces an equivalence of categories if and only if it is essentially surjective on objects and fully faithful.
\end{exer}

\begin{exer}
  Show that the category $\MAT$ is equivalent to the category of finite-dimensional real vector spaces and linear maps.
\end{exer}

\section{Representable Functors}

Representable functors are functors into $\SET$ which are isomorphic to the so-called hom-functors \cref{exer:covariant-hom,exer:contravariant-hom}.
\cref{exer:coproducts-as-representable} illustrates that universal properties can be described in terms of representables.
\cref{exer:forgetful-functor-is-representable} illustrates that free constructions can be described in terms of representables.
\cref{exer:representables-on-posets} illustrates that objects in a category are completely determined by their representable functors.

\begin{exer}
\label{exer:covariant-hom}
Let $\CC$ be a category and $x \in \CC$.
Consider the function $\CC(x,-) : \Ob{\CC} \to \Ob{\SET}$ sending an object $y \in \CC$ to the hom-set $\CC(x,y)$.
For $y, z \in \CC$ and $g \in \CC(y, z)$ define
\[
\CC(x, g) : \CC(x,y) \to \CC(x,z) : f \mapsto g \circ f.
\]
Show that this defines a functor $\CC(x,-) : \CC \to \SET$.
This functor is called the \textbf{covariant hom-functor}.
\end{exer}

Recall that $\op{\CC}(x,y) := \CC(y,x)$.

\begin{exer}
\label{exer:contravariant-hom}
Show that for every $y : \CC$, the function $\CC(-,y) : \Ob{\CC} \to \Ob{\SET}$ sending an object $x \in \CC$ to the hom-set $\CC(x,y)$ induces a functor of type $\CC(-,y) : \op{\CC} \to \SET$.
This functor is called the \textbf{contravariant hom-functor}.
\end{exer}

\begin{exer}
Show that every category $\CC$ induces a functor
\[
\CC(-,-) : \op{\CC} \times \CC \to \SET.
\]
\end{exer}

\begin{exer}
Let $F : \CC \to \DD$ be a functor.
Can you construct a natural transformation from $\CC(-,-)$ to $\DD(-,-) \circ ({\op{F} \times F})$?
\end{exer}

\begin{dfn}
A functor $F : \op{\CC} \to \SET$ is \textbf{representable} if there exists an object $c \in \CC$ and a natural isomorphism $\theta : \CC(-,c) \simeq F$.
We say that $(c, \theta)$ is a \textbf{representation} for $F$.
\end{dfn}

\begin{lemma}
A representation for a functor $F$ into $\SET$ is unique up to isomorphism.
\end{lemma}

\begin{exer}
\label{exer:coproducts-as-representable}
Let $\CC$ be a category and $x,y \in \CC$ objects.
Show that the following are equivalent:
\begin{enumerate}
\item the coproduct of $x$ and $y$ exists;
\item the functor $\CC(-,x) \times \CC(-,y)$ is representable.
\end{enumerate}
\end{exer}

Recall that $\MON$ denotes the category of monoids and monoid homomorphisms.
Observe that for every set $A$, $\List(A)$ has a monoid structure whose multiplication is given by the concatenation of lists and whose unit is the empty/nil list.
See \cref{ch:forgetful-free} for more details.
\begin{exer}[\cref{sol:forgetful-functor-is-representable}]\label{exer:forgetful-functor-is-representable}
Show that the forgetful functor $U : \MON \to \SET$ is representable.
Furthermore, show that for every set $A$ the functor 
\[
\SET(A, U-) : \MON \to \SET,
\]
is representable.
\end{exer}

\begin{exer}[\cref{sol:representables-on-posets}]
\label{exer:representables-on-posets}
Let $L := (L,\leq)$ be a poset and $\mathcal{L} := \POS(L,\leq)$ be its associated category.
Describe the representable functors on $\mathcal{L}$.
\end{exer}

\chapter{Adjunctions}
\label{sec:adjunctions}

\begin{dfn} A pair $(F,G)$ of functors $F : \CC\to\DD, G:\DD\to\CC$ is called an \textbf{adjoint pair} if for every objects $X\in\Ob{\CC}$ and $Y\in\Ob{\DD}$, there exists a bijection 
\[
\alpha_{X,Y} : \CHom{\DD}{F(X)}{Y} \to \CHom{\CC}{X}{G(Y)},
\]
which are moreover natural in both $X$ and $Y$, i.e. for each $f\in\CHom{\CC}{X_1}{X_2}$ and $g\in\CHom{\DD}{Y_1}{Y_2}$, the following diagrams commute:
\begin{eqnarray}
\begin{tikzcd}
\CHom{\DD}{F(X_2)}{Y} \arrow[r, "\alpha_{X_2,Y}"] \arrow[d,swap, "\co{F(f)}{-}"] & \CHom{\CC}{X_2}{G(Y)} \arrow[d, "\co{f}{-}"] \\
\CHom{\DD}{F(X_1)}{Y} \arrow[r,swap, "\alpha_{X_1,Y}"] & \CHom{\CC}{X_1}{G(Y)}
\end{tikzcd}\\
\begin{tikzcd}
\CHom{\DD}{F(X)}{Y_1} \arrow[r, "\alpha_{X,Y_1}"] \arrow[d,swap, "\co{-}{g}"] & \CHom{\CC}{X}{G(Y_1)} \arrow[d, "\co{-}{G(g)}"] \\
\CHom{\DD}{F(X)}{Y_2} \arrow[r,swap, "\alpha_{X,Y_2}"] & \CHom{\CC}{X}{G(Y_2)}
\end{tikzcd}
\end{eqnarray}
If $(F,G)$ is an adjoint pair, we call $F$ the \textbf{left adjoint} of $G$ and we call $G$ the \textbf{right adjoint} of $F$ and we denote $F \dashv G$.
\end{dfn}

\begin{exer} Let $F:\CC\to\DD$ be a functor. Show that if $F$ has a left (resp. right) adjoint $G$, then $G$ must be unique up to isomorphism\footnote{Isomorphism w.r.t the functor category.}.
\end{exer}

\begin{exer} Let $U:\MON\to \SET$ be the forgetful functor (defined in \cref{example:forgetful_montoset}) which maps a monoid to its underlying set and let $F : \SET\to\MON$ be the functor which maps a set to the free monoid of this set (defined in \cref{exa:freemonoids}). Then is $(F,U)$ an adjoint pair. (Hint: Use \cref{prop:UVP_forget_montoset}). 
\end{exer}

\begin{exer}\label{exer:adjunction_homtensor_currying} Let $Y$ be a set. Show that the functor (induced by the following mapping on objects)
\[
- \times Y : \SET\to \SET: X\mapsto X\times Y,
\]
has a right adjoint, which is given by the functor (induced by the following mapping on objects)
\[
\CHom{\SET}{Y}{-} : \SET\to\SET : X\mapsto \CHom{\SET}{Y}{X}.
\]
Is there an analogous statement in the category $\COQ$ instead of $\SET$?
\end{exer}

\begin{thm} Let $(F,G)$ be a pair of functors $F : \CC\to\DD, G: \DD\to\CC$. The following statements are equivalent:
\begin{enumerate}
\item $(F,G)$ is an adjoint pair.
\item There exists natural transformations 
\[
\NatTrans{\eta}{\Id[\CC]}{\co{F}{G}}, \quad \NatTrans{\epsilon}{\co{G}{F}}{\Id[\DD]},
\]
such that for all $X\in\CC$ and $Y\in\DD$ the following diagrams commute:
\[
\begin{tikzcd}
F(X) \arrow[r,"{F(\eta_X)}"] \arrow[rd,swap, "{\Id[F(X)]}"] & F(G(F(X))) \arrow[d, "{\epsilon_{F(X)}}"] \\
& F(X)
\end{tikzcd} \quad 
\begin{tikzcd}
G(Y) \arrow[r,"{\eta_{G(Y)}}"] \arrow[rd,swap, "{\Id[G(Y)]}"] & G(F(G(Y))) \arrow[d, "{G(\epsilon_{Y})}"] \\
& G(Y)
\end{tikzcd}
\]
\end{enumerate}
In case $(F,G)$ satisfies these (equivalent) conditions, we call $\eta$ the \textbf{unit of the adjunction} and $\epsilon$ the \textbf{counit of the adjunction}. The equalities in condition $2$ are called the \textbf{triangle identities}.
\begin{proof}
First, assume that $(F,G)$ is an adjoint pair. We have to define the unit and counit and show that the triangle identities hold:
\begin{itemize}
\item \textbf{Unit}: For each $X\in\Ob{\CC}$, we should first define $\eta_X \in \CHom{\CC}{X}{G(F(X))}$. Since $F \dashv G$, we have a bijection
\[
\alpha_{X,FX} : \CHom{\DD}{FX}{FX} \to \CHom{\CC}{X}{G(F(X))},
\]
hence, we define $\eta_X := \alpha_{X,FX}(\Id[FX])$. We now show that $(\eta_X)_{X\in\Ob{\CC}}$ forms a natural transformation: Assume $f\in\CHom{\CC}{X}{Y}$. We have to show that the following diagram commutes:
\[
\begin{tikzcd}
X \arrow[rr, "{\alpha_{X,FX}(\Id[FX])}"] \arrow[d,swap, "f"] && G(F(X)) \arrow[d, "G(F(f))"] \\
Y \arrow[rr, swap, "{\alpha_{Y,FY}(\Id[FY])}"] && G(F(Y))
\end{tikzcd}
\]
That this square is indeed commutative follows from the following computation:
\begin{equation}\label{eqn:unitnaturality_fromhomsetadj}
\co{f}{\alpha(\Id[FY])} = \alpha(\co{F(f)}{\Id[FY]}) = \alpha(F(f)) =  \alpha(\co{\Id[FX]}{F(f)}) = \co{\alpha(\Id[Fx])}{G(F(f))},
\end{equation}
where the first and last equality hold by naturality of $\alpha$.
\item \textbf{Counit}: For each $Y\in\Ob{\DD}$, we the counit $\epsilon_Y \in \CHom{\DD}{F(G(Y))}{Y}$ is defined as the image of $\Id[G(Y)]$ of the bijection
\[
\alpha^{-1}_{GY,Y} : \CHom{\CC}{GY}{GY}\to \CHom{\DD}{F(G(Y))}{Y}.
\]
That $\epsilon$ indeed forms a natural transformation is analogous to the computation in \cref{eqn:unitnaturality_fromhomsetadj}.
\item \textbf{Triangle identities}: Both the triangle identities are proved analogously, hence we will only show the first triangle identity, i.e. $\Id[FX] = \co{F(\eta_X)}{\epsilon_{FX}}$. Unfolding the definition of $\epsilon$, this is equivalent to showing:
\[
\Id[FX] = \co{F(\eta_X)}{\alpha^{-1}_{GFX,FX}(\Id[GF(X)])}.
\]
Since the components of $\alpha$ are bijections, this is equivalent to showing 
\[
\alpha(\Id[FX]) = \alpha(\co{F(\eta_X)}{\alpha^{-1}_{GFX,FX}(\Id[GF(X)])})
\]
This indeed holds by the following computation:
\[
\alpha(\co{F(\eta_X)}{\alpha^{-1}_{GFX,FX}(\Id[GF(X)])}) = \co{\eta_X}{\alpha(\alpha^{-1}(\Id[GFX]))} = \co{\eta_X}{\Id[GFX]} = \eta_X,
\]
where the first (resp. second) equality holds by naturality (resp. bijectiveness) of $\alpha$.
\end{itemize}
This concludes the proof of $(1)\implies (2)$. Now assume that $(2)$ holds. We have to construct bijections
\[
\alpha_{X,Y} : \CHom{\DD}{F(X)}{Y} \to \CHom{\CC}{X}{G(Y)},
\]
which are natural in $X$ and $Y$. Let $g\in \CHom{\DD}{F(X)}{Y}$. Define $\alpha_{X,Y}(g)$ as the composite:
\[
X \xrightarrow{\eta_X} G(F(X)) \xrightarrow{G(g)} G(Y).
\]
For the other direction, let $f\in \CHom{\CC}{X}{G(Y)}$. Define $\alpha^{-1}_{X,Y}(f)$ as the composite:
\[
FX \xrightarrow{F(f)} F(G(Y)) \xrightarrow{\epsilon_Y} Y.
\]
That $\alpha$ and $\alpha^{-1}$ are inverses of each other follows from the following computation:
\begin{eqnarray*}
\alpha(\alpha^{-1}(f)) = \alpha(\co{F(f)}{\epsilon_Y}) =& \co{\eta_X}{G(\co{F(f)}{\epsilon_Y})} &\text{ by definition, }\\
	=& \co{\eta_X}{(\co{GF(f)}{G\epsilon_Y})} &\text{ by functoriality of $G$},\\
	=& \co{(\co{\eta_X}{GF(f)})}{G\epsilon_Y} &\text{ by associativity},\\
	=& \co{(\co{f}{\eta_{GY})}}{G\epsilon_Y} &\text{ by naturality of $\eta$}\\
	=& \co{f}{(\co{\eta_{GY}}{G\epsilon_Y})} &\text{ by associativity}\\
	=& f &\text{by triangle identity}.
\end{eqnarray*}
The other equality is shown analogous by using functoriality of $F$, naturality of $\epsilon$ and the other triangle identity.

It remains to show the naturality of $\alpha$ in both $x$ and $y$ which is left to the reader as a good exercise on diagram chasing. 
\end{proof}
\end{thm}

\chapter{Monads and Effects}
\label{sec:monads}

In this section, we study the definition of monads as well as instances of monads useful in functional programming.

There are different, equivalent, definitions of monads.
Since these definitions are equivalent, we can use them interchangeably once we have understood them.
But at first we need to distinguish them, and hence need different names for each definition; we choose to call one definition ``Kleisli triple'' or ``extension system'' (see \cref{def:kleisli-triple}) and one ``monad'' (see \cref{def:monad}).
Whenever we want to gloss over the precise definition, we simply say ``monad''.

Recall that a monad (as defined, e.g., in Haskell) is a function |m :: * -> *| together with the additional data of a function |pure :: a -> m a| (for each type |a|) and a function |(>>=) :: m a -> (a -> m b) -> m b| (for types |a| and |b|).
The operations |pure| and |(>>=)| are expected to satisfy the following laws:
\begin{enumerate}
\item |t >>= pure == t| 
\item |pure(x) >>= f == f x| 
\item |(t >>= f) >>= g == t >>= (\x -> f x >>= g)|
\end{enumerate}
However, these laws cannot be enforced in Haskell, since Haskell does not have any infrastructure for equational reasoning.
(The laws could be tested for using Quickcheck or a similar framework.)

In a category, however, we can define monads including the monad laws.
As mentioned above, we give two different definitions of monad;
one called ``Kleisli triple'' (\cref{def:kleisli-triple}), which corresponds to what is called ``monad'' in Haskell,
and one called ``monad'' (\cref{def:monad}). 


\begin{dfn}\label{def:kleisli-triple}
  A \textbf{Kleisli triple} over a category $\CC$ is  consisting of the following data:
\begin{enumerate}
\item A function $T: \Ob{\CC}\to \Ob{\CC}$.
\item For each $X\in\Ob{\CC}$, a morphism $\eta_X \in \CHom{\CC}{X}{T(X)}$.
\item For each $f\in\CHom{\CC}{X}{T(Y)}$, a morphism $f^{*} \in \CHom{\CC}{T(X)}{T(Y)}$.
\end{enumerate}
such that the following properties holds:
\begin{enumerate}[resume]
\item For each $X \in \Ob\CC$, we have $\eta_X^{*} = \Id[T(X)]$.
\item For each $f\in\CHom{\CC}{X}{T(Y)}$, the following diagram commutes:
\begin{center}
\begin{tikzcd}
X \arrow[r, "{\eta_X}"] \arrow[rd,swap,"f"] & T(X) \arrow[d, "f^{*}"] \\
& T(Y)
\end{tikzcd}
\end{center}
\item For each $f\in\CHom{\CC}{X}{T(Y)}$ and $g\in\CHom{\CC}{Y}{T(Z)}$, the following diagram commutes:
\begin{center}
\begin{tikzcd}
T(X) \arrow[r, "f^{*}"] \arrow[rd,swap,"{(\co{f}{g^{*}})^{*}}"] & T(Y) \arrow[d, "g^{*}"] \\
& T(Z)
\end{tikzcd}
\end{center}
\end{enumerate}
We denote a Kleisli triple as $(T,\eta, (-)^{*})$.
\end{dfn}

\begin{exer}
  Convince yourself that the operations and laws of a Kleisli triple correspond, in the category $\HASK$, to the operations and properties of a monad in Haskell.
\end{exer}

\begin{exer}
  Consider the assignment $\Ob{\Maybe} : \Ob{\SET}\to \Ob{\SET}$ with $\Ob{\Maybe}(X) := X + 1$.
  Define a Kleisli triple with this function as the first component.
\end{exer}

\begin{exer}[\cref{sol:kleisli_triple_list}]\label{exer:kleisli_triple_list} Show how the following assignment induces a Kleisli triple over the category $\SET$:
\[
X\mapsto \List(X).
\]
The resulting monad is called the \textbf{List monad}.
\end{exer}

\begin{exer}[\cref{sol:kleisli_triple_bintree}]\label{exer:kleisli_triple_bintree} Show how the following assignment induces a Kleisli triple over the category $\SET$:
\[
X\mapsto \BinTree(X),
\]
where $\BinTree(X)$ is the set of binary trees labelled with elements from $X$ at the leaves, that is the set inductively generated by the constructors $\mathsf{leaf}: X\to \BinTree(X)$ and $\mathsf{branch} : \BinTree(X)\to \BinTree(X)\to \BinTree(X)$.
The resulting monad is called the \textbf{Tree monad}.
\end{exer}

\begin{exer}[\cref{sol:kleisli_triple_exception}]\label{exer:kleisli_triple_exception} Let $E$ be a set (considered as a set of \textit{exceptions}). Show how the following assignment induces a Kleisli triple over the category $\SET$:
\[
X\mapsto (X + E),
\]
The resulting monad is called the \textbf{Exception monad}.
\end{exer}

\begin{exer}[\cref{sol:kleisli_triple_side_effects}]\label{exer:kleisli_triple_side_effects} Let $S$ be a set (considered as a set of states, e.g. a set of stores or a set of input/output sequences). Show how the following assignment induces a Kleisli triple over the category $\SET$:
\[
X\mapsto S \to (X \times S),
\]
The resulting monad is called the \textbf{Monad of side-effects}.
\end{exer}

\begin{exer}[\cref{sol:kleisli_triple_nondeterminism}]\label{exer:kleisli_triple_nondeterminism} Show how the following assignment induces a Kleisli triple over the category $\SET$:
\[
X\mapsto \mathbb{P}_{fin}(X) := \left\{A\subseteq X \mid  A \text{ is finite}\right\}.
\]
The resulting monad is called the \textbf{Monad of nondeterminism}.
\end{exer}

\begin{exer}[\cref{sol:kleisli_triple_continuation}]\label{exer:kleisli_triple_continuation} Let $R$ be a set (considered as a set of \textit{results}). Show how the following assignment induces a Kleisli triple over the category $\SET$:
\[
X\mapsto Cont^R(X) := (X \to R) \to R.
\]
The resulting monad is called the \textbf{Continuation monad}.
\end{exer}

\begin{exer}[\cref{sol:kleisli_triple_familiesofelements}]\label{exer:kleisli_triple_familiesofelements} Let $R$ be a set. Show how the following assignment induces a Kleisli triple over the category $\SET$: 
\[
X \mapsto R \to X
\]
The resulting monad is called the \textbf{Monad of families of elements}.

\end{exer}

\begin{reading*}
  Moggi \cite{DBLP:journals/iandc/Moggi91} defines a syntax, or rather, several syntaxes, for programming languages with categorical semantics in monads such as the ones above.
  The basic idea, taken verbatim from Moggi's paper, is as follows:
  \begin{quote}
    The basic idea behind the categorical semantics below is that, in order to interpret a programming language in a category $\CC$, we distinguish the object $A$ of values (of type $A$) from the object $T A$ of computations (of type $A$), and take as denotations of programs (of type $A$) the \emph{elements} of $T A$.
In particular, we identify the type $A$ with the object of values (of type $A$) and obtain the object
of computations (of type $A$) by applying an unary type-constructor $T$ to $A$. We call $T$ a \emph{notion of computation}, since it abstracts away from the type of values computations may produce. There are many choices for $T A$ corresponding to different notions of computations.
  \end{quote}

  More information about monads and effects is also given in \cite{DBLP:conf/ac/BentonHM00}.
\end{reading*}

The notion of a Kleisli triple can equivalently be described  as follows:
\begin{dfn}\label{def:monad}
A \textbf{monad} over a category $\CC$ consists of the following data:
\begin{itemize}
\item A (endo)functor $T:\CC\to\CC$.
\item A natural transformation (``unit'') $\NatTrans{\eta}{\Id[\CC]}{T}$.
\item A natural transformation (``multiplication'') $\NatTrans{\mu}{\co{T}{T}}{T}$.
\end{itemize}
such that for each $X\in\Ob{\CC}$ the following diagrams commute (``associativity'' and ``unit'' laws):
\begin{center}
\begin{tikzcd}
T^3(X) \arrow[r, "\mu_{T(X)}"] \arrow[d,swap, "T(\mu_X)"] & T^2(X) \arrow[d, "\mu_X"] \\
T^2(X) \arrow[r,swap, "\mu_X"] & T(X)
\end{tikzcd}
\quad
\begin{tikzcd} 
T(X) \arrow[r, "{\eta_{T(X)}}"] \arrow[rd,swap, "{\Id[T(X)]}"] 
& T^2(X) \arrow[d,"{\mu_X}"] & T(X) \arrow[l,swap,"{T(\eta_X)}"] \arrow[ld, "{\Id[T(X)]}"] \\
& T(X) &
\end{tikzcd}

\end{center}
where we denote $T^2 := \co{T}{T}$ and $T^3 := \co{T}{\co{T}{T}}$.
\end{dfn}

\begin{exer} Given a monad, construct a Kleisli triple from it.
Conversely, given a Kleisli triple, construct a monad from it.
\end{exer}

\begin{exer}
  For each of the Kleisli triples above, describe the monad multiplication $\mu$ obtained from it.
\end{exer}

\begin{exer}
A monad is, in particular, a functor $T : \CC \to \CC$, hence an object in $[\CC,\CC]$.
Can you describe a monad in terms of this functor category?
\end{exer}

Every Kleisli triple induces a category:
\begin{dfn} Let $(T,\eta, (-)^{*})$ be a Kleisli triple over $\CC$. The \textbf{Kleisli category}, denoted by $\CC_{T}$, is the category defined by the following data:
\begin{itemize}
\item $\Ob{(\CC_T)} := \Ob{\CC}$.
\item For each $X,Y\in\Ob{(\CC_T)}$, $\CHom{\CC_T}{X}{Y} := \CHom{\CC}{X}{TY}$.
\item The identity on $X\in\Ob{(\CC_T)}$ is $\eta_X$.
\item The composition of $f\in \CHom{\CC_T}{X}{Y}$ and $g\in \CHom{\CC_T}{Y}{Z}$ is $\co{f}{g^{*}}$.
\end{itemize}
\end{dfn}

\begin{exer} Show that for every Kleisli triple, its Kleisli category satisfies the properties of a category.
\end{exer}

\begin{reading*}
  Another use of monads is the mathematical modelling of abstract syntax and \emph{substitution}.
  See, for instance, Altenkirch and Reus' paper \cite{DBLP:conf/csl/AltenkirchR99}.

  An intermediate notion between functors and monads are \emph{applicative functors}. These are described by McBride and Paterson \cite{DBLP:journals/jfp/McbrideP08}.
  To understand applicative functors in full generality, it is necessary to discuss monoidal categories and monoidal functors, as well as adjunctions.
\end{reading*}


\chapter{Conclusions and Further Reading}

We have given a brief, and necessarily incomplete, introduction to category theory.
The choice of topics was guided by the applications to programming we considered.
 Here below, we list a few avenues for further exploration; this list is curated with a view towards pedagogy rather than scientific completeness.
\begin{itemize}
\item Altenkirch and Reus show how the lambda calculus forms a monad.
  (This can be generalized to other languages, including languages with simple types, see, e.g., \cite{DBLP:journals/jfrea/AhrensZ11}.)
\item For \emph{nested} datatypes, the iteration principle of \cref{sec:datatypes-as-initial} is often not enough; Bird and Paterson~\cite{DBLP:journals/fac/BirdP99} develop a \emph{generalized} fold operator for nested datatypes.
\item Some functions cannot be straightforwardly defined using the simple iteration scheme of \cref{sec:datatypes-as-initial}; for instance, the factorial function does not adhere to the required pattern.
  Vene~\cite{vene_phd} studies more general recursion schemes beyond the iteration (catamorphisms) of~\cref{sec:datatypes-as-initial}.
\item Harper~\cite{DBLP:phd/ethos/Harper13} develops a unifying framework for shortcut fusion.

\end{itemize}

\appendix

\chapter{Solutions to Selected Exercises}
\label{sec:solutions}

\begin{solution}[\cref{exer:post_antisymmetry}]\label{sol:post_antisymmetry}
The inequality $x\leq y$ (resp. $y\leq x$) means that we have a (unique) morphism from $f\in \Hom{x}{y}$ (resp. $g\in \Hom{y}{x}$). Consequently, we get a \textit{loop} $\co{f}{g} \in \Hom{x}{x}$. Since the hom-sets are either empty or a singleton, we have $\co{f}{g} = \Id[x]$. Hence, antisymmetry means that if $\Id[X] = \co{f}{g}$ for some $f\in \Hom{x}{y}$ and $g\in \Hom{y}{x}$, we must have 
\[x=y,\quad f = \Id[x] = g.\]
Rephrased a little bit different, we get: The category $\PREtoCAT(X,\leq)$ has no non-trivial loops if $(X,\leq)$ is antisymmetric.
\end{solution}


\begin{solution}[\cref{exer:categories_coming_from_monoids}]\label{sol:categories_coming_from_monoids}
A category $\CC$ is of the form $(M,m,e)$ if and only if $\CC$ has a unique object. Indeed, if $\CC$ has a unique object, lets denote this by $X$, then we can define a monoid $(M,m,e)$ as follows:
\begin{itemize}
\item The underlying set of the monoid is $M := \CHom{\CC}{X}{X}$.
\item The multiplication $m$ is given by $m(f,g) := \co{f}{g}$.
\item The identity element $e$ is given by $e := \Id[X]$.
\end{itemize}
That $(M,m,e)$ is indeed a monoid, i.e. satisfies the monoid laws, is just a translation of the axioms of $\CC$ being a category.
\end{solution}

\begin{solution}[\cref{exer:category_of_monoids}]\label{sol:category_of_monoids}
Since a monoid consists of a set $M$ together with a binary operation $m:M\to M\to M$ (called the \textit{multiplication}) and a \textit{identity} element $e\in M$, a suitable \textit{morphism of monoids}, from $(M_1,m_1,e_1)$ to $(M_2,m_2,e_2)$ should consists of a function $f:M\to N$ which preserves the structure (i.e. the multiplication and the identity element). More precisely:
\begin{itemize}
\item Preservation of the multiplication:
\[
\forall x,y\in M_1: f(m_1(x,y)) = m_2(f(x), f(y)).
\]
\item Preservation of the identity:
\[
f(e_1) = e_2.
\]
\end{itemize}
We will now prove that the monoids (as objects) and morphisms of monoids (as the morphisms) carry the data of a category, i.e. we have to define identity morphisms and the composition of morphisms:
\begin{itemize}
\item Let $(M,m,e)$ be a monoid. The identity morphism is given by the identity function $\Id[M]$ on the underlying set $M$. 
\item Let $(M_i,m_i,e_i)$ be a monoid for $i=1,2,3$ and let 
\[ 
f:(M_1,m_1,e_1)\to (M_2,m_2,e_2),\quad g:(M_2,m_2,e_2)\to (M_3,m_3,e_3)
\] 
be morphisms of monoids. The composition $\co{f}{g}$ of $f$ and $g$ is defined as the composition of the underlying functions.
\end{itemize}
Before we show that this data satisfies the properties of a category, we first have to show that everything is well-defined, i.e. that the identity is a morphism of monoids and that the composition of morphisms of monoids is again a morphism of monoids:
\begin{itemize}
\item That the identity is a morphism of monoids follows by the following calculations:
\begin{eqnarray*}
\forall x,y\in M &:& \Id[M](m(x,y)) = m(x,y) = m(\Id[M](x),\Id[M](y)),\\
&& \Id[M](e) = e.
\end{eqnarray*}
The equalities holds because $\Id[M]$ is the identity function on $M$.
\item That the composition of morphism of monoids is again a morphism of monoids follows by the following calculations:
\begin{itemize}
\item Composition preserves multiplication: for any $x,y \in M_1$, we have
\begin{eqnarray*}
 \co{f}{g}(m_1(x,y)) &=& g\left(f(m_1(x,y))\right)\\ 
	&=& g\left(m_2(f(x),f(y)))\right) \quad \text{ ($f$ preserves mult.)} \\
	&=& m_3(g(f(x)),g(f(y))) \quad \text{ ($g$ preserves mult.)}\\
	&=& m_3(\co{f}{g}(x),\co{f}{g}(y))
\end{eqnarray*}
Composition preserves identity element by:
\[
\co{f}{g}(e_1) = g(f(e_1)) = g(e_2) = e_3,
\]
where the second (resp. third) equality holds since $f$ (resp. $g$) preserves the identity element.
\end{itemize}
\end{itemize}
So everything is indeed well-defined. So we are now ready to show that composition of some morphism of monoids $f$ with the identity morphism is again $f$ (both on the left and right) and that the composition of morphisms of monoids is associative. This follows immediate since everything is defined using functions and we know that functions satisfy these properties.
\end{solution}

\begin{solution}[\cref{exer:opposite}]\label{sol:opposite}
  In order to avoid confusion, we use the following notation: For any $f\in\CHom{\CC}{X}{Y}$ morphism, we denote by $\op{f}$ the corresponding morphism in $\CHom{\op\CC}{Y}{X}$.
  
  Let $X\in \Ob{\op\CC} = \Ob{\CC}$. The identity morphism is defined as the morphism corresponding to the identity, i.e. it is $\op{(\Id[X])}$.
  
  Let $\op{g} \in \CHom{\op\CC}{Z}{Y}, \op{f}\in \CHom{\op\CC}{Y}{X}$. The composition is defined as: $\co{\op g}{\op f} := \op{(\co{f}{g})}$.
  
That this data satisfies the properties of a category, follows because $\CC$ is a category, indeed:
\begin{itemize}
\item That the left unit law holds follows by the right unit law of $\CC$ as follows:
\[
\co{\op{\Id}}{\op{f}} = \op{(\co{f}{\Id})} = \op{f},
\]
where the first equality holds by definition of the \textit{opposite composition} and the second holds by the right unit law of $\CC$.\\
The right unit law holds analogously by the left unit law of $\CC$.
\item That the associativity holds follows by the associativity of $\CC$ as follows:
\begin{eqnarray*}
\co{\op{h}}{(\co{\op g}{\op f})} &=& \co{\op{h}}{\op{(\co{f}{g})}}\\
 	&=& \op{(\co{(\co{f}{g})}{h})}\\ 
 	&=& \op{(\co{f}{(\co{g}{h})})} \text{ by associativity of $\CC$},\\
 	&=& \co{\op{(\co{g}{h})}}{\op f}\\
 	&=& \co{(\co{\op h}{\op g})}{\op f}
\end{eqnarray*}
\end{itemize}
\end{solution}

\begin{solution}[\cref{exer:connection_graphs_preordersets}]\label{sol:connection_graphs_preordersets}
Any preordered set $(X,\leq)$ can be described by a graph where the vertices are given by the elements of $X$ and there exists an edge from $x$ to $y$ if and only if $x \leq y$. Hence, if we denote by $G$ the corresponding graph of $(X,\leq)$, we have that $\POS(X,\leq) = \mathbf{Graph}(G)$.\\
In particular we have that the number of edges is either $0$ or $1$. Hence, a category generated by a graph comes from a preordered set if and only if the number of morphisms in any fixed hom-set is either $0$ or $1$.

If $(X,\leq)$ is a poset, i.e. we have antisymmetry, then we have that the corresponding graph (and consequently the corresponding category) have no (non-trivial) loops.
\end{solution}

\begin{solution}[\cref{exer:categories_with_natural_numbers}]
\label{sol:categories_with_natural_numbers}
All of these categories have natural numbers as their collection of objects. To define these categories, we define the hom-sets, the identity morphisms and the composition of morphisms. We also have to show that the left and right unit laws and associativity of composition hold.  
\begin{enumerate}
\item This category is generated by the preorder given by the $\leq$ relation. This means that the hom-sets, identity morphisms and composition of morphisms of this category are defined as explained in \cref{example:posetcategories}. 
For $m,n \in \NN$, the hom-set $\Hom m n$ consists of a unique element if $m \leq n$ and is empty otherwise. For each $m \in \NN$, by reflexivity we have $m \leq m$. Hence, we have that $\Hom m m$ consists of a unique element, which we take to be the identity.
For each $l, m, n\in \NN$, the composition operator is of the form
\[
\Hom m n \to \Hom l m \to \Hom l n,
\]
which only needs to be defined in case $l \leq m$ and $m \leq n$. In this case, by transitivity we have $l \leq n$, and $\Hom l n$ consists of a unique element, which we take to be the composite. 
\item For $m,n \in \NN$, the hom-set $\Hom m n$ consists of functions from the standard finite set $[m]$ to the standard finite set $[n]$.
For each $m \in \NN$, the identity morphism is a function from $[m]$ to $[m]$ defined as follows: 
\begin{align*}
	\Id[m] :  \{0, \cdots , m-1\} & \to \{0, \cdots , m-1\} \\
	k & \mapsto k.
\end{align*}
For each $l, m, n\in \NN$, the composition operator is of the form
\[
\Hom m n \to \Hom l m \to \Hom l n. 
\]
Given morphisms $f : m \to n$ and $g : l \to m $, i.e. functions $f$ form $[m]$ to $[n]$ and $g$ from $[l]$ to $[m]$, the composition of morphisms $\co{g}{f}$ is defined to be the composition of functions $\co{g}{f}$.
Function composition is associative; hence, so is the composition of morphisms. 
\item Let $n\in \NN$ be a natural number. The identity morphism on $n$ is an element of $\CHom{\CC}{n}{n}$, which is the set of $n\times n$-matrices. 
Hence, we can take $\Id[n]$ to be the identity $n\times n$-matrix, i.e. all elements are zero except on the diagonal where all the elements are $1$.

Let $l,m,n \in\NN$ be natural numbers and $M$ (resp. $N$) be a $l\times m$-matrix (resp. $m\times n$-matrix). The composition $\co{N}{M}$ should be an $l\times n$-matrix, hence we define $\co{N}{M}$ as the matrix multiplication of $M$ and $N$, i.e. $MN$.

Matrix multiplication is associative and the (right or left) multiplication of any matrix $M$ with the identity matrix (of the right size) is again $M$. Hence, this data satisfies the properties of being a category. 

Note that a $0 \times 0$ matrix corresponds to a linear map from the zero vector space to the zero vector space mapping $0$ to $0$.
Similarly for each $n \in \NN$, an $n \times 0$ (resp. $0 \times n$) matrix corresponds to a linear map from the zero vector space (resp. an n-dimensional vector space) to an n-dimensional vector space (resp. the zero vector space).

\end{enumerate}
\end{solution}

\begin{solution}[\cref{exer:category_of_relations}] \label{sol:category_of_relations}
Let $X$ and $Y$ be sets. The hom-set $\Hom X Y$ in $\REL$ consists of subsets of $X \times Y$. For each set $X$, the identity $\Id[X]$ can be defined as follows:
\[ \Id[X] = \{ (x,x) \mid x \in X \} .\]
For sets $X,Y$ and $Z$, relation $R_1 \subseteq X \times Y$ and relation $R_2 \subseteq Y \times Z$, the composition $R_2 \circ R_1$ can be defined as follows:
\[ \{ (x,z) \in X \times Z \mid \exists y \in Y \text{ s.t. } (x,y) \in R_1 \wedge (y,z) \in R_2 \} . \]
Using these definitions, it is straightforward to check that composition is associative and left and right unit laws hold.
\end{solution}

\begin{solution}[\cref{exer:inverse-iso}]\label{sol:inverse-iso}
That $f:a\to b$ is an isomorphism with inverse $g$ means precisely that $\co{f}{g} = \Id[a]$ and $\co{g}{f} = \Id[b]$. But stating that $g$ is an isomorphism with inverse $f$ means precisely those conditions. Hence, this hold by definition.
\end{solution}

\begin{solution}[\cref{exer:inverse_uniqueness}]\label{sol:inverse_uniqueness}
Let $f:a\to b$ be an isomorphism. That $f$ has a unique inverse means that if $g,h : b\to a$ are morphisms such that 
\[
\co{f}{g} = \Id[a], \co{g}{f} = \Id[b], \co{f}{h} = \Id[a], \co{h}{f} = \Id[b]
\]
then we must have $g = h$.\\
So assume $g$ and $h$ satisfy the condition of being an inverse of $f$. Then we have:
\begin{eqnarray*}
g =& \co{g}{\Id[b]} &,\text{ by left unit law},\\
	=& \co{g}{(\co{f}{h})} &, \text{ since $h$ is inverse of $f$},\\
	=& \co{(\co{g}{f})}{h} &, \text{ by associativity},\\
	=& \co{\Id[a]}{h} &, \text{ since $g$ is inverse of $f$}\\
	=& h &, \text{ by right unit law}
\end{eqnarray*}
\end{solution}

\begin{solution}[\cref{exer:compofiso}]\label{sol:compofiso}
Let $f: a\to b$ and $g:b\to c$ be isomorphisms. Denote their (unique) inverses by $f^{-1}$ and $g^{-1}$. We have to show that there exists a morphism $h : c\to a$ such that 
\[
\co{(\co{f}{g})}{h} = \Id[a], \quad \co{h}{(\co{f}{g})} = \Id[c].
\]
We define $h := \co{g^{-1}}{f^{-1}}$. The left equality then holds by the following computation:
\begin{eqnarray*}
\co{(\co{f}{g})}{h} =& \co{(\co{f}{g})}{(\co{g^{-1}}{f^{-1}})} &\\
	=& \co{f^{-1}}{\co{(\co{g}{g^{-1}})}{f}} &\text{ by associativity,}\\
	=& \co{f^{-1}}{\co{\Id[b]}{f}} &\text{ since $g^{-1}$ inverse of $g$,}\\
	=& \co{f^{-1}}{f} &\text{ by unit law,}\\
	=& \Id[a] &\text{ since $f^{-1}$ inverse of $f$.}
\end{eqnarray*}
The right equality holds analogously.
\end{solution}

\begin{solution}[\cref{exer:iso-bool}]\label{sol:iso-bool}
The Haskell datatype |Bool| is given by:
\begin{lstlisting}
data Bool = True | False
\end{lstlisting}
In order to construct a (Haskell) function $f$ from |BW| to |Bool|, it suffices to define $f(Black)$ and $f(White)$.\\
The first isomorphism, denoted by $f_1$, is given by $f_1(Black)=True$ and $f_1(White) = False$. Its inverse (denoted by $g_1$) is given by $g_1(True) = Black$ and $g_1(False) = White$. To show that these are inverse, we have to show 
\[
g_1 (f_1 (White)) = White, \quad g_1 (f_1 (Black)) = Black.
\]
These equalities holds by definition of $f_1$ and $g_1$.\\
The second isomorphism, denoted by $f_2$, is given by $f_2(Black)=False$ and $f_2(White) = True$. Its inverse (denoted by $g_2$) is given by $g_2(False) = Black$ and $g_1(True) = White$. That $g_2$ is the inverse of $f_2$ is also immediate.
\end{solution}

\begin{solution}[\cref{exer:iso_in_sets}]\label{sol:iso_in_sets}
The isomorphisms in $\SET$ are precisely the bijective functions, indeed:
\begin{itemize}
\item Assume $f: X\to Y$ is a bijection, i.e.
\[
\forall y\in Y: \exists! x_{y}\in X: f(x)=y 
\]
We show that the inverse of $f$ is given by:
\[
g : Y\to X: y\mapsto x_y.
\]
So we have to show $\co{f}{g} = \Id[X]$ and $\co{g}{f} = \Id[Y]$. Let $x\in X$, since that $g(f(x))$ is the unique element $z\in X$ such that $f(z) = f(x)$ (and since $x$ satisfies this condition), we have $g(f(x)) = x$. Since this holds for all $x\in X$, we have $\co{f}{g} = \Id[X]$.\\
Let $y\in Y$ and let $x := x_y$  be the unique element in $X$ such that $f(x)=y$. So by definition of $g$, we have $g(y) = x$, hence 
$f(g(y)) = f(x) = y$.
\item Assume $f:X\to Y$ is an isomorphism with inverse $f^{-1}$. Let $y\in Y$, we have to show that there exists a unique $x\in X$ such that $f(x)=y$. Define $x := g(y)$. Since $\co{g}{f} = \Id[Y]$, we have $y = f(g(y)) = f(x)$, hence, this $x$ indeed satisfies the condition. To show that $x$ is unique, let $z\in X$ satisfy $f(z)=y$. That $z=x$ now follows from $\co{f}{g} = \Id[X]$, indeed: 
$z = g(f(z)) = g(y) = x$.
\end{itemize}
\end{solution}

\begin{solution}[\cref{exer:iso_in_pos}]\label{sol:iso_in_pos}
The isomorphisms in $\POS$ are precisely the bijections $f:(X,\leq_X) \to (Y,\leq_Y)$ such that 
\begin{equation}\label{eqn:order_iso}
x_1 \leq_X x_2 \iff f(x_1) \leq_Y f(x_2),
\end{equation}
Indeed:
\begin{itemize}
\item Assume $f$ is a bijection which satisfies \cref{eqn:order_iso}. Since it is a bijection, we know (by the solution to \cref{exer:iso_in_sets}), that there exists a function $g:(Y,\leq_Y)\to (X,\leq_X)$ such that $\co{f}{g}=\Id[(X,\leq_X)]$ and $\co{g}{f}=\Id[(Y,\leq_Y)]$. However, this does not conclude the proof of the first implication, because we do not know a priori, that $g$ is a morphism of posets. So we have to show
\[
\forall y_1,y_2\in Y: y_1\leq_Y y_2 \implies g(y_1)\leq_X g(y_2).
\]
Let $y_1,y_2\in Y$. Since $f$ is bijective, there exist $x_1,x_2 \in X$ such that $f(x_1)=y_1$ and $f(x_2)=y_2$. If $f(x_1) = y_1\leq_Y y_2 = f(x_2)$, then by \cref{eqn:order_iso}, we also have that $x_1 \leq x_2$. But by definition of $g$, we have $g(y_1)=x_1$ and $g(y_2)=x_2$, hence $g(y_1)\leq_X g(y_2)$ which shows that $g$ is an order-preserving morphism, i.e. $g \in \CHom{\POS}{(Y,\leq_Y)}{(X,\leq_X)}$.
\item Assume $f$ is an isomorphism in $\POS$ with inverse $g$. Since $f$ is a function which satisfies $\co{f}{g}=\Id[(X,\leq_X)]$ and $\co{g}{f}=\Id[(Y,\leq_Y)]$, we have (by the same argument as in the solution to \cref{exer:iso_in_sets}), that $f$ is a bijection. Hence, it remains to show that \cref{eqn:order_iso} holds. Let $x_1,x_2\in X$.
  
  If $x_1\leq_X x_2$, then we have $f(x_1)\leq_Y f(x_2)$ since $f \in\CHom{\POS}{(X,\leq_X)}{(Y,\leq_Y)}$.
  
Assume $f(x_1)\leq_Y f(x_2)$. Since $g \in\CHom{\POS}{(Y,\leq_Y)}{(X,\leq_X)}$, we have $g(f(x_1)) \leq_X g(f(x_2))$. But $\co{f}{g} = \Id[X]$, hence $x_1 \leq_X x_2$.
\end{itemize}
\end{solution}

\begin{solution}[\cref{exer:iso_in_posetcategory}]\label{sol:iso_in_posetcategory}
First, let $(X,\leq_Y)$ be a preorder. A morphism $f:x\to y$ is an isomorphism if and only if there exists a morphism $g: y\to x$ such that $\co{f}{g}=\Id[x]$ and $\co{g}{f}=\Id[y]$. But, in a preorder category, each hom-set has a unique element if it is non-empty. So, for any $g\in \Hom{y}{x}$ and $f\in \Hom{x}{y}$, we always have $\co{f}{g}=\Id[x]$ and $\co{g}{f}=\Id[y]$. Hence a morphism $f:x\to y$ in a preorder-category is an isomorphism if and only if there exists a morphism $g:y\to x$. The existence of a morphism $f:x\to y$ means precisely that $x\leq y$. Hence, isomorphisms in a preorder-category corresponds with a pair of elements $(x,y)$ in $X$ such that $x\leq y$ and $y\leq x$.\\
If $(X,\leq_X)$ is a poset, i.e. satisfies antisymmetry, then if $x\leq y$ and $y\leq x$, we must have $x=y$. Consequently, in a poset-category, the only isomorphisms are the identity morphisms (i.e. corresponding with $x\leq x$).
\end{solution}

\begin{solution}[\cref{exer:iso_in_cats_of_nats}]\label{sol:iso_in_cats_of_nats}
	A morphism $f:m\to n$ is an isomorphism if and only if there exists a morphism $g: n\to m$ such that $\co{f}{g}=\Id[m]$ and $\co{g}{f}=\Id[n]$.
	\begin{enumerate}
		\item $\leq$ is a preorder on the natural numbers; hence, as explained in \cref{sol:iso_in_posetcategory}, the isomorphisms in $\POS(\NN,\leq)$ are the identity morphisms.
		\item Let $m$ and $n$ be natural numbers. A morphism $f : m \to n$ in $\SKELFINSET$ is a function from $[m]$ to $[n]$. A function $f$ has an inverse if there exists a function $g$ from $[n]$ to $[m]$ such that $\co{f}{g}$ is the identity function on $[m]$ and $\co{g}{f}$ is the identity function on $[n]$.		
		Such a function $g$ is a morphism from $[n]$ to $[m]$, and this requirement is precisely the requirement for $f$ being an isomorphism in $\SKELFINSET$ are the invertible functions. A function is invertible if and only if it is bijective.
		Thus, we can equivalently say that isomorphisms in $\SKELFINSET$ are bijective functions. 
		\item  Let $m$ and $n$ be natural numbers.
		A morphism $M : m \to n$ in $\MAT$ is an $n \times m$ matrix. A matrix $M$ has an inverse if there exists a matrix $N$ such that $MN = NM = I$, where $I$ is the identity matrix of the appropriate size.
		This requirement is precisely the requirement for $M$ being an isomorphism in $\MAT$. Hence, the isomorphisms in $\MAT$ are the invertible Matrices.
		Note that all invertible matrices are square matrices, which means that all isomorphisms in $\MAT$ are morphisms in $\MAT(n,n)$ for some $n \in \NN$.
	\end{enumerate}
\end{solution}

\begin{solution}[\cref{exer:section-retraction-bool-int}]\label{sol:section-retraction-bool-int}
Consider
\begin{lstlisting}
bool2Int :: Bool -> Int
bool2Int False = 0
bool2Int True  = 1
\end{lstlisting}    

We can go back, so that we get |False| and |True| from |0| and |1|:
\begin{lstlisting}
int2Bool :: Int -> Bool
int2Bool n | n == 0    = False
           | otherwise = True
\end{lstlisting}
However, notice that not only |1| is converted back to |True|, but also everything other than |0| is converted to |True|.

We have
\begin{lstlisting}
   Int2Bool (bool2Int y) = y
\end{lstlisting}
for every |y :: Bool|, but we don't have |bool2Int (int2Bool x) = x| for all |x :: Int|.

We can say that there is enough room in the type integers for it to host a copy of the type of booleans, but there isn't enough room in the type of booleans for it to host a copy of the type of integers.

But notice that there are other ways in which the type |Bool| lives inside the type |Int| as a retract: for example, we can send |False| to |13| and |True| to |17|, and then send back everything bigger than |15| to |True| and everything else to |False|.
\end{solution}

\begin{solution}[\cref{ex:mono-inj}]\label{sol:mono-inj}
The monomorphisms in $\SET$ correspond precisely with the injective functions, i.e. the functions $f:X\to Y$ which satisfy
\[
\forall x_1,x_2\in X: f(x_1)=f(x_2) \implies x_1=x_2.
\]
Indeed:
\begin{itemize}
\item First, we show that if $f$ is injective, then it is a monomorphism. Assume $f$ is injective. Let $g,h: Z\to X$ be functions such that $\co{g}{f}=\co{h}{f}$. We have to show $g=h$, i.e. 
\[
\forall z\in Z: g(z)=h(z).
\]
Since $f$ is injective, it suffices to show 
\[
\forall z\in Z: f(g(z))=f(h(z)).
\]
But this holds by the condition of $g$ and $h$. Hence, $f$ is indeed a monomorphism.
\item Next, we show that if $f$ is a monomorphism, then it is injective. Assume $f$ is a monomorphism. We have to show that for all $x_1,x_2\in X$, $f(x_1) = f(x_2)$ implies $x_1 = x_2$. Let $\mathbf{1} = \{\star\}$ be a singleton set and define
\[
g_1 : \mathbf{1}\to X: \star\mapsto x_1,\quad  g_2 : \mathbf{1}\to X: \star\mapsto x_2. 
\]
Since $f(x_1)=f(x_2)$, we have $\co{g_1}{f} = \co{g_2}{f}$. But $f$ is a monomorphism; hence $g_1 = g_2$ which means $x_1 = g_1(\star) = g_2(\star) = x_2$. Thus, $f$ is indeed injective.
\end{itemize}
\end{solution}

\begin{solution}[\cref{exer:sections_in_set_injective}]\label{sol:sections_in_set_injective}
Let $f:X\to Y$ be a section with a retraction $h:Y\to X$, i.e. $\co{f}{h} = \Id[Y]$. By \cref{ex:mono-inj}, it suffices to show that $f$ is a monomorphism. Let  $g_1, g_2 : Z \to X$ be morphisms in $\SET$ such that $\co{g_1}{f} = \co{g_2}{f}$. We have to show $g_1=g_2$, which follows from the following computation:
\begin{eqnarray*}
g_1 =& \co{g_1}{\Id[Y]} & \text{ by unit law},\\ 
	=& \co{g_1}{(\co{f}{h})} & \text{ since $f$ section},\\ 
	=& \co{(\co{g_1}{f})}{h} & \text{ by associativity},\\
	=&  \co{(\co{g_2}{f})}{h} & \text{ by assumption},\\
	=& \co{g_2}{(\co{f}{h})} & \text{ by associativity},\\ 
	=& \co{g_2}{\Id[Y]} & \text{ since $f$ section},\\
	=& g_2 & \text{ by unit law}.
\end{eqnarray*}
Notice that this proof shows that in an arbitrary category, a section is always a monomorphism.
\end{solution}

\begin{solution}[\cref{exer:iso_to_monoepi}]\label{sol:iso_to_monoepi}
Let $f : a\cong b$ be an isomorphism with inverse $f^{-1}$. There are multiple proofs which one can give, an abstract one (which is \textit{indirect} in the sense that we use another exercise/lemma) and a more \textit{direct} one.
\begin{itemize}
\item \textbf{Indirect proof:} By the solution of \ref{exer:sections_in_set_injective}, we know that any section (from a section-retraction pair) is a monomorphism. An analogous argument shows that any retraction (section-retraction pair) is an epimorphism. Hence it suffices to show that $f$ is both a section and a retraction, but this is immediate because $\co{f}{f^{-1}} = \Id[a]$ and $\co{f^{-1}}{f} = \Id[b]$.
\item \textbf{Direct proof:} We first show that $f$ is a monomorphism. Assume $g_1,g_2 : c\to a$ are morphisms such that $\co{g_1}{f} = \co{g_2}{f}$. We then have that $g_1 = g_2$ because 
\begin{eqnarray*}
g_1 = \co{g_1}{\Id[a]} =& \co{g_1}{(\co{f}{f^{-1}})} & \text{ since $f : a\cong b$},\\ 
	=& \co{(\co{g_1}{f})}{f^{-1}} & \text{ by associativity}, \\ 
	=& \co{(\co{g_2}{f})}{f^{-1}} & \text{ by assumption}, \\
	=& \co{g_2}{(\co{f}{f^{-1}})} & \text{ by associativity} \\
	=& \co{g_2}{\Id[a]} & \text{ since $f : a\cong b$} \\
	=& g_2.
\end{eqnarray*}
That $f$ is also an epimorphism is analogous, indeed: Assume $g_1,g_2 : b\to c$ are morphisms such that $\co{f}{g_1} = \co{f}{g_2}$. We then have that $g_1 = g_2$ because 
\begin{eqnarray*}
g_1 = \co{\Id[b]}{g_1} =& \co{(\co{f^{-1}}{f})}{g_1} & \text{ since $f : a\cong b$},\\ 
	=& \co{f^{-1}}{(\co{f}{g_1})} & \text{ by associativity}, \\ 
	=& \co{f^{-1}}{(\co{f}{g_2})} & \text{ by assumption}, \\
	=& \co{(\co{f^{-1}}{f})}{g_2} & \text{ by associativity} \\
	=& \co{\Id[b]}{g_2} & \text{ since $f : a\cong b$} \\
	=& g_2.
\end{eqnarray*}
\end{itemize}
\end{solution}

\begin{solution}[\cref{exer:counterexample_monoepi_not_iso}]\label{sol:counterexample_monoepi_not_iso}
Let $(X,\leq_X)$ be a preordered set. Any morphism $f \in \POS(X,\leq_X)(x,y)$ is always both a monomorphism and an epimorphism because hom-sets have at most on element. But, by \cref{exer:iso_in_posetcategory}, we know that in a poset (not a preordered set!), the only isomorphisms are the identity morphisms. Hence, if $x\leq y$ but $x\not=y$ (living in a poset), then the corresponding morphism in $\Hom{x}{y}$ is both an epimorphism and monomorphisms but not an isomorphism.\\
A concrete example is given by e.g. the poset of truth values $\{0, 1\}$. We have $0\leq 1$ and those are not equal.
\end{solution}

\begin{solution}[\cref{exer:initial_set}] \label{sol:initial_set}
An initial object (and the only one), is the emptyset $\emptyset$, indeed: Let $X$ be a set. Then there is clearly a unique function $\emptyset\to X$.
\end{solution}

\begin{solution}[\cref{exer:initial_posetcat}]\label{sol:initial_posetcat}
A initial object in $\POS(X,\leq)$ is the minimal object, that is an element $\bot\in X$ such that
\begin{equation}
\forall y\in X: \bot \leq y.
\end{equation}
Indeed: Assume $x$ is an initial object in $(X,\leq)$, i.e. for any other element $y\in X$, there exists a (unique) morphism $x\to y$, i.e. hence, by definition of the hom-sets, we have $x\leq y$. So $x$ is indeed the minimal object.

Conversely, assume $\bot$ is a minimal element, hence, for each $y\in X$, we have $\bot\leq y$. Hence $\Hom{\bot}{y}$ is non-empty. So it must contain exactly one element. This means precisely that it is initial.

A somewhat \textit{more compact} solution is as follows: By definition of $\POS(X,\leq)$, for each $x\in X$, we have:
\[
\forall y\in X: \left(x\leq y \iff \exists! f\in \Hom{x}{y}\right).
\]
Hence an object $x$ is initial if and only if, $x\leq y$ for all $y\in X$, if and only if it is minimal.
\end{solution}

\begin{solution}[\cref{exer:initial-unique}]\label{sol:initial-unique}
Let $A$ and $B$ be initial objects in $\CC$. By initiality of $A$ (resp. $B$), there exists a unique morphism $f \in \CHom{\CC}{A}{B}$ (resp. $g \in \CHom{\CC}{B}{A}$). Both $\Id[A]$ and $\co{f}{g}$ (resp. $\Id[B]$ and $\co{g}{f}$) are in $\CHom{\CC}{A}{A}$ (resp. $\CHom{\CC}{B}{B}$), but we know that there exists a unique morphism in $\CHom{\CC}{A}{A}$ (resp. $\CHom{\CC}{B}{B}$). Hence, $\Id[A] = \co{f}{g}$ (resp. $\Id[B] = \co{g}{f}$), which means that $f$ and $g$ are inverses.
\end{solution}

\begin{solution}[\cref{exer:initiality_preserved_by_iso}]\label{sol:initiality_preserved_by_iso}
Assume $A\in \Ob{\CC}$ is initial, $B\in\Ob{\CC}$ an arbitrary object and $i:A\cong B$ an isomorphism. We have to show that $B$ is initial, i.e. for each $X\in\Ob{\CC}$, there should exists a unique morphism $B\to X$.

Fix such an $X$. By initiality of $A$, there exists a (unique) morphism $f\in \CHom{\CC}{A}{X}$. If we denote the inverse of $i$ by $j$, we have $\co{j}{f} \in\CHom{\CC}{B}{X}$. To show that $\co{j}{f}$ is the unique morphism in this hom-set, let $g\in \CHom{\CC}{B}{X}$. So we have $\co{i}{g} \in \CHom{\CC}{A}{X}$. By initiality of $A$, we have $\co{i}{g} = f$. The claim now follows by the following computation:
\[
\co{j}{f} = \co{j}{(\co{i}{g})} = \co{(\co{j}{i})}{g} = \co{\Id[B]}{g} = g.
\]
\end{solution}

\begin{solution}[\cref{exer:cat-without-initial}]\label{sol:cat-without-initial}
We give three solutions to this exercise.
\begin{itemize}
\item Consider the category generated by the graph: 
\[
\begin{tikzcd}
x & y
\end{tikzcd}
\]
This category can not have an initial object since there is no morphism from $x$ to $y$ or vice versa.
\item Consider the category generated by the graph: 
\[
\begin{tikzcd}
x \arrow[r, bend left, "f"] \arrow[r, bend right, "g"] & y
\end{tikzcd}
\]
This category also can not have an initial object, indeed: There is no morphism from $y$ to $x$, hence $y$ can not be initial. But also $x$ can not be initial since $f$ and $g$ are different morphisms.
\item Consider the category $ \POS(\mathbb Z, \leq) $, i.e., the category
\[
\begin{tikzcd}
	\dots \arrow[r] & -2 \arrow[r] & -1 \arrow[r] & 0 \arrow[r] & 1 \arrow[r] & 2 \arrow[r] & \dots
\end{tikzcd}
\]
This category can not have an initial object: Suppose it has an initial object $ x \in \mathbb Z $. Then we have another object $ x - 1 \in \mathbb Z $. Since $ x $ is initial, we have a morphism $ x \to x - 1 $, which means that $ x \leq x - 1 $, which is absurd. Therefore, this category does not have an initial object.
\end{itemize}
\end{solution}

\begin{solution}[\cref{exer:coproduct-represent}]\label{sol:coproduct-represent}
\newcommand{\CP}{\CC(A+B,X)}
\newcommand{\CQ}{\CC(A,X)\times\CC(B,X)}
\newcommand{\PP}{\CC(X,A\times B)}
\newcommand{\PQ}{\CC(X,A)\times\CC(X,B)}
 We need to construct a morphism \[\alpha:\SET(\CP, \CQ)\] and its inverse \[\beta:\SET(\CQ, \CP)\]
such that
\begin{equation}
\label{coproduct-iso-1}
    \alpha \circ \beta = \Id[\CQ]
\end{equation}
\begin{equation}
\label{coproduct-iso-2}
    \beta \circ \alpha = \Id[\CP]
\end{equation}
Note that $\alpha$ and $\beta$ are morphisms in the category of sets ($\SET$).\\

We define \[\alpha(k):=(\co {\inl} {k}, \co {\inr} {k})\] to be our candidate function $\alpha$. And we define \[\beta(a,b):= \outofcoproduct{a}{b}\] to be our candidate function $\beta$. Here $\outofcoproduct{a}{b}$ is also known as the coproduct map, which is exactly the unique morphism from $A+B$ into $X$ that makes the corresponding coproduct triangles commute, as in the following diagram:

\[
\begin{tikzcd}
A \arrow[r, "\inl"] \arrow[rd, "a"'] & A+B \arrow[d, "{\outofcoproduct{a}{b}}" description] & B \arrow[l, "\inr"'] \arrow[ld, "b"] \\
                                    & X                                    &
\end{tikzcd}
\]

Equivalently, it is the unique morphism s.t.

\begin{equation}
\label{coproduct-iso-inl}
    \co {\inl} {\outofcoproduct{a}{b}} = a
\end{equation}
\begin{equation}
\label{coproduct-iso-inr}
    \co {\inr} {\outofcoproduct{a}{b}} = b
\end{equation}

To show that $\alpha$ is an isomorphism we have to show that equations \cref{coproduct-iso-1} and \cref{coproduct-iso-2} hold.
\textbf{Claim $\co {\beta} {\alpha} = \Id[\CQ]$}(\cref{coproduct-iso-1})
\begin{proof}
Fix $(a',b') \in \CQ$.
\begin{align*}
 &\quad (\co {\beta} {\alpha}) ((a',b'))\\
=&\qquad \text{\{ Function composition \}} \\
 &\quad \alpha(\beta((a',b'))) \\
=&\qquad \text{\{ Definition of $\beta$ \}} \\
 &\quad \alpha(\outofcoproduct{a'}{b'}) \\
=&\qquad \text{\{ Definition of $\alpha$ \}} \\
 &\quad (\co {\inl} {\outofcoproduct{a'}{b'}}, \co {\inr} {\outofcoproduct{a'}{b'}}) \\
=&\qquad \text{\{ Using \ref{coproduct-iso-inl} and \ref{coproduct-iso-inr} \}} \\
 &\quad (a', b') \\
=&\qquad \text{\{ Identity on $\CQ$ \}} \\
 &\quad \Id[\CQ]((a', b')) \\
\end{align*}
\end{proof}
We conclude $\co {\beta} {\alpha} = \Id[\CQ]$ (assuming functional extensionality).

\textbf{Claim $\co {\alpha} {\beta} = \Id[\CP]$} (\cref{coproduct-iso-2}).
\begin{proof}
Fix $k' \in \CP$.
\begin{align*}
 &\quad (\co {\alpha} {\beta}) (k')\\
=&\qquad \text{\{ Function composition \}} \\
 &\quad \beta(\alpha(k')) \\
=&\qquad \text{\{ Definition of $\alpha$ \}} \\
 &\quad \beta((\co {\inl} {k'}, \co {\inr} {k'}))\\
=&\qquad \text{\{ Definition of $\beta$ \}} \\
 &\quad \outofcoproduct{\co {\inl} {k'}}{\co {\inr} {k'}}\\
=&\qquad \text{\{ Uniqueness of $\outofcoproduct{\co {\inl} {k'}}{\co {\inr} {k'}} \in \CC(A+B,X) \}$} \\
 &\quad k'\\
=&\qquad \text{\{ Identity on $\CP$ \}} \\
 &\quad \Id[\CP] (k')
\end{align*}
We conclude $\co {\alpha} {\beta} = \Id[\CP]$ (assuming functional extensionality).
\end{proof}

\end{solution}

\begin{solution}[\cref{exer:coproduct_cats_of_nats}]\label{sol:coproduct_cats_of_nats}
	\begin{enumerate}
		\item We know from \cref{exer:coproduct_posetcat}, that the coproduct of $A$ and $B$ in $\POS(X,\leq)$ is the smallest element that is greater than or equal to $A$ and $B$, for $A,B \in X$.
		Hence, for $m,n \in \NN$, the coproduct $m + n$ in $\POS(\NN,\leq)$ is $\max(m,n)$.
		\item Let $m,n \in \NN$. The coproduct $m + n$ in $\SKELFINSET$ is the sum of the natural numbers $m + n$. The inclusion maps $\iota_l$ and $\iota_r$ are defined as follows:
		\begin{align*}
			\iota_l : [m] &\to [m + n] \\
			k &\mapsto k, \\
			\iota_r : [n] &\to [m + n] \\
			k &\mapsto k + m.
		\end{align*}

		We now need to show that for each $l \in \NN$, $i_l : [m] \to [l]$ and $i_r : [n] \to [l]$, there exists a unique $f : [m + n] \to [l]$ such that $f \circ \iota_l = i_l$ and $f \circ \iota_r = i_r$:
		\[
			\begin{tikzcd}
			  m \ar[r, "\inl"] \ar[rd, "i_l"']
			  &
			  m + n  \ar[d, dashed, "f"]
			  &
			  n \ar[l, "\inr"'] \ar[ld, "i_r"]
			  \\
			  &
			  l.
			\end{tikzcd}
		\]
		We define $f$ as follows: 
		\begin{align*}
			f : [m+n] & \to  [l] \\
			k & \mapsto i_l(k) & \mathrm{if~} k < m \\
			k & \mapsto i_r(k - m)  & \mathrm{if~} m \leq k.
		\end{align*}	
		This $f$ satisfies $f \circ \iota_l = i_l$ as for each $k \in [m]$ we have:
		\[ f (\iota_l(k)) = f(k) = i_l(k),\]
		and satisfies $f \circ \iota_r =  i_r$ as for each $k \in [n]$ we have:
		\[ f(\iota_r (k)) = f(k + m) = i_r(k).\]
		Note that in both cases, we are using functional extensionality.

		To show uniqueness, we need to show that for each $g : [m+n] \to [l]$ such that $g \circ \iota_l = i_l$ and $g \circ \iota_r = i_r$, we have $f = g$. We show that for each $k \in [m+n]$, $f(k) = g(k)$, which using functional extensionality implies $f = g$. 
		Let $k \in [m+n]$. If $k < m$ we have: 
		\[ f(k) = i_l(k) = g (\iota_l (k)) = g(k),\]
		and if $m \leq k$ we have:
		\[ f(k) = i_r(k - m) = g (\iota_r (k - m)) = g(k). \]
		Hence, for each $k \in [l]$, $f(k) = g(k)$, and by functional extensionality $f = g$.
		\item Let $m,n \in \NN$. The coproduct $m + n$ in $\MAT$ is the sum of the natural number $m + n$. The left and right inclusions correspond to the $(m+n) \times m$ matrix $J_l$ and the $(m+n) \times n$ matrix $J_r$ respectively, which are defined as follows:
		\[
		J_l = 
		\begin{pmatrix}
			I_m \\ 
			0_{n \times m}
		\end{pmatrix},
		J_r = 
		\begin{pmatrix}
			0_{m \times n} \\
			I_n
		\end{pmatrix}.
		\]
		The matrix $J_l$ (resp. $J_r$) can be thought of as the embedding from $\mathbb{R}^{m}$ (resp. $\mathbb{R}^{n}$) to the first $m$ (resp. last $n$) dimensions of $\mathbb{R}^{m+n}$.

		We need to show that for each $l \in \NN$, $l \times m$ matrix $K_l$ and $l \times n$ matrix $K_r$, there exists a unique $l \times (m + n)$ matrix $M$ such that $M J_l = K_l$ and $M J_r = K_r$. 
		When multiplied from the right, the matrix $J_l$ picks out the first $m$ columns, and $J_r$ picks out the last $n$ columns. This means that the matrix $M$ defined as follows satisfies $M J_l = K_l$ and $M J_r = K_r$:
		\[
		M = 
		\begin{bmatrix}
			K_l \mid K_r
		\end{bmatrix}.
		\]
		To show uniqueness, we need to show that for each $l \times (m+n)$ matrix $N$ such that $N J_l = K_l$ and $N J_r = K_r$, we have $M=N$.
		From $N J_l = K_l$ we know that the first $m$ columns of $N$ are $K_l$ which is equal to the first $m$ columns of $M$. From $N J_r = K_r$ we know that columns $m+1$ to $m+n$ of $N$ are equal to $K_r$, which are equal to columns $m+1$ to $m+n$ of $M$. Hence, all columns of $M$ and $N$ are equal and $M =N$.
	\end{enumerate}
\end{solution}

\begin{solution}[\cref{exer:swap_binary_coproduct}]\label{sol:swap_binary_coproduct}
We fix objects $A,B\in\Ob{\CC}$. These objects have two binary coproducts in $\CC$: $(A+B,\inl,\inr)$ and $(B+A,\inl',\inr')$, where $\inl,\inr,\inl',\inr'$ are morphisms with types:

\begin{itemize}
    \item $\inl:A \to A+B$
    \item $\inr:B \to A+B$
    \item $\inl':B \to B+A$ 
    \item $\inr':A \to B+A$
\end{itemize}

We need to provide a morphism $f:A+B \to B+A$ and its inverse $g:B+A \to A+B$, such that

\begin{equation}
\label{unit1}
    \co {g} {f} = \Id[B+A]
\end{equation}
\begin{equation}
\label{unit2}
    \co {f} {g} = \Id[A+B]
\end{equation}

We define $f : A + B \to B + A$ via the universal property of $(A + B, \inl, \inr)$, as the unique morphism that makes the corresponding triangles commute --- see \eqref{co_f1} and \eqref{co_f2}.
Analogously, we define $g : B + A \to A + B$; see \eqref{co_g1} and \eqref{co_g2}.

\noindent
We can visualize these coproducts with the following diagram:

\[
\begin{tikzcd}
A \arrow[r, "\inl"] \arrow[rd, "\inr'"'] & A+B \arrow[d, "f"', shift right] & B \arrow[l, "\inr"'] \arrow[ld, "\inl'"] \\
                                      & B+A \arrow[u, "g"', shift right] &                                       
\end{tikzcd}
\]

\noindent
Here $f$ and $g$ are morphisms that exist as a result of the definition of coproducts. These will be our candidates for constructing an isomorphism $A+B\cong B+A$. By the definition of coproducts we also have that the following equations hold:

\begin{equation}
\label{co_f1}
    \co {\inl} {f} = \inr'
\end{equation}
\begin{equation}
\label{co_f2}
    \co {\inr} {f} = \inl'
\end{equation}
\begin{equation}
\label{co_g1}
    \co {\inl'} {g} = \inr
\end{equation}
\begin{equation}
\label{co_g2}
    \co {\inr'} {g} = \inl
\end{equation}

We will use these, along with an additional lemma (Lemma \ref{lemma}), to show that equations \ref{unit1} and \ref{unit2} hold for our $f$ and $g$, and therefore they form an isomorphism.\\

\noindent
\textbf{Claim $\co {g} {f} = \Id[B+A]$ (\ref{unit1})}.
\begin{proof}
Starting from equation \ref{co_f1}, we derive
\begin{align*}
 &\quad \co {\inl} {f} = \inr' \\
\implies&\qquad \text{\{ By \ref{co_g1} \}} \\
 &\quad \co {(\co {\inr'} {g})} {f} = \inr'  \\
\implies&\qquad \text{\{ Associativity \}} \\
 &\quad \co {\inr'} {(\co {g} {f})} = \inr'
\end{align*}
\noindent
And starting from equation \ref{co_f2}, we derive
\begin{align*}
 &\quad \co {\inr} {f} = \inl' \\
\implies&\qquad \text{\{ By \ref{co_g2} \}} \\
 &\quad \co {(\co {\inl'} {g})} {f} = \inl'  \\
\implies&\qquad \text{\{ Associativity \}} \\
 &\quad \co {\inl'} {(\co {g} {f})} = \inl'
\end{align*}
\noindent
Combining $\co {\inr'} {(\co {g} {f})} = \inl'$ and $\co {\inl'} {(\co {g} {f})} = \inl'$ with Lemma \ref{lemma}, we conclude $\co {g} {f}=\Id[B+A]$.
\end{proof}

\noindent
\textbf{Claim $\co {f} {g} = \Id[A+B]$ (\ref{unit2})}.
\begin{proof}
Starting from equation \ref{co_g1}, we derive
\begin{align*}
 &\quad \co {\inl'} {g} = \inr \\
\implies&\qquad \text{\{ By \ref{co_f2} \}} \\
 &\quad \co {(\co {\inr} {f})} {g} = \inr  \\
\implies&\qquad \text{\{ Associativity \}} \\
 &\quad \co {\inr} {(\co {f} {g})} = \inr
\end{align*}
\noindent
And starting from equation \ref{co_g2}, we derive
\begin{align*}
 &\quad \co {\inr'} {g} = \inl \\
\implies&\qquad \text{\{ By \ref{co_f1} \}} \\
 &\quad \co {(\co {\inl} {f})} {g} = \inl  \\
\implies&\qquad \text{\{ Associativity \}} \\
 &\quad \co {\inl} {(\co {f} {g})} = \inl
\end{align*}
\noindent
Combining $\co {\inl} {(\co {f} {g})} = \inl$ and $\co {\inr} {(\co {f} {g})} = \inr$ with Lemma \ref{lemma}, we conclude $\co {f} {g}=\Id[A+B]$.
\end{proof}

\noindent
Since $f : A+B \to B+A$ is an isomorphism with $g : B+A \to A+B$ as its inverse, we have successfully constructed our desired isomorphism $A+B \cong B+A$.
\end{solution}


\begin{solution}[\cref{exer:kleisli_triple_list}]
  \label{sol:kleisli_triple_list}
  For any set $X$, we write
  \[ (+) : \List(X) \times \List(X) \to \List(X)\]
  for list concatenation.

  For each set $X\in\Ob \SET$, we define:
  \[
    \eta_X : X \to \List(X) : x\mapsto [x] := \cons(x,\nil).
  \]
  For each function $f\in\CHom{\Ob \SET}{X}{\List(Y)}$, we define, by list recursion, the following function:
  \begin{align}
    f^{*} : \List(X) &\to \List(Y) 
    \\
    \nil & \mapsto \nil  \label{eq:list-bind-nil}
    \\
    \cons(x,xs) &\mapsto fx + f^*xs  \label{eq:list-bind-cons}
  \end{align}
  
We now show that the properties of a Kleisli triple hold:
\begin{enumerate}
\item For each set $X$, we have to show $\eta_X^{*} = \Id[T(X)]$,
  that is, for a list $\ell \in \List (X)$, we show $\eta_X^{*}(\ell) = \ell$.
  We prove this by structural induction on the list $\ell$.
  
  In case $\ell = \nil$, we have, by \cref{eq:list-bind-nil}, that $\eta_X^{*}(\nil) = \nil$.
  
  In case $\ell = \cons(x,xs)$, we compute
  \begin{align*}
    \eta_X^{*}(\cons(x,xs)) &= \eta_X(x) + \eta_X^{*}(xs) & \text{ by definition of } (-)^{*},
    \\
                            &= \eta_X(x) + \Id[\List(X)](xs)  & \text{ by induction hypothesis}
    \\
                            &= [x] + \Id[\List(X)](xs) & \text{ by definition of } \eta_X
    \\
                            &= [x] + xs
    \\
                            &= \cons(x,xs).
  \end{align*}

\item For each function $f:X\to \List(Y)$, we have to show $f^{*}(\eta_X(x)) = f(x)$, this indeed holds by the following computation:
\[
f^{*}(\eta_X(x)) = f^{*}(\cons(x,\nil)) = \cons(fx, f^{*}(\nil)) = \cons(fx, \nil),
\]
where the first equality holds by definition of $\eta_X$ and the second equality holds by definition of $f^{*}$.
\item Let $f:X\to \List(Y)$ and $g:Y\to \List(Z)$ be functions, we have to show 
\[
g^{*}(f^{*}(\ell)) = (\co{f}{g^{*}})^{*}(\ell),
\] 
for any $\ell \in \List(X)$.
 We prove this by structural induction on the list $\ell$.

\begin{itemize}
\item In case $\ell = \nil$, the equality holds trivially by \cref{eq:list-bind-nil}.

\item In case $\ell := \cons(x,s)$. By definition of $(f)^{*}$, the left hand side of is:
\[
g^{*}(f^{*}(\cons(x,s))) = g^{*}(f(x) + f^{*}(xs)),
\]
and the right hand side is:
\[
(\co{f}{g^{*}})^{*}(\cons(x,s)) = g^{*}(f(x)) + (\co{f}{g^{*}})^{*}(xs) = g^{*}(f(x)) + g^{*}(f^{*}(xs)),
\]
where the first equality holds by definition of $(\co{f}{g^{*}})^{*}$ and the second holds by the induction hypothesis. Hence it remains to show the following equality:
\begin{align}\label{eqn:bind_distributes}
g^{*}(f(x) + f^{*}(xs)) = g^{*}(f(x)) + g^{*}(f^{*}(xs)).
\end{align}

We do a pattern matching on $f(x)$ to show \cref{eqn:bind_distributes}:
\begin{itemize}
\item In case $f(x) = \nil$, we have
\[
g^{*}(\nil + f^{*}(xs)) = g^{*}(f^{*}(xs)) = \nil + g^{*}(f^{*}(xs)) = g^{*}(\nil) + g^{*}(f^{*}(xs)),
\]
where the third equality holds by \cref{eq:list-bind-nil}.
\item In case $f(x) = \cons(y,u)$, with $y\in Y$ and $u\in \List(Y)$, we have:
\begin{eqnarray*}
g^{*}(\cons(f(x),f^{*}(s))) =& g^{*}(\cons(\cons(y,u), f^{*}(xs))), \\
	=& g^{*}(\cons(y, u + f^{*}(xs))),\\
	=& g(y) + g^{*}(u + f^{*}(xs)),\\
	=& g(y) + g^{*}(u) + g^{*}(f^{*}(xs)),\\
	=& g^{*}(\cons(y,u)) + g^{*}(f^{*}(xs)).
\end{eqnarray*}
where the second equality holds by definition of $\cons$, the third and fifth by definition of $g^{*}$ and the fourth by the induction hypothesis.

\end{itemize}
\end{itemize}
\end{enumerate}
\end{solution}

\begin{solution}[\cref{exer:kleisli_triple_bintree}]
\label{sol:kleisli_triple_bintree}
For each set $X\in\Ob \SET$, we define:
\[
\eta_X : X \to \BinTree(X) : x\mapsto \Leaf x.
\]
For each function $f\in\CHom{\Ob \SET}{X}{\BinTree(Y)}$, we define:
\begin{align*}
f^{*} : \BinTree(X) \to \BinTree(Y) : t \mapsto 
\begin{cases}
f(a) &\quad \text{ if } t=\Leaf a,\\
\Branch{f^{*}(t_1)}{f^{*}(t_2)} &\quad \text{ if } t=\Branch{t_1}{t_2}.
\end{cases}
\end{align*}

We now show that the properties of a Kleisli triple hold:
\begin{enumerate}
\item For each set $X$, we have to show $\eta_X^{*} = \Id[\BinTree(X)]$. We show this by pattern matching on $t$:
\begin{itemize}
\item If $t=\Leaf a$, then
\[
\eta_X^{*}(t) = \eta_X^{*}(\Leaf{a}) = \eta_X(a) = \Leaf{a} = t.
\]
\item If $t=\Branch{t_1}{t_2}$, then
\[
\eta_X^{*}(t) = \eta_X^{*}(\Branch{t_1}{t_2}) = \Branch{\eta_X^{*}(t_1)}{\eta_X^{*}(t_2)} = \Branch{t_1}{t_2} = t.
\]
\end{itemize}

\item For each function $f:X\to \BinTree Y$, we have to show $f^{*}(\eta_X(a)) = f(a)$, this indeed holds by the following computation:
\[
f^{*}(\eta_X(a)) = f^{*}(\Leaf{a}) = f(a).
\]

\item Let $f:X\to \BinTree Y$ and $g:Y\to \BinTree Z$ be functions, we have to show 
\[
g^{*}(f^{*}(t)) = (\co{f}{g^{*}})^{*}(t).
\] 
That this equality holds follows by pattern matching:
\begin{itemize}
\item If $t= \Leaf{a}$, then
\begin{align*}
	g^{*}(f^{*}(t)) & = g^{*}(f^{*}(\Leaf{a})) = g^{*}(f(a)) = (\co{f}{g^{*}})(a) = (\co{f}{g^{*}})^{*}(\Leaf{a}) \\
	& = (\co{f}{g^{*}})^{*}(t).	
\end{align*}

\item If $t=\Branch{t_1}{t_2}$, then is the left-hand-side given by
\begin{align*}
	g^{*}(f^{*}(t)) & = g^{*}(f^{*}(\Branch{t_1}{t_2})) = g^{*}\left(\Branch{f^{*}(t_1)}{f^{*}(t_2)}\right) \\
	& = \Branch{g^{*}(f^{*}(t_1))}{g^{*}(f^{*}(t_2))}.	
\end{align*}

The right-hand-side is given by 
\[
(\co{f}{g^{*}})^{*}(t) = (\co{f}{g^{*}})^{*}(\Branch{t_1}{t_2}) = \Branch{(\co{f}{g^{*}})(t_1)}{(\co{f}{g^{*}})(t_2)}.
\]
Hence, by the induction hypothesis, the both sides are equal.
\end{itemize}

\end{enumerate}
\end{solution}

\begin{solution}[\cref{exer:kleisli_triple_exception}]
\label{sol:kleisli_triple_exception}
Before we continue with this exercise, we first fix some notation. Since $X+E$ is the disjoint union of $X$ and $E$, we have the canonical inclusions which we denote by
\[ i^X_l : X\to X + E, \quad i^X_r : E\to X+E. \] 
Hence, a function whose domain is $X+E$ is completely determined by specifying where each $i^X_l(x)$ and each $i^X_r(e)$ are mapped to. \textit{Notice that this is precisely the notation and the universal property of the coproduct (in $\SET$)}.

For each set $X\in\Ob \SET$, we define:
\[
\eta_X : X \to X+E : x\mapsto i^X_l(x).
\]
For each function $f\in\CHom{\Ob \SET}{X}{Y+E}$, we define:
\begin{align*}
f^{*} :X+E \to Y+E : z \mapsto 
\begin{cases}
f(x) &\quad \text{ if } z=i^X_l(x),\\
i^Y_r(e) &\quad \text{ if } z=i^X_r(e).
\end{cases}
\end{align*}

We now show that the properties of a Kleisli triple hold:
\begin{enumerate}
\item For each set $X$, we have to show $\eta_X^{*} = \Id[X+E]$: 
\begin{itemize}
\item If $z=i^X_l(x)$, then 
\[
\eta_X^{*}(z) = \eta_X^{*}(i^X_l(x)) = \eta_X(x) = i^X_l(x) = z = \Id[X+E](z).
\]
\item If $z=i^X_r(e)$, then
\[
\eta_X^{*}(e) = \eta_X^{*}(i^X_r(e)) = i^X_r(e) = z = \Id[X+E](z).
\]
\end{itemize}

\item For each function $f:X\to Y + E$, we have to show $f^{*}(\eta_X(x)) = f(x)$ but this holds directly by the definition of $(-)^{*}$ since $\eta_X(x)=i_l^X(x)$.

\item Let $f:X\to Y + E$ and $g:Y\to Z + E$ be functions, we have to show 
\[
g^{*}(f^{*}(z)) = (\co{f}{g^{*}})^{*}(z).
\] 
To show this, we do pattern matching on $z\in X+E$:
\begin{itemize}
\item If $z=i^X_l(x)$, then
\[
g^{*}(f^{*}(z)) = g^{*}(f^{*}(i_l^X(x))) = g^{*}(f(x)) = (\co{f}{g^{*}})^{*}(i_l^X(x)) = (\co{f}{g^{*}})^{*}(z).
\]
\item If $z=i^X_r(e)$, then
\[
g^{*}(f^{*}(z)) = g^{*}(f^{*}(i_l^X(e))) = g^{*}(i_l^Y(e)) = i_l^Z(e) = (\co{f}{g^{*}})^{*}(i_l^X(e)) = (\co{f}{g^{*}})^{*}(z).
\]
\end{itemize}

\end{enumerate}
\end{solution}

\begin{solution}[\cref{exer:kleisli_triple_side_effects}]
	\label{sol:kleisli_triple_side_effects}
	For each set $X\in\Ob \SET$, we define:
	\[
	\eta_X : X \to S \to X \times S : x\mapsto (\lambda s: S, (x,s)).
	\]
	For each function $f\in\CHom{\Ob \SET}{X}{S \to Y \times S}$, and each $c : S \to X \times S$, we define:
	\begin{align*}
	f^{*} (c) : S & \to Y \times S  \\
	s & \mapsto (f (\pi_l(c(s))) (\pi_r (c(s)))),
	\end{align*}
	where $\pi_l$ and $\pi_r$ are the left and right projections of the product. 
	
	We now show that the properties of a Kleisli triple hold:
	\begin{enumerate}
	\item For each set $X$, we have to show $\eta_X^{*} = \Id[S \to X \times S]$. For each $c : S \to X \times S$ and $s : S$ we have:
	\[\eta^*_X (c) (s)= (\eta_X) (\pi_l (c(s))) (\pi_r (c(s))) = (\pi_l (c(s)), \pi_r (c(s))) = c (s).\]
	
	\item For each function $f:X\to S \to Y \times S$, we have to show $f^{*}(\eta_X(x)) = f(x)$ for all $x \in X$. For each $s : S$ we have: 
	\[ f^* (\eta_X (x)) (s) = f^* (\lambda s: S, (x,s)) (s) = f (\pi_l (x,s)) (\pi_r (x,s)) = f (x)(s). \]
	
	\item Let $f:X\to S \to Y \times S$ and $g:Y\to S \to Z \times S$ be functions, we have to show for each $c : S \to X \times S$:
	\[
	g^{*}(f^{*}(c)) = (\co{f}{g^{*}})^{*}(c).
	\] 
	From the left hand side we have: 
	\begin{align*}
	g^* (f^* (c)) & = g^* (\lambda s : S, f (\pi_l (c(s))) (\pi_r (c(s)))) = (\lambda s : S, g^* (f (\pi_l (c(s))) (\pi_r (c(s))))) \\
	& = (\lambda s : S, (g^* \circ f) (\pi_l (c(s))) (\pi_r (c(s)))),
	\end{align*}
	and from the right hand side we have:
	\[ (g^* \circ f)^* (c) = (\lambda s : S, (g^* \circ f) (\pi_l (c(s))) (\pi_r (c(s)))).\]
	Hence, the left and right hand sides are equal. 
	\end{enumerate}
	\end{solution}

\begin{solution}[\cref{exer:kleisli_triple_nondeterminism}]
  \label{sol:kleisli_triple_nondeterminism}
For each set $X\in\Ob \SET$, we define:
\[
\eta_X : X \to \mathbb{P}_{fin}(X) : x\mapsto \{x\}.
\]
For each function $f\in\CHom{\Ob \SET}{X}{\mathbb{P}_{fin}(Y)}$, we define:
\begin{align*}
f^{*} : \mathbb{P}_{fin}(X) \to \mathbb{P}_{fin}(Y) : A \mapsto \bigcup_{a\in A} f(a).
\end{align*}
First notice that $\eta_X$ and $f^{*}$ are well-defined, indeed: 
\begin{itemize}
\item $\eta_X(x) = \{x\}$ is clearly finite since it only contains one element.
\item Let $A\in \mathbb{P}_{fin}(X)$. By definition of $f$, for each $a\in A$, $f(a)$ is finite. But there are only a finite number of elements in $A$, so $\bigcup_{a\in A} f(a)$ is a finite union of finite sets, hence, it is again finite.
\end{itemize}
We now show that the properties of a Kleisli triple hold:
\begin{enumerate}
\item For each set $X$, we have to show $\eta_X^{*} = \Id[\mathbb{P}_{fin}(X)]$. Let $A\in \mathbb{P}_{fin}(X)$, the claim then follows by the following computation:
\[
\eta_X^{*}(A) = \bigcup_{a\in A} \eta_X(a) = \bigcup_{a\in A} \{a\} = A = \Id[\mathbb{P}_{fin}(X)](A).
\]

\item For each function $f:X\to \mathbb{P}_{fin}(Y)$, we have to show $f^{*}(\eta_X(x)) = f(x)$ but this holds directly by the definition of $(-)^{*}$ since
\[
f^{*}(\eta_X(x)) = f^{*}(\{x\}) = \bigcup_{a\in \{x\}} f(a) = f(x).
\]

\item Let $f:X\to \mathbb{P}_{fin}(Y)$ and $g:Y\to \mathbb{P}_{fin}(Z)$ be functions, we have to show 
\[
g^{*}(f^{*}(A)) = (\co{f}{g^{*}})^{*}(A).
\] 
Let $A\in \mathbb{P}_{fin}(X)$, the left-hand-side is given as:
\[
g^{*}(f^{*}(A)) = g^{*}\left( \bigcup_{a\in A} f(a) \right) = \bigcup_{b \in \bigcup_{a\in A} f(a)} g(f(a)) = \bigcup_{a\in A} \bigcup_{b\in f(a)} g(f(a)).
\]
The right-hand-side is given as:
\[
(\co{f}{g^{*}})^{*}(A) = \bigcup_{a\in A} (\co{f}{g^{*}})(a) = \bigcup_{a\in A} g^{*}(f(a)) = \bigcup_{a\in A} \bigcup_{b\in f(a)} g(f(a)).
\]
Hence, both sides are equal.

\end{enumerate}
\end{solution}

\begin{solution}[\cref{exer:kleisli_triple_continuation}]
\label{sol:kleisli_triple_continuation}
For each set $X\in\Ob \SET$, we define:
\[
\eta_X : X \to (X\to R)\to R : x\mapsto (\lambda f, f(x)).
\]
For each function $f\in\CHom{\Ob \SET}{X}{Cont^R(Y)}$, we define:
\begin{align*}
f^{*} : Cont^R(X) \to Cont^R(Y) : i \mapsto \lambda (j:Y\to R), i(f(-)(j)).
\end{align*}
Notice that this is indeed well-defined: Let $i\in Cont^R(X)$, i.e. $i:(X\to R)\to R$. Then $f^{*}(i) : (Y\to R)\to R$. Let $j:Y\to R$. Then $f(-)(j) : X\to R$, hence we can apply it to $i$ and we have $i\left(f(-)(j)\right)\in R$.

Let $i\in Cont^R(X)$. We now show that this data satisfies the properties of a Kleisli triple:
\begin{enumerate}
\item For each set $X$, we have to show $\eta_X^{*} = \Id[Cont^R(X)]$. Let $x\in X$. The claim then follows by the following computation:
\[
\eta_X^{*}(i) = \lambda j, i\left(\eta_X(-)(j)\right) = \lambda j, i\left(\lambda x, j(x)\right) = i.
\]

\item For each function $f:X\to Cont^R(Y)$, we have to show $f^{*}(\eta_X(x)) = f(x)$, this follows by the following computation:
\[
f^{*}(\eta_X(x)) = f^{*}(\lambda g,g(x)) = \lambda j, \left(\left(\lambda g,g(x)\right)(f(-)(j))\right) = \lambda j, (f(x)(j)) = f(x).
\]

\item Let $f:X\to Cont^R(Y)$ and $g:Y\to Cont^R(Z)$ be functions, we have to show 
\[
g^{*}(f^{*}(i)) = (\co{f}{g^{*}})^{*}(i).
\] 
The left-hand-side is given as:
\begin{align*}
	g^{*}(f^{*}(i)) & = g^{*}\left(\lambda j, i\left(f(-)(j)\right)\right) = \lambda \tilde{j}, \left(\lambda j, i\left(f(-)(j)\right)\right)\left(g(-)(\tilde{j})\right) \\
	& = \lambda \tilde{j}, i\left(f(-)\left(g(-)(\tilde{j})\right)\right).
\end{align*}

The right-hand-side is given as:
\[
(\co{f}{g^{*}})^{*}(i) = \lambda j, i\left((\co{f}{g^{*}})(-)(j)\right).
\]
So to show that both sides are equal, it suffices to show that for each $j$, we have 
\[
f(-)\left(g(-)(j)\right) = (\co{f}{g^{*}})(-)(j).
\]
Notice that these are functions $X\to R$. Hence we will show this pointwise for each $x\in X$. The left-hand-side is given by: 
\[
f(-)\left(g(-)(j)\right)(x) = f(x)\left(g(-)(j)\right).
\]
The right-hand-side is given by:
\begin{eqnarray*}
(\co{f}{g^{*}})(-)(j)(x) &=& (\co{f}{g^{*}})(x)(j)\\ 
	&=& \left( (\co{f}{g^{*}})(x) \right)(j)\\ 
	&=& \left(g^{*}(f(x))\right)(j)\\ 
	&=& \left( \lambda k, f(x)\left(g(-)(k)\right) \right)(j)\\ 
	&=& f(x)\left(g(-)(j)\right).
\end{eqnarray*}
Hence, both sides are equal.

\end{enumerate}
\end{solution}

\begin{solution}[\cref{exer:kleisli_triple_familiesofelements}]\label{sol:kleisli_triple_familiesofelements}
For each set $X\in\Ob \SET$, we define:
\[
\eta_X : X \to (R \to X) : x\mapsto (\lambda \_, x).
\]
For each function $f\in\CHom{\Ob \SET}{X}{R\to Y}$, we define:
\begin{align*}
f^{*} : (R\to X) \to (R\to Y) : g \mapsto \lambda r, f(g(r))(r).
\end{align*}
Notice that this is indeed well-typed: Let $g : R\to X$ and $r\in R$. Then $g(r)\in X$, therefore, $f(g(r)) : R\to Y$ and consequently $f(g(r))(r)\in Y$.

We now show that this data satisfies the properties of a Kleisli triple:
\begin{enumerate}
\item For each set $X$, we have to show $\eta_X^{*} = \Id[R\to X]$. Let $r\in R$ and $g\in R\to X$. The claim then follows by the following computation (using functional extensionality):
\[
\eta_X^{*}(g)(r) = \eta_X(g(r))(r) = (\lambda \_, g(r))(r) = g(r).
\]
\item Let $f: X\to (R\to Y)$. We have to show that $\co{\eta_X}{f^{*}} = f$, which follows from the following computation using functional extensionality. For each $x\in X$ and $r\in R$:
\begin{align*}
	f^{*}(\eta_X(x))(r) & = f^{*}(\lambda \_,x)(r) = \left(\lambda r', f((\lambda \_, x)(r'))(r')\right)(r) = f((\lambda \_,x)(r))(r) \\
	& = f(x)(r).
\end{align*}

\item Let $f: X\to (R\to Y)$ and $g: Y\to (R\to Z)$. We have to show, by function extensionality, that for any $\phi\in R\to X$, we have 
\[
(\co{f}{g^{*}})^{*}(\phi) = g^{*}(f^{*}(\phi)).
\]
This indeed follows, yet again by functional extensionality. Let $r\in R$. We calculate both the left and right-hand side:
\begin{align*}
LHS =& (\co{f}{g^{*}})^{*}(\phi)(r)\\ 
	=& \co{g^{*}}{f}(\phi(r))(r)\\ 
	=& g^{*}\left(f(\phi(r))\right)(r)\\ 
	=& g\left(f(\phi(r))(r)\right)(r).
\end{align*}
\begin{align*}
RHS =& g^{*}(f^{*}(\phi))(r)\\ 
	=& g^{*}\left(\lambda s, f(\phi(s))(s)\right)(r)\\
	=& g( f(\phi(r))(r) )(r).
\end{align*}
Thus, the left and right hand sides compute to the same term, hence they are equal.
\end{enumerate}
\end{solution}

\begin{solution}[\cref{exer:initial_pointset}]\label{sol:initial_pointset}
	An initial object is a one-element set $ (\{ \star \}, \star) $. Let $ (X, x) $ be a pointed set. Then we have a unique function $ \star \to X $ that sends $ \star $, the chosen (and only) point of $ \{ \star \} $, to $ x $, the chosen point of $ X $, namely $ f: \star \mapsto x $.
\end{solution}

\begin{solution}[\cref{exer:initial_cats_of_nats}]\label{sol:initial_cats_of_nats}
\begin{enumerate}
	\item $0$ is the initial object in $\POS(\NN,\leq)$. As explained in \cref{sol:initial_posetcat}, the initial object in $\POS(X,\leq)$ is the minimal object in $X$, which in this case is $0$.
	\item $0$ is the initial object in $\SKELFINSET$. Let $n \in \NN = \SKELFINSET_0$. There is a unique function from $[0]$ to $[n]$, as $[0]$ is $\emptyset$ and there is a unique function of the form $\emptyset \to \{0, 1, \dots, n-1\}$.
	\item $0$ is the initial object in $\MAT$. Let $n \in \NN = \MAT_0$. There is a unique morphism from $0$ to $n$ as there is a unique $n \times 0$ matrix. 
     Note that an $n \times 0$ matrix is a linear map from the zero vector space to an n-dimensional vector space, which is unique and maps 0 to 0. 
\end{enumerate}
\end{solution}

\begin{solution}[\cref{exer:initial_rel}] \label{sol:initial_rel}
	The initial object in $\REL$ is the empty set. For each set $X$, there is a unique relation from $\emptyset$ to $X$, the empty relation $\emptyset \subseteq \emptyset \times X = \emptyset$.
\end{solution}

\begin{solution}[\cref{exer:terminal_set}]\label{sol:terminal_set}
	A terminal set in the category of sets is a one-element set $ \{ \star \} $. Given any set $ X $, we have a unique function $ X \to \star $, since any element of $ X $ must be sent to $ \star $.
\end{solution}

\begin{solution}[\cref{exer:terminal_posetcat}]\label{sol:terminal_posetcat}
	Given a poset $ (X, \leq) $, a terminal object in the category $ \POS(X, \leq) $ is exactly a maximal element in $ X $. Indeed, given such a maximal element $ x \in X = \POS(X, \leq)_0 $, we have for all $ y \in \POS(X, \leq)_0 $, since $ x $ is maximal, that $ y \leq x $. Therefore, we have a morphism $ f: x \to y $. By the definition of the hom-sets in $ \POS(X, \leq) $, $ f $ is unique and we conclude that $ x $ is a terminal object.
	Conversely, unfolding the definition of terminal object shows that a terminal object yields a maximal element.
\end{solution}

\begin{solution}[\cref{exer:terminal-unique}]\label{sol:terminal-unique}
	Suppose that we have a category $ \CC $ and two terminal objects $ B, B^\prime \in \Ob\CC $. Since $ B^\prime $ is terminal, we have a morphism $ f: B \to B^\prime $ and since $ B $ is terminal, we have a morphism $ g: B^\prime \to B $. Note that we have two morphisms from $ B $ to $ B $, namely $ g \circ f $ and $ \Id[B] $. Also, because $ B $ is terminal, there exists a unique morphism $ B \to B $. Therefore, $ g \circ f = \Id[B] $. In the same way, we have $ f \circ g = \Id[B^\prime] $. Therefore, $ f $ is the isomorphism (with inverse $g$) between $ B $ and $ B^\prime $ that we are looking for.
\end{solution}

\begin{solution}[\cref{exer:terminality_preserved_by_iso}]\label{sol:terminality_preserved_by_iso}
	Let $ \CC $ be a category and take $ B, B^\prime \in \Ob\CC $ objects in $ \CC $. Suppose that we have an isomorphism $ i: B \cong B^\prime $ and that $ B $ is a terminal object in $ \CC $. We have to show that $ B^\prime $ is terminal. In other words, for all $ A \in \Ob\CC $, we have to show that there exists a unique morphism $ f: A \to B^\prime $.

	Now, given such an object $ A \in \Ob\CC $, we have a morphism $ f: A \to B $ by terminality of $ B $. Therefore, we have a morphism $ g = i \circ f: A \to B^\prime $. This proves existence. For uniqueness, suppose that we also have another morphism $ h: A \to B^\prime $. Then we have morphisms $ \Inv{i} \circ g $ and $ \Inv{i} \circ h $ from $ A $ to $ B $. Since $ B $ is terminal, there exists only one morphism from $ A $ to $ B $, so $ \Inv{i} \circ g = \Inv{i} \circ h $. Therefore, we have 
	\[ g = i \circ (\Inv{i} \circ g) = i \circ (\Inv i \circ h) = h \]
	and this concludes the proof.
\end{solution}

\begin{solution}[\cref{exer:terminal_iff_initial_op}]\label{sol:terminal_iff_initial_op}
  Let $ \CC $ be a category. Suppose that $ \CC $ has a terminal object $ B \in \Ob\CC $. Note that $ \Ob\CC = \Ob{\op\CC} $. We will show that $ B $ is an initial object in $ \op\CC $. That is, for all $ A \in \Ob{\op\CC} $, we will show that $ \op\CC $ has a unique morphism from $ B $ to $ A $.
  Let $ A \in \Ob{\op\CC} $ be an arbitrary object. Since $ B $ is terminal in $ \CC $, we have that $ \CC(A, B) $ contains exactly one element. Then $ \op\CC(B, A) $ contains exactly one element as well, because $ \CC(A, B) = \op\CC(B, A) $. Therefore, $ B $ is an initial object in $ \op\CC $.
  
  Conversely, suppose that $ \op\CC $ has an initial object $ B \in \Ob{\op\CC} = \Ob\CC $. Given any object $ A \in \Ob\CC = \Ob{\op\CC}  $, since $ B $ is initial in $ \op\CC $, $ \op\CC(B, A) $ contains exactly one element $ f $. Then $ \CC(A, B) $ contains exactly one element (this is $ f $ again) as well. Therefore, $ B $ is a terminal object in $ \CC $.
\end{solution}

\begin{solution}[\cref{exer:cat-without-terminal}]\label{sol:cat-without-terminal}
We give three solutions to this exercise.
\begin{itemize}
\item Consider the category generated by the graph: 
\[
\begin{tikzcd}
x & y
\end{tikzcd}
\]
This category can not have a terminal object since there is no morphism from $x$ to $y$ or vice versa.
\item Consider the category generated by the graph: 
\[
\begin{tikzcd}
x \arrow[r, bend left, "f"] \arrow[r, bend right, "g"] & y
\end{tikzcd}
\]
This category also can not have a terminal object, indeed: There is no morphism from $y$ to $x$, hence $x$ can not be terminal. But also $y$ can not be terminal since $f$ and $g$ are different morphisms.
\item Consider the category $ \POS(\NN, \leq) $, i.e., the category
\[
\begin{tikzcd}
	0 \arrow[r] & 1 \arrow[r] & 2 \arrow[r] & \dots
\end{tikzcd}
\]
This category can not have a terminal object: Suppose it has a terminal object $ n \in \NN $. Then we have another object $ n + 1 \in \NN $. Since $ n $ is terminal, we have a morphism $ n + 1 \to n $, which means that $ n + 1 \leq n $, which is absurd. Therefore, this category does not have a terminal object.
\end{itemize}
\end{solution}

\begin{solution}[\cref{exer:terminal_cats_of_nats}]\label{sol:terminal_cats_of_nats}
	\begin{enumerate}
		\item As explained in \cref{sol:cat-without-terminal}, $\POS(\NN,\leq)$ does not have a terminal object.
		\item $1$ is the terminal object in $\SKELFINSET$. Let $n \in \NN = \SKELFINSET_0$. There is a unique function from $[n]$ to $[1]$, as $[1]$ is $\{0\}$, a set with one element. There is a unique function of the form $\{0, 1, \dots, n-1\} \to \{ 0 \}$, since all the elements in $[n]$ must be mapped to $0$.
		\item $0$ is the terminal object in $\MAT$. Let $n \in \NN = \MAT_0$. There is a unique morphism from $n$ to $0$ as there is a unique $0 \times n$ matrix. 
		Note that an $0 \times n$ matrix is a linear map from an n-dimensional vector space to the zero vector space, which is unique.
	\end{enumerate}
\end{solution}

\begin{solution}[\cref{exer:terminal_rel}]\label{sol:terminal_rel}
	The terminal object in $\REL$ is the empty set. For each set $X$, there is a unique relation from $X$ to $\emptyset$, the empty relation $\emptyset \subseteq X \times \emptyset = \emptyset$.
\end{solution}

\begin{solution}[\cref{exer:product-represent}]\label{sol:product-represent}
\newcommand{\CP}{\CC(A+B,X)}
\newcommand{\CQ}{\CC(A,X)\times\CC(B,X)}
\newcommand{\PP}{\CC(X,A\times B)}
\newcommand{\PQ}{\CC(X,A)\times\CC(X,B)}
	
This means we are after the morphisms \[\alpha:\SET(\PP, \PQ)\] and its inverse \[\beta:\SET(\PQ, \PP)\]

Note that the source and target of the hom-sets are reversed compared to the problem statement for coproducts. This is not a coincidence, as coproducts in one category have a direct correspondence with products in the opposite category -- which is the category with all morphisms reversed. This fact alone might be convincing enough to conclude that the morphisms we are after are the following:

\[\alpha(k):=(\co {k} {\projl}, \co {k} {\projr})\] and \[\beta(a,b) := \intoproduct{a}{b}\] where $\intoproduct{a}{b}$, also known as the product map, is the unique morphism that makes the corresponding product triangles commute, as in the following diagram:

\[
\begin{tikzcd}
A & A\times B \arrow[r, "\pi_r"] \arrow[l, "\pi_l"']                    & B \\
	& X \arrow[u, "{<a,b>}" description] \arrow[ru, "b"'] \arrow[lu, "a"] &  
\end{tikzcd}
\]

Or equivalently, the unique morphism s.t. 

\begin{equation}
\label{ga2}
	\co {\intoproduct{a}{b}} {\projl} = a
\end{equation}
\begin{equation}
\label{gb2}
	\co {\intoproduct{a}{b}} {\projr} = b
\end{equation}

Note that the order of composition is reversed w.r.t. the order of composition in the coproducts case.

To fully convince ourselves, without relying on the opposite category, we can construct a proof by applying the following substitutions to the proof for coproducts:

\begin{itemize}
	\item \textit{coproduct} becomes \textit{product}
	\item $A+B$ becomes $A\times B$
	\item $\CQ$ becomes $\PQ$
	\item $\CP$ becomes $\PP$
	\item $\co {\inl} {\_}$ becomes $\co {\_} {\projl}$
	\item $\co {\inr} {\_}$ becomes $\co {\_} {\projr}$
	\item $\outofcoproduct{a}{b}$ becomes $\intoproduct{a}{b}$
\end{itemize}
\end{solution}

\begin{solution}[\cref{exer:product_set}]\label{sol:product_set}
Given $ A, B \in \Ob \SET $, we claim that the cartesian product $ A \times B $ with the left projection $ \projl: (a, b) \mapsto a $ and right projection $ \projr: (a, b) \mapsto b $ is a product of $ A $ and $ B $.

Indeed, given an object $ Q \in \Ob \SET $ with morphisms $ l: Q \to A $ and $ r: Q \to B $, we have a morphism $ f := \langle l, r \rangle : Q \to A \times B $, given by $ q \mapsto (l(q), r(q)) $. For all $ q \in Q $, we have $ \projl \circ f(q) = \projl(l(q), r(q)) = l(q) $ and $ \projr \circ f(q) = \projr(l(q), r(q)) = r(q) $, so the diagram commutes, which proves existence.

Now, for uniqueness, suppose that we also have another morphism $ g: Q \to A \times B $ that makes the diagram commute. Note that we can write all elements $ p \in A \times B $ as $ p = (\projl(p), \projr(p)) $. Then we have, for all $ q \in Q $,
\[ g(q) = (\projl(g(q)), \projr(g(q))) = (\projl \circ g(q), \projr \circ g(q)) = (l(q), r(q)) = f(q), \]
which completes the proof.
\end{solution}

\begin{solution}[\cref{exer:product_posetcat}]\label{sol:product_posetcat}
	For a poset $ (X, \leq) $, note that morphisms between objects always are unique if they exist, so uniqueness conditions on morphisms (and whether diagrams commute or not) are not relevant here.
	Furthermore, any diagram commutes automatically.
	
	For objects $ A, B \in \Ob{\POS(X, \leq)} $, a product of $ A $ and $ B $ is an object $ C \in \Ob{\POS(X, \leq)} $ with morphisms $ C \to A $ and $ C \to B $ such that for any object $ D \in \Ob{\POS(X, \leq)} $ with morphisms $ D \to A $ and $ D \to B $, we have a morphism $ D \to C $.

	This means that we need an object $ C $ such that $ C \leq A $ and $ C \leq B $, and for all objects $ D $ such that $ D \leq A $ and $ D \leq B $, we also have $ D \leq C $. Therefore, a product of $ A $ and $ B $ is the (unique up to isomorphism) greatest element that is less than or equal to $ A $ and $ B $, if it exists.

	\begin{rem}
		The uniqueness up to isomorphism is necessary. For example, consider the poset-category generated by the following diagram:
		
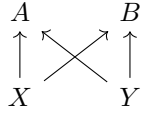
\begin{center}
			\begin{tikzcd}
				A & B\\
				X \arrow[u] \arrow[ru] & Y \arrow[u] \arrow[lu]
			\end{tikzcd}
		\end{center}
		For the product of $ A $ and $ B $, the candidates are $ X $ and $ Y $, since we have both $ X \leq A $ and $ X \leq B $, and $ Y \leq A $ and $ Y \leq B $. However, $ X $ cannot be the product, since if we are given $ D = Y $, we need $ Y \leq X $, which is not true. In the same way, $ Y $ cannot be the product, because $ X \leq Y $ does not hold.
	\end{rem}
\end{solution}

\begin{solution}[\cref{exer:product_cats_of_nats}]\label{sol:product_cats_of_nats}
	\begin{enumerate}
		\item As explained in \cref{sol:product_posetcat}, the product of $A$ and $B$ in $\POS(X,\leq)$ is the greatest element that is less than or equal to $A$ and $B$, for $A,B \in X$.
		Hence, for $m,n \in \NN$, the product $m \times n$ in $\POS(\NN,\leq)$ is $\min(m,n)$.
		\item Let $m,n \in \NN$. The product $m \times n$ in $\SKELFINSET$ is the product of the natural numbers $mn$. We define the projection maps separately for the cases where neither $m$ nor $n$ is zero and where at least one of them is zero. 
		
		\textbf{Case 1: $m\neq 0$ and $n \neq 0$} 
		
		Without loss of generality, we can assume $m \geq n$. The projection maps are defined as follows: 
		\begin{align*}
			\pi_l : [mn] &\to [m] \\
			k &\mapsto \lfloor k / n \rfloor, \\
			\pi_r : [mn] &\to [n] \\
			k &\mapsto k \mod n.
		\end{align*}

		We now need to show that for each $l \in \NN$, $q_l : [l] \to [m]$ and $q_r : [l] \to [n]$, there exists a unique $f : [l] \to [mn]$ such that $\pi_l \circ f = q_l$ and $\pi_r \circ f = q_r$:
		\[
			\begin{tikzcd}
			  &
			  l \ar[ld, "q_1"'] \ar[rd, "q_2"] \ar[d, dashed, "f"]
			  &
			  \\
			  m
			  &
			  mn \ar[l, "\projl"] \ar[r, "\projr"']
			  &
			  n.
			\end{tikzcd}
	    \]
		We define $f$ as follows: 
		\begin{align*}
			f : [l] &\to  [mn] \\
			k & \mapsto n q_l (k) + q_r(k).
		\end{align*}	
		$n q_l (k) + q_r(k)$ is in $[mn]$ since $q_l(k) < m$ and $q_r(k) < n$. This $f$ satisfies $\pi_l \circ f = q_l$ as for each $k \in [l]$ we have:
		\[ \pi_l (f(k)) = \lfloor \frac{n q_l(k) + q_r(k)}{n} \rfloor = q_l(k), \]
		and satisfies $\pi_r \circ f = q_r$ as for each $k \in [l]$ we have: 
		\[ \pi_r(f(k)) = n q_l(k) + q_r(k) \mod n = q_r(k). \] 
		Note that in both cases, we are using functional extensionality.

		To show uniqueness, we need to show that for each $g : [l] \to [mn]$ such that $\pi_l \circ g = q_l$ and $\pi_r \circ g = q_r$, we have $f = g$. We show that for each $k \in [l]$, $f(k) = g(k)$, which using functional extensionality implies $f = g$. 
		Let $k \in [l]$. The goal is to show that $g(k) = nq_l(k) + q_r(k)$. From $\pi_l \circ g = q_l$, we get $\lfloor \frac{g(k)}{n} \rfloor = q_l(k)$. This means $q_l(k) \leq \frac{g(k)}{n} < q_l(k) + 1$, hence, $n q_l(k) \leq g(k) < n q_l(k) + n$. Thus, we have: 
		\begin{equation} \label{eq:prod_cats_of_nats_1}
			g(k) = nq_l(k) + r,
		\end{equation}
		for some $r < n$. From $\pi_r \circ g = q_r(k)$ we get $g(k) \mod n = q_r(k)$. This means that:
		\begin{equation} \label{eq:prod_cats_of_nats_2}
			g(k) = nq + q_r(k),
		\end{equation}
		for some $q \in \NN$. From the division theorem we know that for each $a, b \in \NN$ there exist unique $q \in \NN$ and $r \in \NN$ such that $r < n$ that satisfy $a = bq + r$. Using \cref{eq:prod_cats_of_nats_1,eq:prod_cats_of_nats_2} and $q_r(k) \in [n]$, we get $q = q_l(k)$, $r = q_r(k)$. This implies $g(k) = n q_l(k) + q_r(k)$.
		Hence, $g(k) = f(k)$ and by functional extensionality $g = f$.
		
		\textbf{Case 2: $m = 0$ or $n = 0$} 
		
		In this case the product is 0, and all the morphisms involved are the empty function out of the empty set. This gives us a product, as $l$ (from the figure above) can only be 0 and it holds trivially that there exists a unique function of the form $[0] \to [0]$ that makes the corresponding diagrams commute.
		\item Let $m,n \in \NN$. The product $m \times n$ in $\MAT$ is the sum of the natural number $m + n$. The left and right projections correspond to the $m \times (m+n)$ matrix $P_l$ and the $n \times (m + n)$ matrix $P_r$ respectively, which are defined as follows:
		\[
		P_l = 
		\begin{bmatrix}
			I_m \mid 0_{m \times n} \\
		\end{bmatrix},
		P_r = 
		\begin{bmatrix}
			0_{n \times m} \mid I_n \\
		\end{bmatrix}.
		\]
		The matrix $P_l$ (resp. $P_r$) can be thought of as the projection from $\mathbb{R}^{m+n}$ to the first $m$ (resp. last $n$) dimensions.

		We need to show that for each $l \in \NN$, $m \times l$ matrix $Q_l$ and $n \times l$ matrix $Q_r$, there exists a unique $(m + n) \times l$ matrix $M$ such that $P_l M = Q_l$ and $P_r M = Q_r$. 
		When multiplied from the left, the matrix $P_l$ picks out the first $m$ rows, and $P_r$ picks out the last $n$ rows. This means that the matrix $M$ defined as follows satisfies $P_l M = Q_l$ and $P_r M = Q_r$:
		\[
		M = 
		\begin{pmatrix}
			Q_l \\
			Q_r
		\end{pmatrix}.
		\]
		To show uniqueness, we need to show that for each $(m+n) \times l$ matrix $N$ such that $P_l N = Q_l$ and $P_r N = Q_r$, we have $M=N$.
		From $P_l N = Q_l$ we know that the first $m$ rows of $N$ are $Q_l$ which is equal to the first $m$ rows of $M$. From $P_r N = Q_r$ we know that rows $m+1$ to $m+n$ of $N$ are equal to $Q_r$, which are equal to rows $m+1$ to $m+n$ of $M$. Hence, all rows of $M$ and $N$ are equal and $M =N$.
	\end{enumerate}
\end{solution}

\begin{solution}[\cref{exer:product_rel}]\label{sol:product_rel}
	Let $X$ and $Y$ be sets. The product $X \times Y$ in $\REL$ is the disjoint union $X + Y$ of sets defined as:
	\[ X + Y = \{ (x,0) \mid x \in X \} \cup  \{ (y,1) \mid y \in Y \}. \]
	The left and right projections are defined as follows:
	\[ \pi_l : \{ ((x,0),x) \mid x \in X \} \subseteq (X + Y) \times X, \]
	\[ \pi_r : \{ ((y,1),y) \mid y \in Y \} \subseteq (X + Y) \times X. \]
	Now we need to show that for each set $Z$ and relations $q_l \subseteq Z \times X$ and $q_r \subseteq Z \times Y$, there exists a unique relation $f \subseteq Z \times (X + Y)$ such that $\pi_l \circ f = q_l$ and $\pi_r \circ f = q_r$.
	We define $f$ as follows:
	\begin{align*}
		f = & \{ (z,(x,0)) \mid z \in Z \wedge x \in X \wedge (z,x) \in q_l \} \cup \\
			& \{ (z,(y,1)) \mid z \in Z \wedge y \in Y \wedge (z,y) \in q_r \} \subseteq Z \times (X + Y) .
	\end{align*}
	Using the definition of composition in $\REL$, we have that $f$ satisfies $\pi_l \circ f = q_l$ and $\pi_r \circ f = q_r$. 

	To show uniqueness, we need to show that for each relation $g \subseteq Z \times (X + Y)$ such that $\pi_l \circ g = q_l$ and $\pi_r \circ g = q_r$. 
	Using the definition of composition in $\REL$ we have: 
	\begin{align*}
		\pi_1 \circ g & = \{ (z,x) \in Z \times X \mid
		\begin{aligned}[t]
			& (\exists x' \in X \text{ s.t. } (z, (x',0)) \in g \wedge ((x',0),x) \in \pi_l) \vee \\ 
			& (\exists y \in Y \text{ s.t } (z,(y,1)) \in g \wedge ((y,1),x) \in \pi_l) \} 
		\end{aligned} \\
		& = \{ (z,x) \in Z \times X \mid \exists x' \in X \text{ s.t. } (z, (x',0)) \in g \wedge ((x',0),x) \in \pi_l \} \\
		& = \{ (z,x) \in Z \times X \mid (z, (x,0)) \in g \}. 
	\end{align*}
	Since $\pi_1 \circ g = q_l$, $q_l = \{ (z,x) \in Z \times X \mid (z, (x,0)) \in g \}$. Similarly, from $\pi_2 \circ g = q_r$, we get $q_r = \{ (z,y) \in Z \times Y \mid (z, (y,1)) \in g \}$. By substituting these in the definition of $f$, we get $f = g$.
\end{solution}

\begin{solution}[\cref{exer:product-unique}]\label{sol:product-unique}
	In a category $ \CC $, given two objects $ A, B \in \Ob \CC $, suppose that we have two products $ C, C^\prime \in \Ob \CC $, with projections $ \projl: C \to A $, $ \projr: C \to B $, $ \pi^\prime_l: C^\prime \to A $ and $ \pi^\prime_r: C^\prime \to B $.

	Since $ C^\prime $ is a product, and we have the morphisms $ \projl $ and $ \projr $, the universal property gives a morphism $ f: C \to C^\prime $ such that $ \projl = \pi^\prime_l \circ f $ and $ \projr = \pi^\prime_r \circ f $. In the same way, we have a morphism $ g: C^\prime \to C $ such that $ \pi^\prime_l = \projl \circ g $ and $ \pi^\prime_r = \projr \circ g $.
	
	Since $ C $ is a product, the universal property gives that there is exactly one morphism $ h: C \to C $ such that $ \projl \circ h = \projl $ and $ \projr \circ h = \projr $. The morphism $ h = \Id[C] $ satisfies this property. However, we also have the morphism $ g \circ f: C \to C $, such that $ \projl \circ g \circ f = \pi^\prime_l \circ f = \projl $ and $ \projr \circ g \circ f = \pi^\prime_r \circ f = \projr $. Therefore, $ g \circ f = h = \Id[C] $. In the same way, $ f \circ g = \Id[C^\prime] $, so $ f: C \cong C^\prime $ is an isomorphism, with inverse $ g $.
\end{solution}

\begin{solution}[\cref{exer:product_preserved_by_iso}]\label{sol:product_preserved_by_iso}
	Let $ \CC $ be a category and take objects $ A, B \in \Ob\CC $. Let $ (P, \projl: P \to A, \projr: P \to B) $ be a product of $ A $ and $ B $ in $ \CC $. Let $ P^\prime \in \Ob\CC $ be another object, and suppose that we have an isomorphism $ f: P \cong P^\prime $.

	We claim that $ (P^\prime, \projl^\prime, \projr^\prime) $, with $ \projl^\prime = \projl \circ \Inv{f} $ and $ \projr^\prime = \projr \circ \Inv{f} $ is also a product of $ A $ and $ B $. Now, given any object $ Q \in \Ob\CC $ with morphisms $ l: Q \to A $ and $ r: Q \to B $, we have to show that there exists a unique morphism $ g: Q \to P^\prime $ such that $ \projl^\prime \circ g = l $ and $ \projr^\prime \circ g = r $. See also the following diagram:

	
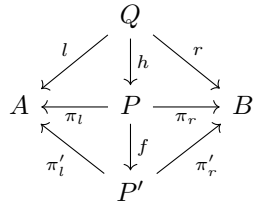
\begin{center}
		\begin{tikzcd}
		& Q \arrow[ld, "l"'] \arrow[rd, "r"] \arrow[d, "h"]               &   \\
		A & P \arrow[l, "\projl"] \arrow[r, "\projr"'] \arrow[d, "f"]         & B \\
		& P^\prime \arrow[lu, "\projl^\prime"] \arrow[ru, "\projr^\prime"'] &  
		\end{tikzcd}
	\end{center}
	
	By the universal property of the product, we have a morphism $ h: Q \to P $ such that $ \projl \circ h = l $ and $ \projr \circ h = r $. We take $ g = f \circ h $. We have 
	\[ \projl^\prime \circ g = (\projl \circ \Inv{f}) \circ (f \circ h) = \projl \circ h = l  \]
	and
	\[ \projr^\prime \circ g = (\projr \circ \Inv{f}) \circ (f \circ h) = \projr \circ h = r.  \]
	This proves existence.

	For uniqueness, suppose that we have two morphisms, $ g, g^\prime: Q \to P^\prime $ such that $ \projl^\prime \circ g = l = \projl^\prime \circ g^\prime $ and $ \projr^\prime \circ g = r = \projr^\prime \circ g^\prime $. By the universal property of the product, there exists exactly one morphism $ h: Q \to P $ such that $ \projl \circ h = l $ and $ \projr \circ h = r $. Since
	\[ \projl \circ (\Inv{f} \circ g) = (\projl \circ \Inv{f}) \circ g = \projl^\prime \circ g = l \]
	and
	\[ \projr \circ (\Inv{f} \circ g) = (\projr \circ \Inv{f}) \circ g = \projr^\prime \circ g = r, \]
	we have $ h = \Inv{f} \circ g $. In the same way, we have $ h = \Inv f \circ g^\prime $. Therefore,
	\[ g = f \circ \Inv f \circ g = f \circ h = f \circ \Inv f \circ g^\prime = g^\prime, \]
	which concludes the proof.
\end{solution}

\begin{solution}[\cref{exer:product_with_terminal}]\label{sol:product_with_terminal}
	Let $ \CC $ be a category, let $ A \in \Ob\CC $ be an object and let $ T \in \CC $ be the terminal object.

	We claim that $ (A, \projl, \projr) $ (with $ \projl = \Id[A] $ and $ \projr $ the unique morphism $ A \to T $) is a product of $ A $ and $ T $.

	We have to prove that for all $ B \in \Ob\CC $ with morphisms $ l: B \to A $ and $ r: B \to T $, there exists a unique morphism $ f: B \to A $ such that $ \projl \circ f = l $ and $ \projr \circ f = r $. To that end, let $ B \in \Ob\CC $ be such an object with morphisms $ l $ and $ r $.
	
	To prove existence, we take $ f = l $. Then we have $ \projl \circ f = \Id[A] \circ l = l $. Since $ T $ is a terminal object, the morphism $ B \to T $ is unique, so $ l = \projr \circ f $.

	To prove uniqueness of $ f $, suppose that there exists also another morphism, $ f^\prime: B \to A $ such that $ \projl \circ f^\prime = l $. Then we have
	\[ f^\prime = \Id{A} \circ f^\prime = \projl \circ f^\prime = l = f, \]
	which concludes the proof.
\end{solution}

\begin{solution}[\cref{exer:product_iff_terminal_in_subcategory}]\label{sol:product_iff_terminal_in_subcategory}
	Let $ \CC $ be a category and $ A, B \in \Ob \CC $ objects. Let us call the category mentioned in the exercise $ \CC_{A \times B} $.

	Suppose that we have a terminal object $ (C, \projl, \projr) \in \Ob{\CC_{A \times B}} $. We claim that this is a product of $ A $ and $ B $ in $ \CC $. Indeed, given any object $ D \in \Ob \CC $ with morphisms $ l: D \to A $, $ r: D \to B $, we have an object $ (D, l, r) \in \Ob{\CC_{A \times B}} $. Since $ (C, \projl, \projr) $ is a terminal object, there exists a unique morphism $ f: (D, l, r) \to (C, \projl, \projr) $. By the definition of morphisms in $ \CC_{A \times B} $, $ f $ is the unique morphism $ f: D \to C $ such that $ \projl \circ f = l $ and $ \projr \circ f = r $.

	Conversely, suppose that we have a product $ C \in \Ob \CC $ of $ A $ and $ B $ with morphisms $ \projl: C \to A $ and $ \projr: C \to B $. This gives an object $ (C, \projl, \projr) \in \Ob{\CC_{A \times B}} $. We claim that this is a terminal object. Indeed, given any object $ (D, l, r) \in \Ob{\CC_{A \times B}} $, since $ C $ is a product in $ \CC $, we have a unique morphism $ f: D \to C $ such that $ \projl \circ f = l $ and $ \projr \circ f = r $. Therefore, $ f $ is a unique morphism in $ \CC_{A \times B}((D, l, r), (C, \projl, \projr)) $. Since this holds for any $ (D, l, r) \in \Ob{\CC_{A \times B}} $, $ (C, \projl, \projr) $ is a terminal object in $ \CC_{A \times B} $, which concludes the proof.
\end{solution}

\begin{solution}[\cref{exer:product_of_morphisms}]\label{sol:product_of_morphisms}
	Let $\CC$ be a category with a choice of product $(A\times B, \projl, \projr)$ for any two objects $A,B\in \Ob{\CC}$. Take objects $ A, B, C, D \in \Ob \CC $ and morphisms $ f: A \to C $ and $ g: B \to D $.

	\begin{center}
		\begin{tikzcd}
		A \arrow[d, "f"'] & A \times B \arrow[l, "\projl"'] \arrow[r, "\projr"] \arrow[ld, "f \circ \projl"'] \arrow[rd, "g \circ \projr"] \arrow[d, "h"] & B \arrow[d, "g"] \\
		C                 & C \times D \arrow[l, "\projl^\prime"] \arrow[r, "\projr^\prime"']                                                             & D               
		\end{tikzcd}
	\end{center}

	We have the products $ (A \times B, \projl, \projr) $ and $ (C \times D, \projl^\prime, \projr^\prime) $. We have morphisms $ f \circ \projl: A \times B \to C $ and $ g \circ \projr: A \times B \to D $. By the universal property of the product $ C \times D $, there exists a (unique) morphism $ h: A \times B \to C \times D $ that makes the diagram commute. This is the morphism we are looking for.
\end{solution}

\begin{solution}[\cref{exer:swap_binary_product}]\label{sol:swap_binary_product}
  Let $\CC$ be a category with a choice of product $(A\times B, \projl, \projr)$ for any two objects $A,B\in \Ob{\CC}$. Let $A, B \in \Ob\CC$ be objects.

	\begin{center}
		\begin{tikzcd}
		& A \times B \arrow[ld, "\projr"'] \arrow[rd, "\projl"] \arrow[dd, "f", bend left]               &   \\
		B &                                                                                               & A \\
		& B \times A \arrow[ru, "\projr^\prime"'] \arrow[lu, "\projl^\prime"] \arrow[uu, "g", bend left] &  
		\end{tikzcd}
	\end{center}

	We have the products $ (A \times B, \projl, \projr) $ and $ (B \times A, \projl^\prime, \projr^\prime) $. By the universal properties of the products $ A \times B $ and $ B \times A $, we get morphisms $ f: A \times B \to B \times A $ and $ g: B \times A \to A \times B $, which make the diagram commute.
	
	By the universal property of the product $ A \times B $, there exists a unique morphism $ h: A \times B \to A \times B $ such that $ \projl \circ h = \projl $ and $ \projr \circ h = \projr $. Since $ \Id[A \times B] $ satisfies this, we have $ h = A \times B $. We also have the morphism $ g \circ f: A \times B \to A \times B $. We have
	\[ \projr \circ g \circ f = \projl^\prime \circ f = \projr \]
	and in the same way, we have $ \projl \circ g \circ f = \projl $. Therefore, $ g \circ f = h = \Id[A \times B] $. In the same way, we have $ f \circ g = \Id[B \times A] $. Therefore, we conclude that $ f $ is the isomorphism we are looking for, with inverse $ g $.
\end{solution}

\begin{solution}[\cref{exer:natlist_is_initial}]\label{sol:natlist_is_initial}
		We construct a category $\Cat{L}$ as follows
		\begin{itemize}
			\item objects are triples consisting of a set $X$, an element $x \in X$ and a function $l : \NN \times X \to X$
			\item morphisms are functions $f: X \to X'$, such that they make the following diagrams commute
      \[
        \begin{tikzcd}
          1 & X && {\NN \times X} & X \\
          & {X'} && {\NN \times X'} & {X'}
          \arrow["x", from=1-1, to=1-2]
          \arrow["{x'}"', from=1-1, to=2-2]
          \arrow["f"{description}, from=1-2, to=2-2]
          \arrow["l"{description}, from=1-4, to=1-5]
          \arrow["{l'}"{description}, from=2-4, to=2-5]
          \arrow["f"{description}, from=1-5, to=2-5]
          \arrow["{1_\NN \times f}"{description}, from=1-4, to=2-4]
        \end{tikzcd}
      \]
			\item the identity morphisms $1_{(X, x, l)}$ is the identity function $1_X : X \to X$
			\item morphism composition is just function composition
		\end{itemize}
		\begin{thm}
			The identity function $1_X : X \to X$ makes the following diagrams commute
			\[
				\begin{tikzcd}
					1 & X && {\NN \times X} & X \\
					& X && {\NN \times X} & X
					\arrow["x", from=1-1, to=1-2]
					\arrow["{x}"', from=1-1, to=2-2]
					\arrow["{1_X}"{description}, from=1-2, to=2-2]
					\arrow["l"{description}, from=1-4, to=1-5]
					\arrow["{l}"{description}, from=2-4, to=2-5]
					\arrow["{1_X}"{description}, from=1-5, to=2-5]
					\arrow["{1_\NN \times 1_X}"{description}, from=1-4, to=2-4]
				\end{tikzcd}
			\]
			That is, the following equations hold:
			\begin{align}
				\label{goal:id-x-x} 1_X \circ x & = x                                     \\
			  \label{goal:id-l-l-id} 1_X \circ l & = l \circ (1_\NN \times 1_X)
			\end{align}
		\end{thm}
		\begin{proof}
			\eqref{goal:id-x-x} holds since $1_X$ follows the \textit{left-identity} law.

			\begin{lemma}
				\label{lem:id-id}
				Given a category $\CC$, two objects $A, B \in \CC_0$, and their product $(A \times B, \pi_1, \pi_2)$,
        \[
					1_{A \times B} = 1_A \times 1_B.
				\]
			\end{lemma}
			\begin{proof}
				We know that $1_A \times 1_B$ is a \textit{unique} morphism that makes the following diagram commute
				\[
					\begin{tikzcd}
						A & {A \times B} & B \\
						A & {A \times B} & B
						\arrow["{\pi_1}"{description}, from=1-2, to=1-1]
						\arrow["{\pi_2}"{description}, from=1-2, to=1-3]
						\arrow["{1_A \times 1_B}"{description}, dashed, from=2-2, to=1-2]
						\arrow["{1_B}"{description}, from=2-3, to=1-3]
						\arrow["{1_A}"{description}, from=2-1, to=1-1]
						\arrow["{\pi_1}"{description}, from=2-2, to=2-1]
						\arrow["{\pi_2}"{description}, from=2-2, to=2-3]
					\end{tikzcd}
				\]
				To prove our goal, we just need to show that $1_{A \times B}$ also makes the diagram commute.
				For the left side we have
				\begin{align*}
					1_A \circ \pi_1 & = \pi_1                      & \text{by the \textit{left-identity} law}  \\
					                & = \pi_1 \circ 1_{A \times B} & \text{by the \textit{right-identity} law}
				\end{align*}
				We can argue similarly for the right side.
			\end{proof}

			Now, proving \eqref{goal:id-l-l-id} is simply
			\begin{align*}
				1_X \circ l & = l                        & \text{by the \textit{left-identity} law}  \\
				            & = l \circ 1_{A \times B}   & \text{by the \textit{right-identity} law} \\
				            & = l \circ (1_A \times 1_B) & \text{by \cref{lem:id-id}}
			\end{align*}
		\end{proof}
		\begin{thm}
			Given three objects $(X,x,l), (X',x',l'), (X'', x'', l'') \in \Cat{L}_0$ and two morphisms $f \in \Cat{L}((X', x', l'), (X'', x'', l''))$ and $g \in \Cat{L}((X, x, l), (X', x', l'))$
			The function $f \circ g$ satisfies
			\[
				\begin{tikzcd}
          1 & X && {\NN \times X} & X \\
            & X'' && {\NN \times X''} & X''
					\arrow["x", from=1-1, to=1-2]
					\arrow["x''"', from=1-1, to=2-2]
					\arrow["{f \circ g}"{description}, from=1-2, to=2-2]
					\arrow["l"{description}, from=1-4, to=1-5]
					\arrow["l''"{description}, from=2-4, to=2-5]
					\arrow["{f \circ g}"{description}, from=1-5, to=2-5]
					\arrow["{1_\NN \times (f \circ g)}"{description}, from=1-4, to=2-4]
				\end{tikzcd}
			\]
			that is
			\begin{align}
				(f \circ g) \circ x & = x'' \label{goal:fgx}                                         \\
				(f \circ g) \circ l & = l'' \circ (1_\NN \times (f \circ g)) \label{goal:fgl}
			\end{align}
		\end{thm}
		\begin{proof}
			\eqref{goal:fgx} is just
			\begin{align*}
				(f \circ g) \circ x & = f \circ (g \circ x) & \text{by \textit{function associativity}} \\
				                    & = f \circ x'          & \text{since $g$ is a morphism}            \\
				                    & = x''                 & \text{since $f$ is a morphism}
			\end{align*}
			For \eqref{goal:fgl} we need
			\begin{lemma}
				\label{lem:id-fg}
				Given category $\CC$,
				objects $A,X,Y,Z \in \CC_0$
				with products $(A \times X), (A \times Y), (A \times Z) \in \CC_0$,
				and morphisms $f \in \CC(Y, Z)$, $g \in \CC(X, Y)$,
				it holds that
				\[
					1_A \times (f \circ g) = (1_A \times f) \circ (1_A \times g).
				\]
			\end{lemma}
			\begin{proof}
				Here is an illustration of the situation
				\[
					\begin{tikzcd}
						A && {A \times X} && X \\
						\\
						A && {A \times Y} && Y \\
						\\
						A && {A \times Z} && Z
						\arrow["{1_A}"{description}, from=1-1, to=3-1]
						\arrow["{1_A}"{description}, from=3-1, to=5-1]
						\arrow["g"{description}, bend left, from=1-5, to=3-5]
						\arrow["f"{description}, bend left, from=3-5, to=5-5]
						\arrow["{f \circ g}"{description}, bend left, shift left=2, from=1-5, to=5-5]
						\arrow["\pi_{11}"{description},from=1-3, to=1-1]
						\arrow["\pi_{12}"{description},from=1-3, to=1-5]
						\arrow["\pi_{21}"{description},from=3-3, to=3-1]
						\arrow["\pi_{22}"{description},from=3-3, to=3-5]
						\arrow["\pi_{31}"{description},from=5-3, to=5-1]
						\arrow["\pi_{32}"{description},from=5-3, to=5-5]
						\arrow["{1_A \times g}"{description}, bend left, dashed, from=1-3, to=3-3]
						\arrow["{1_A \times f}"{description}, bend left, dashed, from=3-3, to=5-3]
						\arrow["{1_A \times (f \circ g)}"{description, pos=0.6}, bend left, shift left=5, dashed, from=1-3, to=5-3]
						\arrow["{(1_A \times f) \circ (1_B \times g)}"{description, pos=0.3}, shift right=4, bend right, from=1-3, to=5-3]
					\end{tikzcd}
				\]
				We know that $1_A \times f$, $1_A \times g$ and $1_A \times (f \circ g)$ are unique, and satisfy
				\begin{align}
					1_A \circ \pi_{21} = \pi_{31} \circ (1_A \times f)           & \wedge f \circ \pi_{22} = \pi_{32} \circ (1_A \times f) \label{Hf}          \\
					1_A \circ \pi_{11} = \pi_{21} \circ (1_A \times g)           & \wedge g \circ \pi_{12} = \pi_{22} \circ (1_A \times g) \label{Hg}          \\
					1_A \circ \pi_{11} = \pi_{31} \circ (1_A \times (f \circ g)) & \wedge (f \circ g) \circ \pi_{12} = \pi_{32} \circ (1_A \times (f \circ g))
				\end{align}

				Since $1_A \times (f \circ g)$ is unique, to prove our statement it is enough to show that
				\begin{align}
					1_A \circ \pi_{11}         & = \pi_{31} \circ ((1_A \times f) \circ (1_B \times g)) \label{goal:left}  \\
					(f \circ g) \circ \pi_{12} & = \pi_{32} \circ ((1_A \times f) \circ (1_B \times g)) \label{goal:right}
				\end{align}

				First, we show \eqref{goal:left}
				\begin{align*}
					1_A \circ \pi_{11} & = \pi_{21} \circ (1_A \times g)                          & \text{by \eqref{Hg}}                      \\
					                   & = ( 1_A \circ \pi_{21} ) \circ (1_A \times g)            & \text{by the \textit{left-identity} law}  \\
					                   & = ( \pi_{31} \circ (1_A \times f) ) \circ (1_A \times g) & \text{by \eqref{Hf}}                      \\
					                   & = \pi_{31} \circ ((1_A \times f) \circ (1_B \times g))   & \text{by \textit{function associativity}}
					.
				\end{align*}

				Finally, we show \eqref{goal:right}
				\begin{align*}
					(f \circ g) \circ \pi_{12} & = f \circ (g \circ \pi_{12})                           & \text{by \textit{function associativity}}  \\
					                           & = f \circ (\pi_{22} \circ (1_A \times g))              & \text{by \eqref{Hg}}                       \\
					                           & = (f \circ \pi_{22}) \circ (1_A \times g)              & \text{by \textit{function associativity}}  \\
					                           & = (\pi_{32} \circ (1_A \times f)) \circ (1_A \times g) & \text{by \eqref{Hf}}                       \\
					                           & = \pi_{32} \circ ((1_A \times f) \circ (1_B \times g)) & \text{by \textit{function associativity}}.
				\end{align*}
			\end{proof}

			Now we can show \eqref{goal:fgl} as follows
			\begin{align*}
				(f \circ g) \circ l & = f \circ (g \circ l)                                               & \text{by \textit{function associativity}} \\
				                    & = f \circ (l' \circ (1_\NN \times g))                        & \text{since $g$ is a morphism}            \\
				                    & = (f \circ l') \circ (1_\NN \times g)                        & \text{by \textit{function associativity}} \\
				                    & = (l'' \circ (1_\NN \times f)) \circ (1_\NN \times g) & \text{since $f$ is a morphism}            \\
				                    & = l'' \circ ((1_\NN \times f) \circ (1_\NN \times g)) & \text{by \textit{function associativity}} \\
				                    & = l'' \circ (1_\NN \times (f \circ g))                       & \text{by \cref{lem:id-fg}}.
			\end{align*}
		\end{proof}

		Since the category $\Cat{L}$ uses functions as morphisms,
		it satisfies the \textit{left-}, \textit{right-identity} and \textit{associativity} laws.

		We can now finally start the proof for ($\mathsf{NatList}, \nil, \cons$) being an initial object.

		\begin{proof}
			To prove ($\mathsf{NatList}, \nil, \cons$) is initial we have to show
			\[
				\forall (X, x, l) \in \Cat{L}_0, \exists! f \in \Cat{L}((\mathsf{NatList}, \nil, \cons), (X, x, l)).
			\]
			First, we show that $f$ exists.
			\[
				f := nl \mapsto \begin{cases}
          x             & \text{if $nl = \nil$ }             \\
					l (n, f (ns)) & \text{if $nl = \cons (n, ns)$}
				\end{cases}
			\]
			To fully show that this is indeed a morphism, we must show that the following diagram commutes.
			\[
				\begin{tikzcd}
					1 & \mathsf{Natlist} && {\NN \times \mathsf{Natlist}} & \mathsf{Natlist} \\
					& {X} && {\NN \times X} & {X}
          \arrow["{\nil}"{description}, from=1-1, to=1-2]
					\arrow["x"', from=1-1, to=2-2]
					\arrow["f"{description}, from=1-2, to=2-2]
          \arrow["{\cons}"{description}, from=1-4, to=1-5]
					\arrow["l"{description}, from=2-4, to=2-5]
					\arrow["f"{description}, from=1-5, to=2-5]
					\arrow["{1_\NN \times f}"{description}, from=1-4, to=2-4]
				\end{tikzcd}
			\]
			For the left diagram we have
			\begin{align*}
				f \circ \nil & = x & \text{by definition of $f$}.
			\end{align*}
			For the right diagram we have
			\begin{align*}
				(f \circ \cons) (n, ns) & = f (\cons (n, ns))                 & \text{by definition of $\circ$} \\
				                                & = l (n, f(ns))                              & \text{by definition of $f$}     \\
				                                & = l ((1_\NN \times f) (n, ns))       &                                 \\
				                                & = (l \circ (1_\NN \times f)) (n, ns) & \text{by definition of $\circ$}
			\end{align*}

			Second, we have to show that it is unique.
			If we have another function $g$ that is in the same hom set as $f$, we will show that it is the same function.
			We will prove this using induction.
			Base case is the mapping of $\nil$, since they both satisfy the condition for the morphisms, we know that
			\[
				f ( \nil ) = x = g ( \nil ).
			\]
			For the inductive case we must show that
			\[
				\forall\ \cons (n, ns) \in \texttt{natlist},
				f\ ns = g\ ns \to f (\cons (n, ns)) = g (\cons (n, ns)).
			\]
			\begin{align*}
				f (\cons (n, ns)) & = (f \circ \cons) (n, ns)           & \text{by definition of $\circ$} \\
				                          & = (l \circ (1_\NN \times f)) (n, ns) & \text{since $f$ is a morphism}  \\
				                          & = l (n, f\ ns)                              & \text{by definition of $\circ$} \\
				                          & = l (n, g\ ns)                              & \text{by induction hypothesis}  \\
				                          & = (l \circ 1_\NN \times g) (n, ns)   & \text{by definition of $\circ$} \\
				                          & = (g \circ \cons) (n, ns)           & \text{since $g$ is a morphism}  \\
				                          & = g (\cons (n, ns))                 & \text{by definition of $\circ$}
			\end{align*}

      With that we have proven that $f$ and $g$ are the same morphism, meaning $f$ is indeed unique.
		\end{proof}
\end{solution}

\begin{solution}[\cref{exer:functor_prod_on_left}]\label{sol:functor_prod_on_left}
	Given two sets, denoted $A$ and $X$, we can form a new set $A \times X$ (the set of all pairs).
	Hence, by fixing $A$ and varying $X$, we have a function from the collection of all sets, to itself:
	\[
		(A \times -) : \Ob{\SET} \to \Ob{\SET} : X \mapsto (A \times X).	
	\]
	In this exercise, we are asked to make this function into a functor.
	(Observe that the function/action on objects of the functor is given by the exercise!)
	The data remaining, is to define the action of the morphisms.
	That is, for every two sets (objects), denoted $X$ and $Y$, we have to construct a function
	\[
		\CHom{\SET}{X}{Y} \to \CHom{\SET}{A \times X}{A \times Y}.
	\]
	In order to define such a function, we need to specify to each element in the source (that is, a function from $X$ to $Y$), to an element in the target (that is a function from $A \times X$ to $A \times Y$).
	Thus, we fix a function $f : \CHom{\SET}{X}{Y}$ (also denoted $f : X \to Y$).
	Now, we have to construct/specify a function $(A \times f) : \CHom{\SET}{A \times X}{A \times Y}$.
	That is, given an element in $(A \times X)$, i.\,e.,\, a pair $(a, x)$ with $a \in A$ and $x \in X$, we have to give an element in $A \times Y$.
	An element in $A \times Y$ is a pair consisting of an element in $A$ and element in $Y$.
	By assumption, $a \in A$. 
	Furthermore, since $f$ is a function whose input type is $X$, we can apply $f$ to $x$.
	That is $f(x) \in Y$.
	Hence, $(a, f(x)) \in A \times Y$.
	Thus, we have defined a function:
	\[
		(A \times f) : A \times X \to A \times Y : (a, x) \mapsto (a, f(x)).
	\]
	(Observe that $(a, f(x)) = (\Id{a}, f(x))$.)
	As searched for, this defines a function
	\[
		\CHom{\SET}{X}{Y} \to \CHom{\SET}{A \times X}{A \times Y} : f \mapsto (A \times f).
	\]
	This concludes the (pieces of) data of a functor $(A \times -) : \SET \to \SET$.
	Now, it remains to prove (check) the preservation properties.
	\begin{enumerate}
		\item The first preservation property says that the composition of functions (morphisms in $\SET$), is preserved.
		That is, for every triple of sets $X, Y, Z \in \Ob{\SET}$, and functions $f : \SET(X, Y)$ and $g : \SET(Y, Z)$, we have to show the following:
		On the one hand, we can apply our functor (or the data at least) to the composite $f \circ g$.
		On the other, we can first apply our functor to $f$ (that is $A \times f = (A \times -)(f)$), and then (pre)compose with $A \times g = (A \times -)(g)$.
		Diagrammatically, we have:
		\[
			\begin{tikzcd}
				{A \times X} 
					\arrow[rr, bend left, "{A \times (f \circ g)}"]
					\arrow[rd, bend right, swap, "{A \times f}"] 
					& & {A \times Z} \\
				& {A \times Y} \arrow[ru, bend right, swap, "{A \times g}"] &
			\end{tikzcd}	
		\]
		The preservation axiom asks that this diagram commutes; that is, these functions (morphisms) should be equal (coincide).
		To proof that two functions are equal, it suffices to show that they are equal pointwise (or objectwise).
		(This is called \textit{function extensionality}.)
		That is, it suffices to check that these functions are equal after applying those functions to the same input arguments.
		(Observe that the input argument is a pair.)
		More formally, we have to show:
		\[
			\forall (a, x) \in {A \times X}, (A \times (f \circ g))(a, x) = \left((A \times g) \circ (A \times f)\right)(a, x).
		\]
		By unfolding the definition of $(A \times -)$ (on morphisms), we easily see that these are equal, indeed:
		the left hand side reduces to:
		\begin{eqnarray*}
			(A \times (f \circ g))(a, x) 
				&=& (a , (f \circ g)(x)), \quad \text{by definition of $A \times -$} \\
				&=& (a, g(f(x))), \quad \text{by definition of function composition};
		\end{eqnarray*}
		the right hand side reduces to:
		\begin{eqnarray*}
			\left((A \times g) \circ (A \times f)\right)(a, x)
				&=& (A \times g)\left((A \times f)(a, x)\right), \quad \text{by definition of function composition}\\
				&=& (A \times g)\left(a, f(x))\right), \quad \text{by definition of $A \times f$}\\
				&=& (a, g(f(x)), \quad \text{by definition of $A \times g$}.
		\end{eqnarray*}
		Hence, we see that both the left and right hand side reduce to the same expression.
		Since $x$ is arbitrary, this concludes the preservation of the composition.
		\item The second preservation property says that the identity is preserved.
		That is, for every set $X \in \Ob{\SET}$, we have to show the following:
		On the one hand, we can apply our functor to $\Id{X}$, which gives a function $A \times \Id{X} : \CHom{\SET}{A \times X}{A \times X}$
		On the other, we have the identity function on $A \times X$.
		The preservation law says that both functions are equal. 
		That is, the following diagram commutes:
		\[
			\begin{tikzcd}
				{A \times X} \arrow[r, bend left, "{A \times \Id{X}}"] \arrow[r, bend right, swap, "\Id{(A \times X)}"] & {A \times X}
			\end{tikzcd}	
		\]
		Again, we apply function extensionality (pointwise equality) to show that these functions are equal.
		Indeed, for every pair $(a, x) \in A \times X$, we have:
		\begin{eqnarray*}
			(A \times \Id{X})(a, x) 
				&=& (a, \Id{X}(x)), \quad \text{by definition $A \times -$}\\
				&=& (a, x), \quad \text{by definition of the identity function $\Id{X}$}\\
				&=& \Id{A \times X}(a, x), \quad \text{by definition of the identity function $\Id{(A \times X)}$}.
		\end{eqnarray*}
	\end{enumerate}
	This concludes the construction of the functor (and in particular, the proofs of preservation).
	Observe in particular, that the main challenge is the construction of $(A \times f)$ (hence, the action of the morphisms.)
	Indeed, the preservation laws follow by definition.
\end{solution}

\begin{solution}[\cref{exer:in_catamorphism_id}]\label{sol:in_catamorphism_id}
	Let $ F: \CC \to \CC $ be an endofunctor and let $ (\Initalg F, \In) $ be an initial algebra. By definition, $ \catam{\In} $ is the unique morphism $ \Initalg F \to \Initalg F $ that makes the following diagram commute
	\begin{center}
		\begin{tikzcd}
			F (\Initalg F) \arrow[r, "\In"] \arrow[d, "F \catam \In"] & \Initalg F \arrow[d, "\catam \In"]\\
			F (\Initalg F) \arrow[r, "\In"] & \Initalg F\\
		\end{tikzcd}
	\end{center}
	Since $ F(\Id[\Initalg F]) = \Id[F(\Initalg F)] $, the morphism $ \Id[\Initalg F] $ satisfies this property, and therefore $ \catam{\In} = \Id[\Initalg F] $.
\end{solution}

\begin{solution}[\cref{exer:fusion-property}]\label{sol:fusion-property}
	Let $ F : \CC \to \CC $ be an endofunctor and let $ (\Initalg F, \In) $ be an initial algebra. Given $F$-algebras $(C,\phi)$ and $(D,\psi)$ and $ f \in \CHom{\CC}{C}{D} $, we have the following commutative diagrams:
	\begin{center}
		\begin{tikzcd}
			F \Initalg F \arrow[r, "\In"] \arrow[d, "F( \catam{\phi} )"] & \Initalg F \arrow[d, "\catam{\phi}"]\\
			F C \arrow[r, "\phi"] & C
		\end{tikzcd}
		\quad
		\begin{tikzcd}
			F \Initalg F \arrow[r, "\In"] \arrow[d, "F( \catam{\psi} )"] & \Initalg F \arrow[d, "\catam{\psi}"]\\
			F D \arrow[r, "\psi"] & D
		\end{tikzcd}
	\end{center}
	Now, suppose that $ f \circ \phi = \psi \circ F(f) $.

	By definition, $ \catam \psi $ is the unique morphism from $ \Initalg F $ to $ D $ such that 
	\[ \catam{\psi} \circ \In = \psi \circ F(\catam{\psi}). \]
	Therefore, to show that $ \catam \psi = f \circ \catam \phi $, it suffices to show that
	\[ (f \circ \catam{\phi}) \circ \In = \psi \circ F(f \circ \catam{\phi}). \]
	And indeed, we have
	\begin{align*}
		f \circ \catam{\phi} \circ \In &= f \circ \phi \circ F(\catam{\phi})\\
		&= \psi \circ F(f) \circ F(\catam{\phi})\\
		&= \psi \circ F(f \circ \catam{\phi}).
	\end{align*}
	Therefore, $ \catam{\psi} = f \circ \catam{\phi} $.
\end{solution}

\begin{solution}[\cref{exer:initialalg_for_idfun_with_initialob}]\label{sol:initialalg_for_idfun_with_initialob}
	Let $\CC$ be a category with an initial object $\bot$. Let $ F: \CC \to \CC $ be the identity endofunctor.

	Note that $ (\bot, \Id[\bot]) $ is an $ F $-algebra since $ \Id[\bot]: \bot \to \bot $ and $ \bot = F(\bot) $. Now, to show that it is initial, suppose we have another $ F $-algebra $ (X \in \Ob \CC, f: F X \to X) $. By initiality of $ \bot $ in $ \CC $, we have a unique morphism $ g \in \CC(\bot, X) $. To show that $ g $ is a $ F $-algebra morphism, we have to show that the following diagram commutes:
	\begin{center}
		\begin{tikzcd}
			\bot = F(\bot) \arrow[r, "{\Id[\bot]}"] \arrow[d, "g = F(g)"] & \bot \arrow[d, "g"]\\
			X = F(X) \arrow[r, "f"] & X
		\end{tikzcd}
	\end{center}
	Because $ \bot $ is initial, there exists exactly one morphism from $ \bot $ to $ X $. Therefore, $ g = f \circ g $ and $ g $ is an $ F $-algebra morphism. It is also unique, because $ \bot $ is initial in $ \CC $. Therefore, $ (\bot, \Id[\bot]) $ is an initial $ F $-algebra.
\end{solution}


\begin{solution}[\cref{exer:list-concat-nil}]\label{sol:list-concat-nil}
  We assume that list concatenation |(++) :: [a] -> [a] -> [a]| was defined by recursion on the \emph{first} argument.
  This means that, for any |l :: [a]|, we define |(++ l) :: [a] -> [a]| as |(++ l) = fold cons l|.

  The equation |nil ++ l = l| then follows by definition: |nil ++ l = (fold cons l) nil = l|.

  Let's show |l ++ nil = l|.
  We have that |l ++ nil = (fold cons nil) l| by the definition of |(++ nil)|.
  We know that $[\cons,\nil] = \In$, the initial algebra for the list functor.
  Thus, by \cref{exer:in_catamorphism_id}, |(fold cons nil) = id| is the identity function on lists.
  We thus have |l ++ nil = (fold cons nil) l = id l = l|.

  When |(++) : [a] -> [a] -> [a]| is defined by recursion on the \emph{second} argument, both equations hold as well, with their proofs swapped compared to those given above.
\end{solution}

\begin{solution}[\cref{exer:initialalg_for_bifunctor_functor}]\label{sol:initialalg_for_bifunctor_functor}
	Let $F:\CC\to\CC\to\CC$ be a bifunctor such that for any object $A\in\CC$, the initial algebra $ (\Initalg{F_A}, \In_A) $ for the functor induced by 
	\[ F_A : \CC\to\CC : X \mapsto F(A,X), \]
	exists.
	
	We define a functor $ G: \CC \to \CC $ by $ G(A) = \Initalg{F_A} $. For $ A, B \in \Ob\CC $, we have an initial $ F_A $-algebra $ (\Initalg{F_A}, \In_A: F_A(\Initalg{F_A}) \to \Initalg{F_A}) $ and an initial $ F_B $-algebra $ (\Initalg{F_B}, \In_B: F_B(\Initalg{F_B}) \to \Initalg{F_B}) $. Now, for all $ f \in \CC(A, B) $, we have by the bifunctoriality of $ F $, a morphism
	\[ F(\_, \Initalg{F_B})(f): F_A(\Initalg{F_B}) = F(A, \Initalg{F_B}) \to F(B, \Initalg{F_B}) = F_B(\Initalg{F_B}). \]
	Then $ (\Initalg{F_B}, \In_B \circ F(\_, \Initalg{F_B})(f): F_A(\Initalg{F_B}) \to \Initalg{F_B}) $ is an $ F_A $-algebra, so by the initiality of $ \Initalg{F_A} $, we have a unique morphism $ f^\prime: \Initalg{F_A} \to \Initalg{F_B} $ such that 
	\[ \In_B \circ F(\_, \Initalg{F_B})(f) \circ F_A(f^\prime) = f^\prime \circ \In_A \]
	and we set $ G(f) = f^\prime $.
	
	We now check that $ G $ is a functor. To show that, we need to show that for all $ A \in \Ob \CC $, $ G(\Id[A]) = \Id[\Initalg{F_A}] $ and for all $ f \in \CC(A, B) $ and $ g \in \CC(B, C) $, $ G(g \circ f) = G(g) \circ G(f) $.

	For the first property, given an object $ A \in \Ob \CC $, $ G(\Id[A]): \Initalg{F_A} \to \Initalg{F_A} $ is the unique morphism such that
	\[ \In_A \circ F(\_, \Initalg{F_A})(\Id[A]) \circ F_A(G(\Id[A])) = G(\Id[A]) \circ \In_A. \]
	Hence, to show that $ G(\Id[A]) = \Id[\Initalg{F_A}] $ holds, it suffices to show 
	\[ \In_A \circ F(\_, \Initalg{F_A})(\Id[A]) \circ F_A(\Id[\Initalg{F_A}]) = \Id[\Initalg{F_A}] \circ \In_A. \]
	This indeed follows by the following computation:
	\begin{align*}
		\In_A \circ F(\_, \Initalg{F_A})(\Id[A]) \circ F_A(\Id[\Initalg{F_A}]) &= \In_A \circ \Id[F_A(\Initalg{F_A})] \circ \Id[F_A(\Initalg{F_A})]\\
		&= \In_A\\
		&= \Id[\Initalg{F_A}] \circ \In_A,
	\end{align*}

	For the second property, given objects $ A, B, C \in \Ob \CC $, and $ f: \CC(A, B) $ and $ g: \CC(B, C) $, we have that
	\[ \In_B \circ F(\_, \Initalg{F_B})(f) \circ F_A(G(f)) = G(f) \circ \In_A \]
	and
	\[ \In_C \circ F(\_, \Initalg{F_C})(g) \circ F_B(G(g)) = G(g) \circ \In_B, \]
	and we know that $ G(g \circ f): \Initalg{F_A} \to \Initalg{F_C} $ is the unique morphism such that
	\[ \In_C \circ F(\_, \Initalg{F_C})(g \circ f) \circ F_A(G(g \circ f)) = G(g \circ f) \circ \In_A. \]
	Note that $ F $ is a bifunctor, and therefore, the following diagram commutes:
	\begin{center}
		\begin{tikzcd}
			F(A, \Initalg{F_B}) \arrow[rr, "{F(\_, \Initalg{F_B})(f)}"] \arrow[d, "{F(A, \_)(G(g))}"'] && F(B, \Initalg{F_B}) \arrow[d, "{F(B, \_)(G(g))}"]\\
			F(A, \Initalg{F_C}) \arrow[rr, "{F(\_, \Initalg{F_C})(f)}"] && F(B, \Initalg{F_C})
		\end{tikzcd}
	\end{center}
	Then we have the following equality:
	\begin{align*}
		& \In_C \circ F(\_, \Initalg{F_C})(g \circ f) \circ F_A(G(g) \circ G(f))\\
		=& \In_C \circ F(\_, \Initalg{F_C})(g) \circ F(\_, \Initalg{F_C})(f) \circ F_A(G(g)) \circ F_A(G(f))\\
		=& \In_C \circ F(\_, \Initalg{F_C})(g) \circ F_B(G(g)) \circ F(\_, \Initalg{F_B})(f) \circ F_A(G(f))\\
		=& G(g) \circ \In_B \circ F(\_, \Initalg{F_B})(f) \circ F_A(G(f))\\
		=& G(g) \circ G(f) \circ \In_A.
	\end{align*}
	Therefore, $ G(g \circ f) = G(g) \circ G(f) $.
\end{solution}

\begin{solution}[\cref{exer:sum_plus_one_as_fold}]\label{sol:sum_plus_one_as_fold}
  We also write $\sum$ for |sum|. Note that |sum| is defined as the catamorphism $\catam{[\Zero, (+)]}$.
  The situation is summarized in the following diagram
  
  \[
    \begin{tikzcd}[row sep = huge, column sep = huge]
      \{*\} + \NN \times \List(\NN)
      \ar[d, "\Id + \Id \times \sum" description]
      \ar[r, "{[\nil, \cons]}"]
      &
      \List(\NN)
      \ar[d, "\sum" description]
      \ar[dd, bend left, "\catam{[1, (+)]}"]
      \\
      \{*\} + \NN \times \NN
      \ar[r, "{[\Zero, (+)]}"]
      \ar[d, "{\Id + \Id \times (+1)}" description]
      &
      \NN
      \ar[d, "{(+1)}" description]
      \\
      \{*\} + \NN \times \NN
      \ar[r, "{[1, (+)]}"]
      &
      \NN 
    \end{tikzcd}
  \]

  In order to show that |(+1) . sum = fold (+) 1|, in the diagram expressed as $\co{\sum}{(+1)} = \catam{[1, (+)]}$, we need to show that the lower rectangle commutes---this is what  \cref{exer:fusion-property} says.

  There are two cases to consider: the case of $*$, and the case of a pair $(m,n) \in \NN\times\NN$.
  In the case of $*$, we obtain
  \[
    (+1)~0 = 1
  \]
  and in the case of $(m,n)$, we obtain
  \[
    (+1)~ ((+)~ m~ n) = (+)~ m~ ((+1)~n)
  \]
  both of which hold.
  
\end{solution}

\begin{solution}[\cref{exer:conatural_numbers_terminal_coalgebra}] \label{sol:conatural_numbers_terminal_coalgebra}
	Let $ F: \SET \to \SET $ be the functor induced by $ X \mapsto 1 + X $ with $ 1 = \{ \star \} $.
	We claim that $ (\mathbb N^c, f) $ is a terminal coalgebra for this functor, with $ \mathbb N^c = \mathbb N + \{ \infty \} $ and $ f: \mathbb N + \{ \infty \} \to 1 + \mathbb N + \{ \infty \} $ given by
	\begin{align*}
		0 &\mapsto \star\\
		s(n) &\mapsto n\\
		\infty &\mapsto \infty
	\end{align*}
	We need to show that for all $ F $-coalgebras $ (X, g) $, there exists a unique $ F $-coalgebra morphism $ \varphi: (X, g) \to (\mathbb N^c , f) $. To this end, take an arbitrary $ F $-coalgebra $ (X, g) $.

	We first show existence. Borrowing some notation from monads, given the morphism $ g: X \to F X $, we write $ g^* $ for the induced morphism $ g^*: F X \to F X $ and we write $ (g^*)^n = g^* \circ g^* \circ \dots \circ g^* $. We define $ \varphi: X \to \mathbb N^c $ as follows: For $ x \in X $, if for some $ n \in \mathbb N $, $ (g^*)^n(x) \in X $ and $ (g^*)^{s(n)}(x) = \star $, we set $ \varphi(x) = n $. Else, we set $ \varphi(x) = \infty $. We need to show that the following diagram commutes:
	\begin{center}
		\begin{tikzcd}
			X \arrow[r, "g"] \arrow[d, "\varphi"] & 1 + X \arrow[d, "{\Id[1] + \varphi}"]\\
			\mathbb N + \{ \infty \} \arrow[r, "f"] & 1 + \mathbb N + \{ \infty \}
		\end{tikzcd}
	\end{center}
	To that end, take $ x \in X $. Suppose that $ (\Id[1] + \varphi)(g(x)) = \star $. Then $ g^1(x) = \star $, so $ \varphi(x) = 0 $ and $ f(\varphi(x)) = \star $. Suppose that $ (\Id[1] + \varphi)(g(x)) = \infty $. Then (by the definition of $ \varphi $) for all $ n \in \mathbb N $, $ (g^*)^{s(n)} = (x)(g^*)^n(g(x)) \in X $, so $ \varphi(x) = \infty $ and $ f(\varphi(x)) = \infty $. Lastly, suppose that $ (\Id[1] + \varphi)(g(x)) = n $. Then (by the definition of $ \varphi $), $ (g^*)^{s(n)}(x) = (g^*)^{n}(g(x)) \in X $ and $ (g^*)^{s(s(n))}(x) = (g^*)^{s(n)}(g(x)) = \star $. By the definition of $ \varphi $, we then have $ \varphi(x) = s(n) $. Then $ f(\varphi(x)) = n $ and we conclude that the diagram commutes and $ \varphi $ is a $ F $-coalgebra morphism.

	For uniqueness, suppose that we also have a morphism $ \psi: X \to \mathbb N^c $ that makes the diagram commute.
	
	Suppose that for some $ x \in X $, $ g(x) = \star $. Then, to make the diagram commute, we must have $ f(\psi(x)) = (\Id[1] + \psi)(\star) = \star $, so by the definition of $ f $, we must have $ \psi(x) = 0 $.

	By induction on $ n $, we will show that for all $ x \in X $, if $ (g^*)^n(x) \in X $ and $ (g^*)^{s(n)} = \star $, we have $ \psi(x) = n $ (which equals $ \varphi(x) $). Since we just showed the case for $ n = 0 $, we will now show the induction step: suppose that it holds for some $ n $. Take $ x \in X $. If $ (g^*)^n(g(x)) = (g^*)^{s(n)}(x) \in X $ and $ (g^*)^{s(n)}(g(x)) = (g^*)^{s(s(n))}(x) = \star $, we have $ \psi(g(x)) = n $. Then we have $ f(\psi(x)) = (\Id[1] + \psi)(g(x)) = n $. Therefore, $ f(\psi(x)) = s(n) $, which proves the induction step.

	Now, take $ x \in X $ and suppose that for all $ n \in \mathbb N $, $ (g^*)^n(x) \in X $. If $ \psi(x) = n $ for some $ n $, then (since the diagram commutes), we have $ \psi(g(x)) = f(\psi(x)) = f(n) = n - 1 $. Repeating this $ n $ times, we have $ (\Id[1] + \psi)((g^*)^n(x)) = 0 $. We take $ x^\prime = (g^*)^n(x) \in X $. We have $ \psi(x^\prime) = 0 $ and $ f(x^\prime) = \star $. However, $ g(x^\prime) \in X $, so $ (\Id[1] + \varphi)(g(x^\prime)) \not = \star $ and the diagram does not commute. Therefore, we cannot have $ \psi(x) \in \mathbb N $, so $ \psi(x) = \infty = \varphi(x) $.

	Therefore, $ \psi = \varphi $ and we conclude that $ \varphi $ is unique, so $ (\mathbb N^c, f) $ is a terminal $ F $-coalgebra.
\end{solution}

\begin{solution}[\cref{exer:stream-of-nats}]\label{sol:stream-of-nats}

	We wish to define $h,t : \NN\to \NN$ such that the following diagram commutes:
	\begin{center}
	\begin{tikzcd}
	\NN \arrow[rr, "{\langle h, t \rangle}"] \arrow[d, swap, "\nats"] && \NN \times \NN \arrow[d, "\Id \times \nats"] \\
	\Stream(\NN) \arrow[rr, swap, "{\langle \head, \tail \rangle}"] && {\NN\times\Stream(\NN)}
	\end{tikzcd}
	\end{center}
	
	This means:
	\begin{align}
	\co{\nats}{\head} &= h \label{eqn:streams_nats_eq1} \\
	\co{\nats}{\tail} &= \co{t}{\nats} \label{eqn:streams_nats_eq2} 
	\end{align}
	
	By definition of $\nats$, \cref{eqn:streams_nats_eq1} means
	\[
	\forall n : \NN: n = \head(\nats(n)) = h(n),
	\]
	and \cref{eqn:streams_nats_eq2} means
	\[
	\forall n : \nats(n+1) = \tail(\nats(n)) = \nats(t(n)).
	\]
	Hence, we define (as candidates)
	\[
	h = \Id[\NN], \quad t = \Succ.
	\]
	By construction, we have that \cref{eqn:streams_nats_eq1} and \cref{eqn:streams_nats_eq2} indeed holds for $h := \Id[\NN]$ and $t := \Succ$, which shows that the sought function $\nats : \NN \to \Stream(\NN)$ is the unique solution of the equation system
  \begin{align*}
    \co {\nats} \head  &= \Id
    \\
    \co {\nats} \tail &= \co \Succ {\nats}
  \end{align*}  
  
  and thus can be defined as the anamorphism $\anam{ \langle \Id, \Succ \rangle }$.
\end{solution}

\begin{solution}[\cref{exer:zip}]\label{sol:zip}

We wish to define $h : \Stream(A)\times\Stream(B)\to A\times B$ and $t: \Stream(A)\times\Stream(B)\to \Stream(A)\times \Stream(B)$ such that the following diagram commutes:
	\begin{center}
	\begin{tikzcd}
	\Stream(A)\times \Stream(B)
		\arrow[rr, "{\langle h, t \rangle}"] 
		\arrow[d, swap, "\zip"] 
	&& 
	A\times B\times \Stream(A)\times \Stream(B)
		\arrow[d, "\Id \times \zip"] \\
	\Stream(A\times B) 
		\arrow[rr, swap, "{\langle \head, \tail \rangle}"] 
	&& 
	{A \times B \times \Stream(A\times B)}
	\end{tikzcd}
	\end{center}
	
	This means:
	\begin{align}
	\co{\zip}{\head} &= h \label{eqn:streams_zip_eq1} \\
	\co{\zip}{\tail} &= \co{t}{\zip} \label{eqn:streams_zip_eq2} 
	\end{align}
	
	By definition of $\zip$, \cref{eqn:streams_zip_eq1} means
	\[
	\forall as : \Stream(A), bs : \Stream(B): (\head(as),\head(bs)) = \head(\zip(as,bs)) = h(as,bs),
	\]
	and \cref{eqn:streams_zip_eq2} means
	\[
	\forall as : \Stream(A), bs : \Stream(B): \zip (\tail(as),\tail(bs)) = \tail(\zip(as,bs)) = \zip (t(as,bs)).
	\]
	Hence, we define (as candidates)
	\[
	h = \head\times\head, \quad t = \tail\times\tail.
	\]
	By construction, we have that \cref{eqn:streams_zip_eq1} and \cref{eqn:streams_zip_eq2} indeed holds for $h := \head\times\head$ and $t := \tail\times\tail$, 

  which shows that the sought function $\zip : \Stream(A) \times \Stream(B) \to \Stream(A\times B)$ is the unique solution to the equation system
  \begin{align*}
    \co \zip \head = \head\times\head
    \\
    \co \zip \tail = \co{(\tail\times\tail)}{\zip}
  \end{align*}
  This function can, therefore, be defined as $\anam{ {\langle \head \times \head, \tail \times \tail \rangle} }$.
\end{solution}

\begin{solution}[\cref{exer:forgetful-functor-is-representable}]\label{sol:forgetful-functor-is-representable}
Observe that the representability of $U$ is a special case of the furthermore statement, indeed: every set $A$ is isomorphic to the set $\bullet \to A$ of functions from the singleton/unit set into $A$.

Fix a set $A$.
We have to construct a monoid $N$ such that $\MON(N,-) \cong \SET(A,U-)$.
The claim is that $N$ is the free monoid $\List(A)$ generated by $A$.

First, we construct a pointwise isomorphism between $\MON(N,-)$ and $\SET(A,U-)$:
\[
\phi_M : \MON(\List(A), M) \to \SET(A, UM), \quad \forall M \in \MON.
\]
This function is given by precomposing with $\eta_A : A \to List(A)$ and forgetting the monoid structure, i.e., $\phi_M(g) := g \circ \eta_A$.

To show that $\phi_M$ is invertible, we can either construct an inverse $\psi_M : \SET(A, UM) \to \MON(\List(A), M)$, or we can show that $\phi_M$ is injective and surjective.

\textbf{Injectivity}. Let $g, h \in \List(A)$ and assume $\phi_M(g) = \phi_M(h)$.
That $g = h$ follows from the following:
\begin{eqnarray*}
\phi_M(g) = \phi_M(h) &\Leftrightarrow& g \circ \eta_A = h \circ \eta_A \\
&\Leftrightarrow& \forall a \in A, g(\eta(a)) = h(\eta(a)) \\
&\Leftrightarrow& \forall a \in A, g([a]) = h([a])\\
&\Leftrightarrow& \forall l \in \List(A), g(l) = h(l)\\
&\Leftrightarrow& g = h.
\end{eqnarray*}

\textbf{Surjectivity}.
The surjectivity follows from the universal property of $\List(A)$ as the free monoid (see \cref{ch:forgetful-free}).
Let $g' \in \SET(A, UM)$.
We have to construct a $g \in \MON(\List(A), M)$ such that $g' = \phi_M(g) := g \circ \eta_A$, i.e., $\forall a \in A, g'(a) = g([a])$.
Hence, $g$ acts on singleton lists as $g'$ which uniquely determines a function from $\List(A)$ to $M$.
Indeed, by case analysis (recursion), we can define $g'(\nil)$ as the unit element in $M$ and $g'([a::l])$ as $[g(a) :: g'(l)]$.
Observe that $g'$ is indeed a monoid homomorphism.

It remains to verify the naturality of $\phi_M$ in $M$, i.e., that for every $f \in \MON(M_0,M_1)$ the following diagram commutes:
\begin{center}
\begin{tikzcd}
{\MON(\List(A), M_0)} \arrow[r, "\phi_{M_0}"] \arrow[d,swap, "f \circ -"] & 
{\SET(A, UM_0)} \arrow[d, "U(f) \circ -"]\\
{\MON(\List(A), M_1)} \arrow[r,swap, "\phi_{M_1}"] & {\SET(A, UM_1)}
\end{tikzcd}
\end{center}
Let $g \in \MON(\List(A), M_0)$ and $f \in \MON(M_0,M_1)$.
That the diagram commutes follows because both morphisms unfold to $f \circ g \circ \eta_A$.
\end{solution}

\begin{solution}[\cref{exer:representables-on-posets}]
\label{sol:representables-on-posets}
Recall that a downward closed subset of $L$ is given by a subset $S$ of $L$ which is closed under the following property:
\[
\forall x \in S, \forall a \in L, a \leq x \Rightarrow a \in S.
\]
We claim that the representable functors correspond to the downward closed subsets of $\mathcal{L}$ with a maximal element.

Let $F : \op{\mathcal{L}} \to \SET$ be a representable functor.
Then, $F \cong \mathcal{L}(-,x)$ for some $x \in L$.
By definition of $\mathcal{L}$, we have that $F(x)$ is the (po)set of elements in $\mathcal{L}$ which are less then $x$, i.e., $F(x) \cong \{ a \in L \mid a \leq x \}$.
Observe that $F(x)$ contains $x$ as a maximal element.
Conversely, if $D$ is a downward closed subset with a maximal element $x$, we obtain a representable functor $F := \mathcal{L}(-,x)$.
\end{solution}


\chapter{Forgetful and Free Functors}
\label{ch:forgetful-free}

A lot of (mathematical) structures are defined as some other kind of mathematical structure, but where extra structure is added. An example of this is the following:\\
Recall that a monoid is a set $M$ together with a binary operation $m:M\to M\to M$ which is associative and such that there is an identity element $e$ (see \cref{monoidcategory}). In particular, any monoid has an underlying set and any morphism of monoids has an underlying function (between those sets). So forgetting the binary operation and identity element defines a functor from $\MON$ to $\SET$ which is called a \textit{forgetful functor}:
\begin{exa}\label{example:forgetful_montoset} The \textbf{forgetful functor from $\MON$ to $\SET$} is the functor specified by the following data:
\begin{itemize}
\item The function on objects is given by 
\[
\Ob{\MON}\to \Ob{\SET}: (M,m,e)\mapsto M.
\]
\item The function on objects is given by
\[
\CHom{\MON}{(M_1,m_1,e_1)}{(M_2,m_2,e_2)} \to \CHom{\SET}{M_1}{M_2} : f\mapsto f.
\]
\end{itemize}
\end{exa}
Notice that if one defines a category whose objects are sets $M$ together with an associative binary operation $m:M\to M\to M$, then one could analogously also define a forgetful functor from $\MON$ to this category by only forgetting the neutral element.

\begin{lemma} The forgetful functor from $\MON$ to $\SET$ satisfies the properties of a functor.
\begin{proof}
We clearly have that everything is well-defined since the codomain is $\SET$.\\
That the identity morphism is preserved holds by definition because the identity morphism of $(M,m,e)$ (in $\MON$) is given by the identity function and the identity morphism of $M$ (in $\SET$) is also given by the identity function.\\
That the composition of morphisms is preserved also holds by definition because the composition of morphisms (in $\MON$) is given by the composition of the underlying functions which is also the composition in $\SET$.
\end{proof}
\end{lemma}

The forgetful functor $Forget$ from $\MON$ to $\SET$ forgets the \textit{algebraic} structure of a monoid and since there are multiple monoid structures on the same set (given an example of this), hence we do not have that there exists some functor $G: \SET\to \MON$ such that $Forget\Comp G$ is the identity on $\MON$. However, to each set, one can define a monoid which satisfies an important property (this is \cref{prop:UVP_forget_montoset}). The associated monoid is called the \textit{free monoid}:
\begin{exa} Let $X$ be a set. The \textbf{free monoid generated by $X$}, denoted by $Free(X)$, is specified by the following data:
\begin{itemize}
\item The underlying set consists of all finite sequences/strings of elements in $X$ (including the empty sequence).
\item The multiplication is defined by concatenating the sequences, i.e. 
\[
m\left((x_1,\cdots,x_n),(y_1,\cdots,y_m)\right) := (x_1,\cdots,x_n,y_1,\cdots,y_m).
\]
\item The identity element is given by the empty sequence.
\end{itemize}
\end{exa}

\begin{exa}\label{exa:freemonoids} The \textbf{free functor from $\SET$ to $\MON$} is specified by the following data:
\begin{itemize}
\item The function on objects is given by 
\[
\Ob{\SET}\to \Ob{\MON}: X\mapsto Free(X).
\]
\item The function on morphisms is given as follows: A morphism $f \in \CHom{\SET}{X}{Y}$ (i.e. a function) is mapped to the monoidal morphism which is given by pointwise application of $f$, i.e.
\[
Free(f)(x_1,\cdots,x_n) := (f(x_1),\cdots, f(x_n)).
\]
\end{itemize}
\end{exa}

\begin{exer} Show that $Free$ satisfies the properties of a functor.
\end{exer}

For any set $X$, we have the \textit{canonical function} 
\[
Free^{X}_{!}: X\to Free(X): x\mapsto (x).
\]
This function satisfies the property that any function from $X$ to an arbitrary monoid corresponds with a unique morphism (of monoids) from the free monoid generated by $X$ to that monoid:
\begin{prop}\label{prop:UVP_forget_montoset} Let $(M,m,e)$ be a monoid and $X$ be a set. For any morphism $f \in \CHom{\SET}{X}{M}$ (i.e. a function), there exists a unique  morphism $\phi^{f} \in \CHom{\MON}{Free(X)}{(M,m,e)}$, such that $f = Free^{X}_{!}\Comp \phi^{f}$.
\begin{proof}
For elements $a,b\in M$, we denote their multiplication by $a\times b := m(a,b)$ (note that by associativity we have that $a_1\times a_2\times\cdots\times a_n)$ is well-defined). Let $f\in \CHom{\SET}{X}{M}$ be a function. Define 
\[
\phi^{f}: Free(X)\to (M,m,e): (x_1\cdots,x_n)\mapsto f(x_1) \times\cdots \times f(x_n),
\] 
We have to define separately what happens with the empty sequence. The empty sequence we map to the identity element $e$, so in particular we have that the identity element is preserved under $\phi^{f}$, so in order to conclude that $\phi^{f}$ is a morphism of monoids, it remains to show that it preserves the binary operation but this is clear by definition.\\
That $f = Free^{X}_{!}\Comp \phi^{f}$ holds follows immediately by the definition of $Free^{X}_{!}$ and $\phi^{f}$, indeed:
\[
\left(Free^{X}_{!}\Comp \phi^{f}\right)(x) = \phi^{f}\left((x)\right) = f(x).
\]
So it only remains to show uniqueness. Assume that $\psi$ also satisfies $Free^{X}_{!}\Comp \psi = f$. The claim now follows because $\psi$ is a morphism of monoids, indeed: Since $\psi$ is morphism is monoids, we have that it preserves the multiplication, but the multiplication is given by concatenation, hence, we have that $\psi$ is uniquely determined by the images of the sequences of length $1$ (and length $0$, but this sequence of length $0$ should be mapped under $\psi$ to $e$). But a sequence of length $1$ is of the form $(x) = Free^{X}_{!}(x)$. So the claim indeed follows by the following computation: 
\[
\psi((x)) = \psi(Free^{X}_{!}(x)) = f(x) = \phi^{f}(Free^{X}_{!}(x)) = \phi^{f}((x)).
\]
\end{proof}
\end{prop}

The following exercise shows that there is a special connection between the forgetful functor $\mathit{forget}: \MON\to\SET$ and the free functor $\mathit{free}: \SET\to\MON$. This connection expresses that these form a so-called \textit{adjoint pair} (see \cref{sec:adjunctions}).
\begin{exer}\label{exer:preadjunction_monset}
Show that for any set $X$ and monoid $(M,m,e)$, there exist bijections between the hom-sets:
\[
\alpha_{X}^{(M,m,e)} : \CHom{\MON}{Free(X)}{(M,m,e)} \to \CHom{\SET}{X}{forget(M,m,e)}.
\]
Hint: use \cref{prop:UVP_forget_montoset}.
\end{exer}

\begin{exer} Define a forgetful functor from the category $\POS$ of posets (defined in \cref{example:poset}) to $\SET$ analogous to the the forgetful functor from $\MON$ to $\SET$ (defined in \cref{example:forgetful_montoset}).
\end{exer}
\begin{rem} The story about free monoids can not be repeated for posets, i.e. there is no free poset structure on all sets. But in order to prove this one needs more machinery.
\end{rem}


\chapter{Contravariant Functors}

A variation on functors are \textbf{contravariant functors}.
A contravariant functor consists of a map on objects, just like a functor.
However, the \textbf{map on morphisms turns the morphisms around}.
We give the formal definition:

\begin{dfn} Let $\CC$ and $\DD$ be categories. A \textbf{contravariant functor} $F$ from $\CC$ to $\DD$ consists of the following data:
\begin{itemize}
\item A function 
\[
\Ob{\CC} \to \Ob{\DD},
\]
written as $X\mapsto F(X)$.
\item For each $X,Y\in \Ob{\CC}$, a function
\[
\CHom{\CC}{X}{Y} \to \CHom{\DD}{F(Y)}{F(X)},
\]
written as $f\mapsto F(f)$.
\end{itemize}
Moreover, this data should satisfy the following properties:
\begin{itemize}
\item (\textbf{Preserves composition}) For $f\in \Hom[\CC]{X}{Y}$ and $g\in \Hom[\CC]{Y}{Z}$, we have $F(\co f g) =  \co {F(g)}{F(f)}$.
\item (\textbf{Preserves identity}) For $X\in\CC$, we have $F(\Id[X]) = \Id[F(X)]$.
\end{itemize}
\end{dfn}

\begin{exer} Notice that the preservation of composition has now changed; why is this the case?
\end{exer}

An example of a contravariant functor is given by the powerset-functor:
\begin{exa} \label{example:powersetfunctor} Recall that the powerset of a set $X$, denoted by $\mathbb{P}(X)$, is the set of all subsets of $X$, i.e. 
\[
\mathbb{P}(X) :=  \left\{A \mid A\subseteq X\right\}.
\]
The contravariant \textbf{powerset-functor}\footnote{Since a function is mapped to the inverse-image function, one also calls the powerset-functor, an inverse image-functor} (on sets), denoted by $\mathbb{P}$, is the functor from $\SET$ to $\SET$ defined by the following data:
\begin{itemize}
\item The function on objects is given by:
\[
\Ob{\SET}\to \Ob{\SET}: X\mapsto \mathbb{P}(X).
\]
\item For each $X,Y\in\SET$, the function on morphisms is given by
\[
\CHom{\SET}{X}{Y} \to \CHom{\SET}{\mathbb{P}(Y)}{\mathbb{P}(X)}: f\mapsto f^{-1},
\]
where $f^{-1}$ given by
\[
f^{-1}:\mathbb{P}(Y)\to \mathbb{P}(X): B\mapsto f^{-1}(B) := \left\{x\in X \mid f(x)\in B\right\}.
\]
\end{itemize}
\end{exa}

\begin{exer} Show that $\mathbb{P}$, defined in \cref{example:powersetfunctor} satisfies the properties of a contravariant functor.
\end{exer}


\printbibliography

\end{document}